\documentclass[12pt]{iopart}
\usepackage{amssymb,enumerate}
\usepackage{multirow}
\usepackage{bm}
\usepackage{epsfig}
\usepackage{subfigure}
\usepackage{color}
\usepackage{multicol}        
\usepackage{url}
\usepackage{cite} 
\usepackage{cleveref}
\usepackage{enumerate}

\usepackage{iopams}  
\begin{document}

\title[Identifying long-term precursor]{Identifying long-term precursors of financial market crashes using correlation patterns}

\author{Hirdesh K. Pharasi$^1$,  Kiran Sharma$^2$, Rakesh Chatterjee$^{1,3}$, Anirban Chakraborti$^2$, Francois Leyvraz$^{1,4}$, and Thomas H. Seligman$^{1,4}$}

\address{$^1$ Instituto de Ciencias F\'{i}sicas, Universidad Nacional Aut\'{o}noma de M\'{e}xico, Cuernavaca-62210, M\'{e}xico}
\address{$^2$ School of Computational and Integrative Sciences, Jawaharlal Nehru University, New Delhi-110067, India}
\address{$^3$ School of Mechanical Engineering and Sackler Center for Computational Molecular and Materials Science, Tel Aviv University, Tel Aviv-6997801, Israel}
\address{$^4$ Centro Internacional de Ciencias, Cuernavaca-62210, M\'{e}xico}
\ead{hirdeshpharasi@gmail.com; anirban@jnu.ac.in}

\vspace{10pt}
\begin{indented}
\item[]\today
\end{indented}
\begin{abstract}
The study of the critical dynamics in complex systems is always interesting yet challenging. Here, we choose financial market as an example of a complex system, and do a comparative analyses of two stock markets -- the S\&P 500 (USA) and Nikkei 225 (JPN). Our analyses are based on the evolution of cross-correlation structure patterns of short time-epochs for a 32-year period (1985-2016). We identify ``market states'' as clusters of similar correlation structures, which occur more frequently than by pure chance (randomness). The dynamical transitions between the correlation structures reflect the evolution of the market states. Power mapping method from the random matrix theory is used to suppress the noise on correlation patterns, and an adaptation of the intra-cluster distance method is used to obtain the ``optimum'' number of market states. We find that the USA is characterized by four market states and JPN by five. We further analyze the co-occurrence of paired market states; the probability of remaining in the same state is much higher than the transition to a different state. The transitions to other states mainly occur among the immediately adjacent states, with a few rare intermittent transitions to the remote states. The state adjacent to the critical state (market crash)  may serve as an indicator or a ``precursor'' for the critical state and this novel method of identifying the long-term precursors may be very helpful for constructing the early warning system in financial markets, as well as in other complex systems.
\end{abstract}

\vspace{2pc}
\section{Introduction}
A financial market is a highly complex and continuously evolving system~\cite{Vemuri_1978, Gellmann_1995, Yaneer_2002}. To understand the statistical behavior of the financial market and its constituent sectors~\cite{Mantegna_2007,Bouchaud_2003,Sinha_2010,Chakraborti_2011a,Chakraborti_2011b,Chakraborti_2015}, researchers focused their attention on the information of co-movements and correlations among the stocks of the market. 
It is well known that the mean correlation among the stocks assumes much higher values during market crashes than in normal business periods \cite{Chakraborti_2018}. Similarly, certain correlation structures seem to occur more frequently than by pure chance (randomness), specially when markets approach a critical period or crash~\cite{Sornette_2004,Buchanan_2000}. However, to identify such similar (clusters)  correlation patterns, referred as ``market states'', as was previously attempted by Munnix et al. \cite{Munnix_2012,Desislava_2014}, is rather challenging due to many factors. The first factor is that financial time series is non-stationary; second factor is that there is always noise present in the correlations computed over finite length time series data \cite{Pharasi_2018}, and it is essential to suppress the corresponding noise in correlation matrices to reveal the actual
correlations. To tackle the first factor of non-stationarity, we work with short time series so that the number of time steps over which we compute the correlations can be considered as reasonably stationary. However, with short time series the correlation matrices become highly singular \cite{Schafer_2013,Laloux_1999,Plerou_1999}. To tackle the second factor of noise-reduction, various techniques \cite{Guhr_2003,Bouchaud_2000} are available. Here, we shall use a recent and efficient one, namely the power map method \cite{Guhr_2003,Schmitt_2016,vinayak_2014}, for noise reduction as well as breaking the degeneracy in the eigenvalues so that the correlation matrices are no longer singular. 
Furthermore, the problem of finding similar clusters (groups) of the correlation patterns is a daunting task by itself.  To go beyond the simple quantification of financial market states in terms of the average correlation, clustering techniques seem promising as does the study of eigenvalues of the correlation matrix of the corresponding time series \cite{Pharasi_2018}. In the research of clustering, the $k$-means method has had some success for top-to-down clustering, but it suffers from one major drawback: the number of clusters and thus the number of states is largely arbitrary (or \textit{ad hoc}). Earlier, Munnix et al. \cite{Munnix_2012} had provided a scheme where all the correlation frames at different time-epochs were initially regarded as a single cluster and then divided into sub-clusters by a procedure based on the $k$-means algorithm. They stopped the division process when the average distance from each cluster center to its members became smaller than a certain threshold. Based on the top-to-down hierarchical clustering method and the threshold at 0.1465, which represented the best ratio of the distances between clusters and their intrinsic radii, Munnix et al. had determined the number of markets states for USA to be eight.
In the present paper, for determining the ``optimal'' number of clusters, we use multidimensional scaling (MDS) technique~\cite{borg_1997} with two/three-dimensional representations, which are comparatively easier for visualization and studying time-evolution. So, using multidimensional scaling map, we apply $k$-means clustering to divide the clusters of similar correlation patterns into $k$ groups. We propose a new way, based on the cluster radii, of estimating the number of clusters $k$, which is fairly robust and stable. We thus have a considerable degree of confidence in determining the ``optimal'' number of market states identified by the new prescription. For our research, we have used adjusted closure price data from Yahoo finance~\cite{Yahoo_finance} for the S\&P 500 (USA) and Nikkei 225 (JPN) stock exchanges, for the 32-year period (1985-2016). The stock list has been filtered such that we have stocks which were included in the market index for the entire period of 32 years. Among others, our main finding is that there exist four market states in USA and five in JPN. We then study the dynamical transitions between the market states, in a probabilistic manner; we also analyze the co-occurrence of paired market states and find that the probability of remaining in the same state is much higher than jumping to another state. The transitions mainly occur among adjacent states, with a few rare intermittent transitions to the remote states. The state adjacent to the critical state may indicate a ``precursor'' to the critical state (market crash) and this novel method of identifying the long-term precursors may be very helpful for constructing the early warning system in financial markets, and in other complex systems.

The paper is organized as follows: We present briefly the methodology and the data description. Then we present the main part of data analyses along with the above mentioned findings. Finally, we present summary and concluding remarks.

\section{Data Description, Methodology and Results}

\subsection{Data description}
\label{Sec:Materials}
We have used the database of Yahoo finance \cite{Yahoo_finance}, for the time series of adjusted closure price for two countries: United States of America (USA) S\&P 500 index and Japan (JPN) Nikkei 225 index, for the period 02-01-1985 to 30-12-2016, and for the corresponding stocks as follows:
\begin{itemize}
\item USA --- 02-Jan-1985 to 30-Dec-2016 ($T=8068$ days); Number of stocks $N=194$;
\item JPN --- 04-Jan-1985 to 30-Dec-2016 ($T=7998$ days); Number of stocks $N=165$,
\end{itemize}
\noindent where we have included the stocks which are present in the indices for the entire duration. The sectoral abbreviations are given in Table~\ref{Table:sectoral_index_sp500}.

\begin{table}[h]
\centering
\caption{Abbreviations of different sectors for S\&P 500 and Nikkei 225 markets\\}
\label{Table:sectoral_index_sp500}
\begin{tabular}{|l|l|l|l|}
\hline
\textbf{Labels}  & \textbf{Sectors}        & \textbf{Labels} & \textbf{Sectors}  \\ \hline
\textbf{CD}      & Consumer Discretionary  & \textbf{ID}     & Industrials       \\ \hline
\textbf{CS}      & Consumer Staples        & \textbf{IT}     & Information Technology\\ \hline
\textbf{CP}      & Capital  Goods          & \textbf{MT}     & Materials              \\ \hline
\textbf{CN}      & Consumer Goods          & \textbf{PR}     & Pharmacuticles          \\ \hline
\textbf{EG}      & Energy                  & \textbf{TC}     & Technology              \\ \hline
\textbf{FN}      & Financials              & \textbf{UT}     & Utilities             \\ \hline
\textbf{HC}      & Health Care             &   &            \\ \hline
\end{tabular}
\end{table}
The list of stocks (along with the sectors) for the two markets are given in the Tables S1 and S2 in Supplementary Information.

\subsection{Cross-correlation matrix and power mapping method}
\begin{figure}[]
\centering
\includegraphics[width=0.45\linewidth]{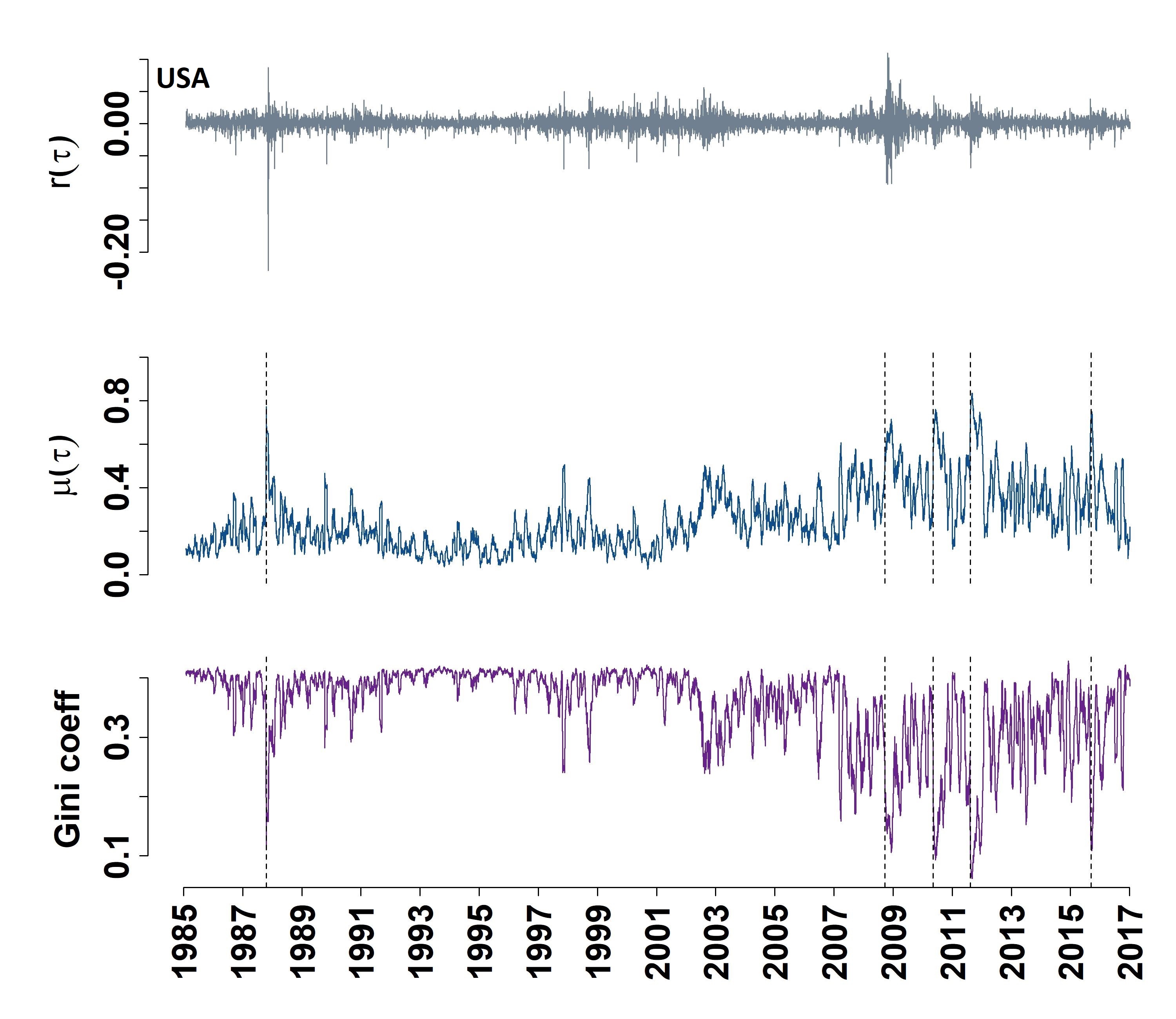}\llap{\parbox[b]{2.8in}{(\textbf{a})\\\rule{0ex}{2.2in}}}
\includegraphics[width=0.45\linewidth]{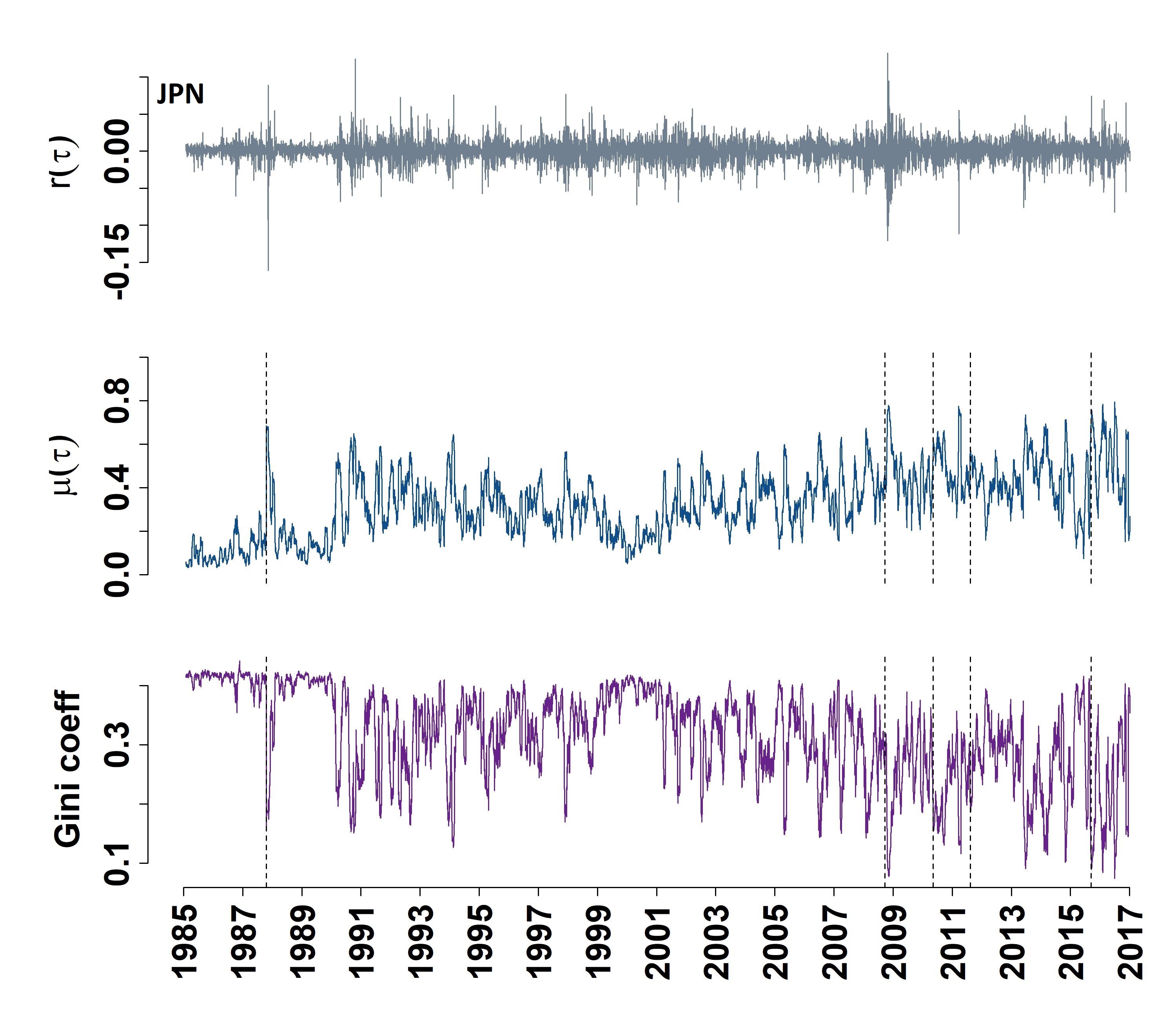}\llap{\parbox[b]{2.8in}{(\textbf{b})\\\rule{0ex}{2.2in}}}
\caption{\textbf{Results of market evolution for (a) USA and (b) JPN, respectively.} The top row shows the returns of the respective market indices. The middle row shows the mean market correlation (averaged over all the cross-correlation coefficients) of the respective markets. The bottom row shows the inequality in the distribution of the cross-correlation coefficients, as characterized by the Gini coefficient. Evidently, whenever there is a market crash, the mean market correlation becomes very high and the Gini coefficient becomes very low, indicating that all the stocks behave very similarly.}
\label{fig:corr_gini}
\end{figure}

We present a study of time evolution of the cross-correlation structures of return time series for $N$ stocks, and determination of the optimal number of market states (correlation patterns that exist more frequently then by pure chance or randomness); also, the dynamical evolution of the market states over different time-epochs. The daily return time series is constructed as $r_k(t)=\ln P_k(t)-\ln P_k(t -1)$ , where $P_k(t)$ is the adjusted closing price of the $k$-th stock at time $t$ (trading day). Then, the cross-correlation matrix is constructed using equal-time Pearson cross-correlation coefficients, $ C_{ij}(\tau) = (\langle r_i r_j \rangle - \langle r_i \rangle \langle r_j \rangle)/\sigma_i\sigma_j$, where $i,j=1, \dots, N$ and $\tau$ indicates the end date of the time-epoch of size $M$ days. Here, we computed daily return cross-correlation matrix $\boldsymbol{C}(\tau)$ computed over the short time-epoch of $M=20$ days, for (a) USA with $N=194$ stocks of S\&P 500 for a return series of $T=8060$ days, and (b) JPN with $N=165$ stocks of Nikkei 225 for $T=7990$ days, during the calendar period  1985-2016. We use time-epochs of 20 days, such that there is a  balance between choosing short time-epochs for detecting changes and long ones for reducing fluctuations.
In figure~\ref{fig:corr_gini}, we show the time evolution of the return of the market index, $r(\tau)$, along with the mean market correlation (average of all the elements of the cross-correlation matrix), $\mu(\tau)$, and the Gini coefficient that characterizes the inequality in the distribution of the correlation coefficients. Evidently, whenever there is a market crash (fall in the $r(\tau)$), the mean market correlation $\mu(\tau)$ rises a lot, and the Gini coefficient falls drastically, indicating that market is extremely correlated and all the stocks behave similarly (see Ref.~\cite{Chakraborti_2018}).
Since the assumption of stationarity manifestly fails for longer return time series, it is often useful to break the long time series of length $T$, into shorter $n$ time-epochs of size $M$ (such that $T/M=n$). The assumption of stationarity then improves for the shorter time-epochs used. However, if there are $N$ return time series such that $N > M$, then this implies that the correlation matrices  are highly singular with $N-M+1$ zero eigenvalues, leading to poor statistics. As mentioned in the introduction, we thus use the power map technique \cite{Guhr_2003,Schmitt_2016,vinayak_2014} to suppress the noise present in the correlation structure of short time series.  In this method, a non-linear distortion is given to each cross-correlation coefficient within an epoch by: $ C_{ij} \rightarrow (\mathrm{sign} ~~C_{ij}) |C_{ij}|^{1+\epsilon}$, where $\epsilon$ is the noise-suppression parameter. This also gives rise to an ``emerging spectrum'' of eigenvalues, arising from the breaking of the degeneracy of the zero eigenvalues (see Ref.~\cite{Pharasi_2018} for a recent review).

\subsection{Noise-suppression in a short time cross-correlation frame}
First, we study the effect of noise-suppression parameter $\epsilon$ on the cross-correlation matrix and its eigenvalue spectrum within a time-epoch.  The cross-correlation structure can be visualized easily through a two/three dimensional map of coordinates generated through a multidimensional scaling algorithm. The MDS is a tool of non-linear dimensional reduction to visualize the similarity of the data set in a $D$-dimensional space. Each object is assigned to a coordinate space in $D$-dimensional space keeping the between-object distance preserved, as close as possible. The choice of $D=2$ or $D=3$ is for optimizing the object location to two/three-dimensional scatter plot or map. As an input to the MDS algorithm, we provide the distance matrix~\cite{Mantegna_1999}, generated from the correlation matrix, using the non-linear transformation:
$$d_{ij}=\sqrt{2(1-C_{ij})}.$$
The effect of the variation of the parameter $\epsilon$ on noise reduction and determining the optimal number of market states, can thus be better captured through the MDS. The question is \textit{what should be the ideal choice of the noise-suppression parameter $\epsilon$}? A very small value of $\epsilon$, say $\epsilon=0.01$, surely breaks the degeneracy of eigenvalues (giving rise to an ``emerging spectrum'' with interesting properties  \cite{Chakraborti_2018}) but does not contribute much to noise-suppression. On the other hand, a large value, say $\epsilon=0.5$, suppresses the noise in the correlation pattern and helps in clustering better way; however, the emerging spectrum approaches towards the main Mar\u{c}enko-Pastur distribution \cite{Marcenko_1967}. In this paper, we are more interested in noise-suppression in the cross-correlation matrix within a single time-epoch rather than properties of the emerging spectrum; hence,  we use $\epsilon=0.6$ and this choice of a high value is based on the robustness and finding distinct clusters of stocks using MDS. The effect can be clearly seen through the supplementary figures S2 and S3. Further, our main aim is to find the optimal number of market states, based on correlation structures which are similar and appear more frequently. Hence, we formulate a similarity measure between different cross-correlation matrices at different time-epochs $\tau$, and then find similar groups of correlation frames across different time-epochs. We find that with $\epsilon=0.6$, the noise suppressed cross-correlation structures can be grouped well into similar clusters, as we will describe below. However, we find that the number of market states  is not very sensitive to the noise-suppression parameter. A higher value of $\epsilon$  lowers the mean of the cross-correlation coefficients, $\mu$ (see supplementary figure S1) and the maximum eigenvalue $\lambda_{max}$ of the cross-correlation matrix.

\begin{figure}[t!]
\centering
\includegraphics[width=0.292\linewidth]{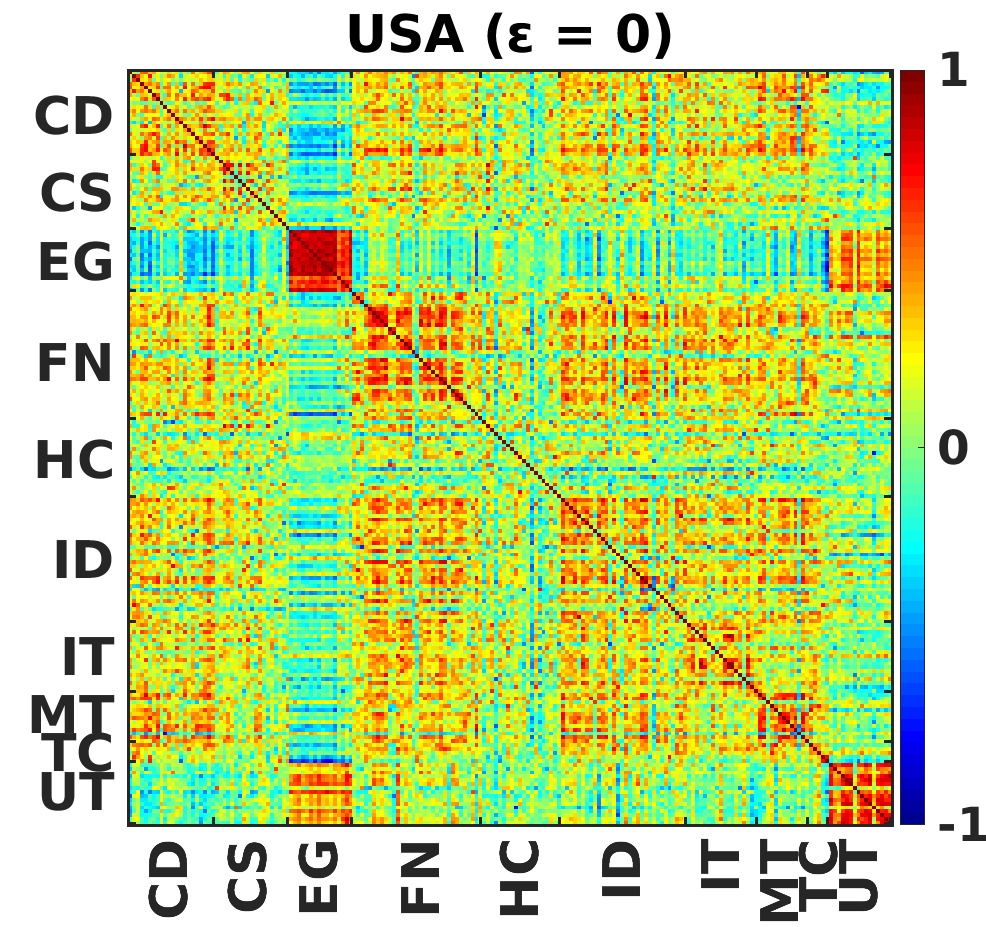}\llap{\parbox[b]{1.9in}{(\textbf{a})\\\rule{0ex}{1.5in}}}
\includegraphics[width=0.3\linewidth]{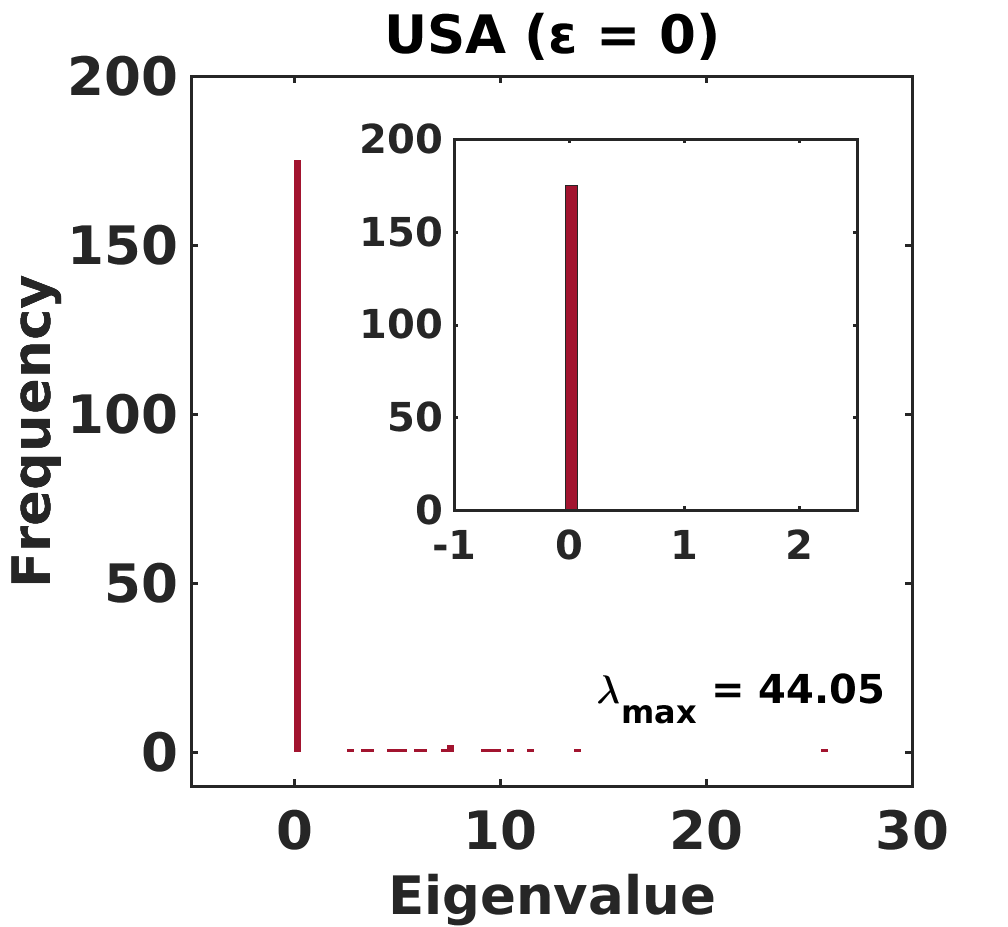}\llap{\parbox[b]{1.9in}{(\textbf{b})\\\rule{0ex}{1.55in}}}
\includegraphics[width=0.3\linewidth]{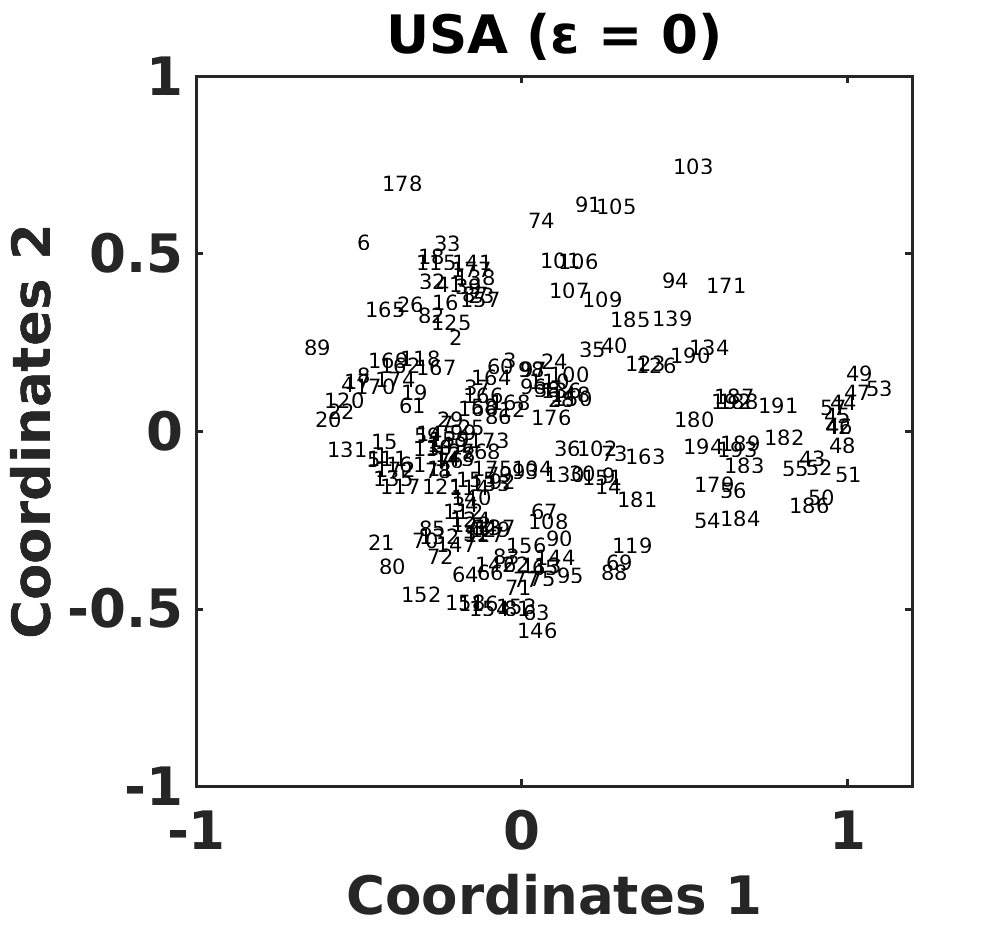}\llap{\parbox[b]{1.9in}{(\textbf{c})\\\rule{0ex}{1.5in}}}\\
\includegraphics[width=0.292\linewidth]{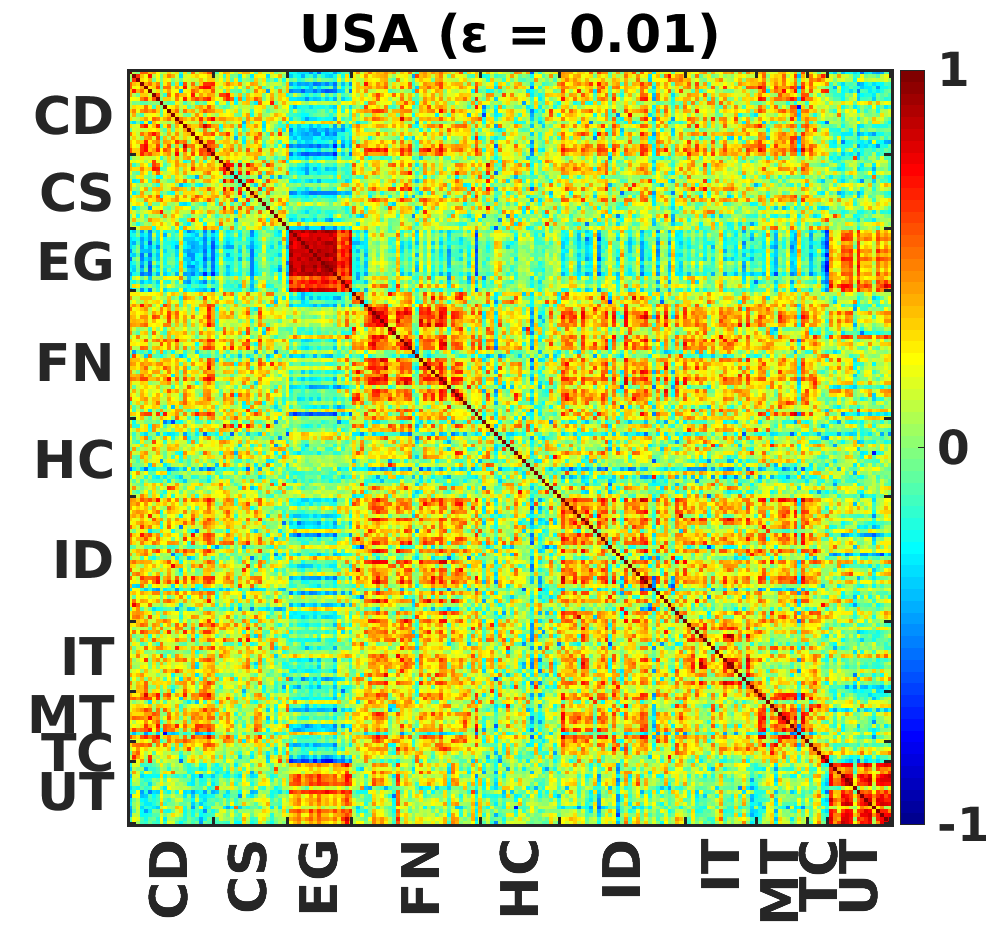}\llap{\parbox[b]{1.9in}{(\textbf{d})\\\rule{0ex}{1.5in}}}
\includegraphics[width=0.3\linewidth]{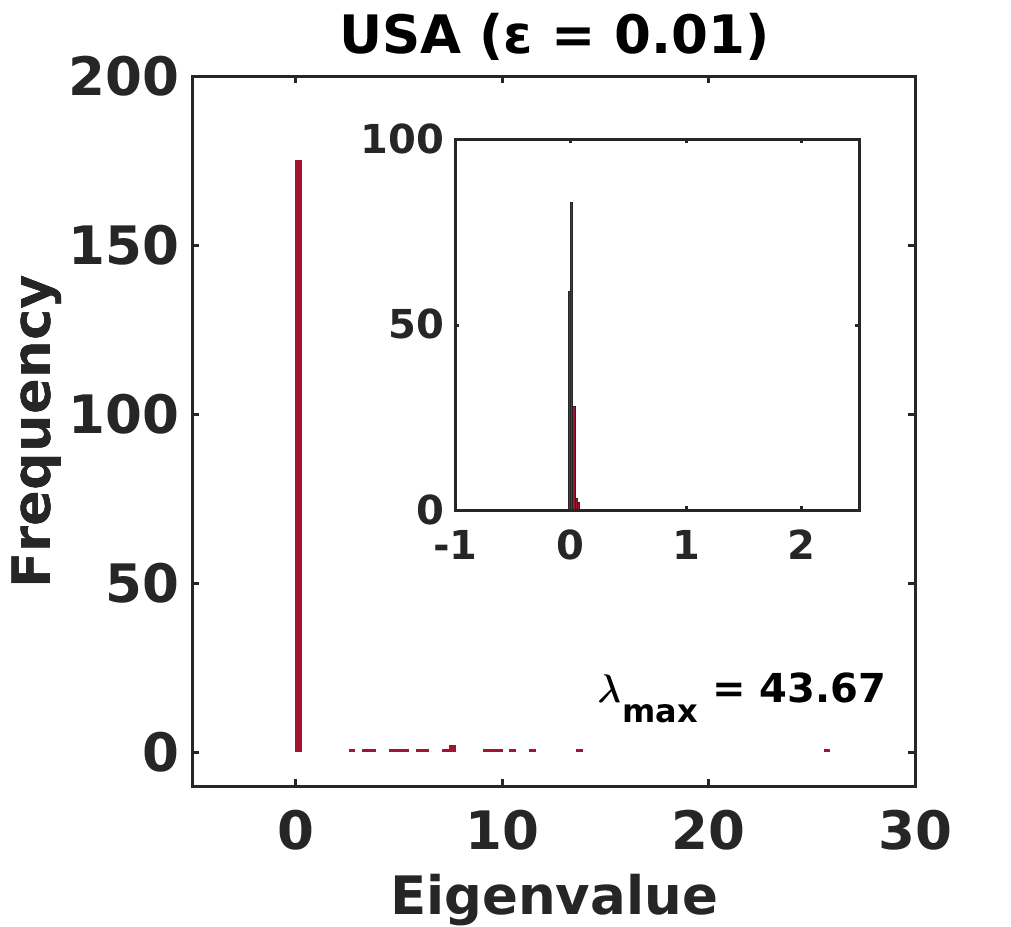}\llap{\parbox[b]{1.9in}{(\textbf{e})\\\rule{0ex}{1.55in}}} 
\includegraphics[width=0.3\linewidth]{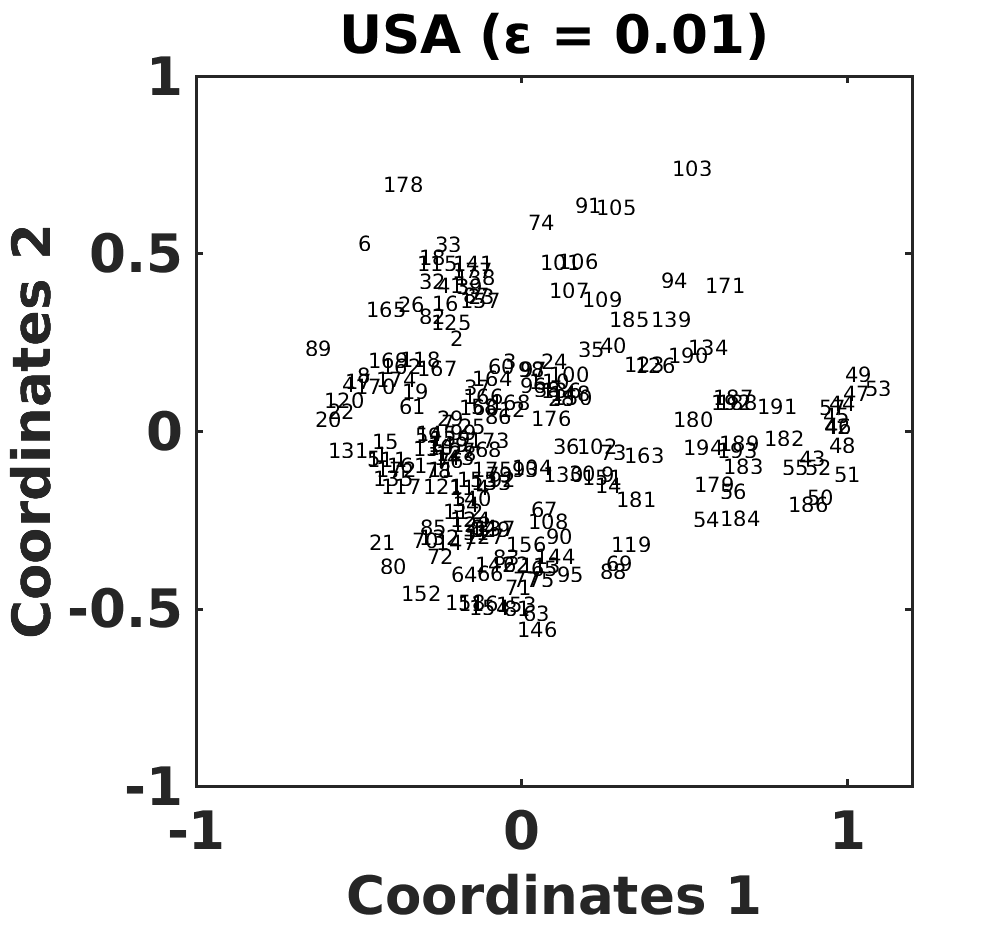}\llap{\parbox[b]{1.9in}{(\textbf{f})\\\rule{0ex}{1.5in}}}\\
\includegraphics[width=0.292\linewidth]{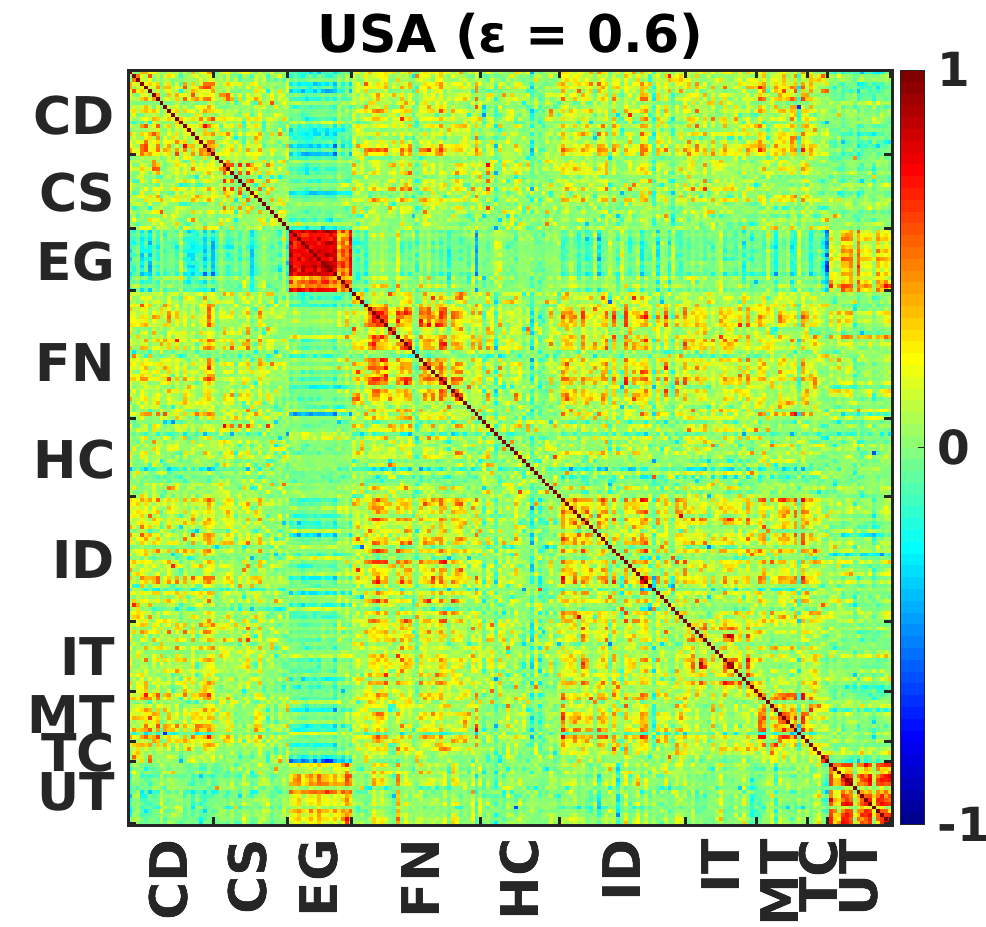}\llap{\parbox[b]{1.9in}{(\textbf{g})\\\rule{0ex}{1.5in}}}
\includegraphics[width=0.3\linewidth]{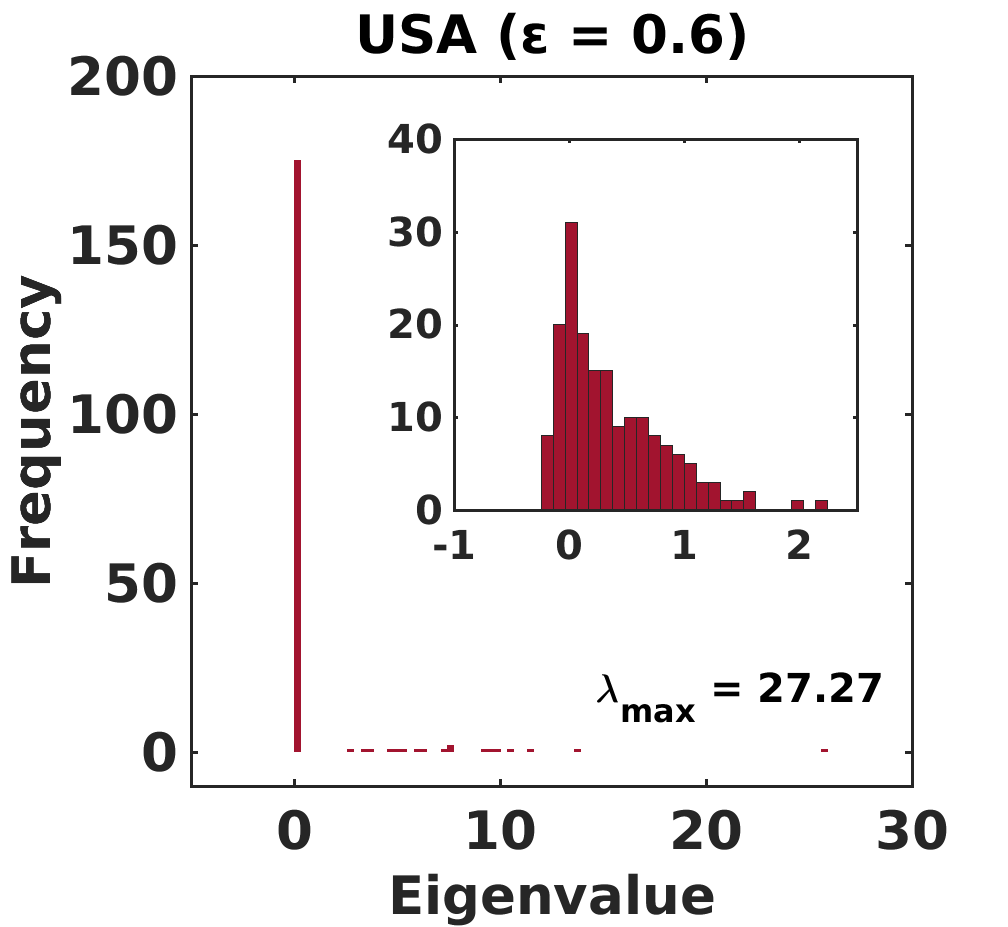}\llap{\parbox[b]{1.9in}{(\textbf{h})\\\rule{0ex}{1.55in}}}
\includegraphics[width=0.3\linewidth]{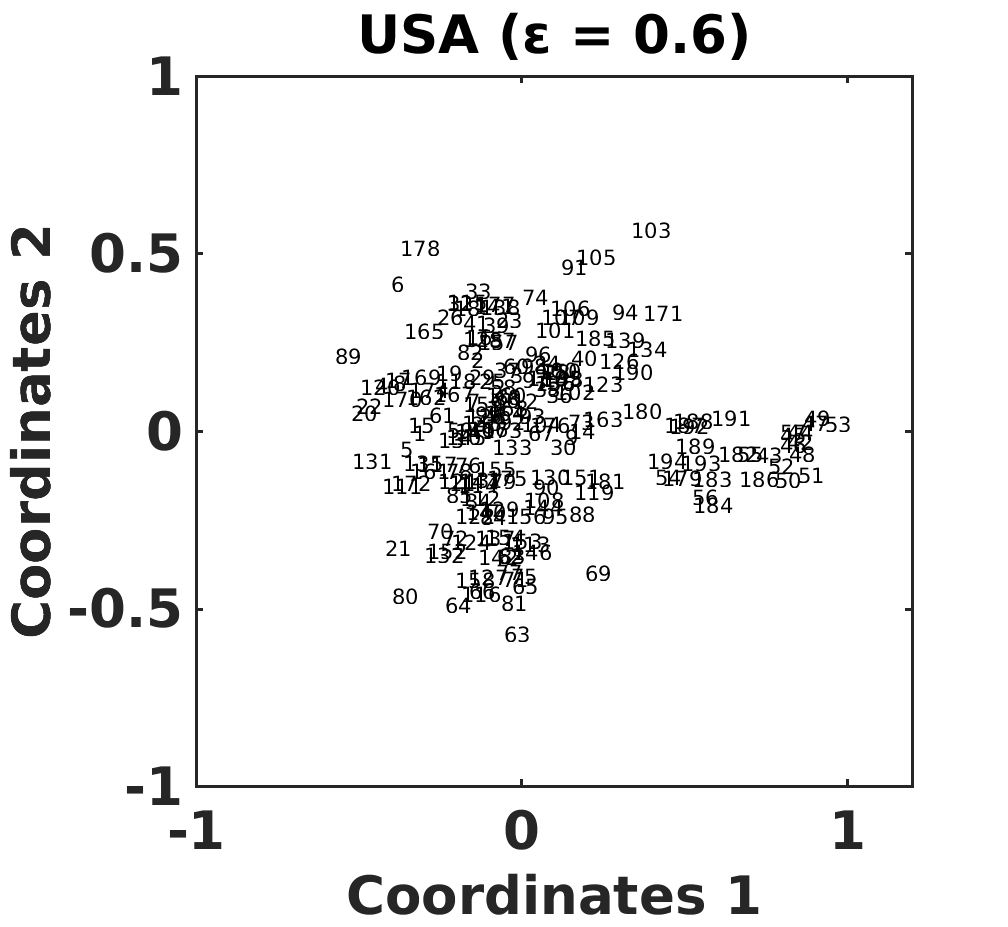}\llap{\parbox[b]{1.9in}{(\textbf{i})\\\rule{0ex}{1.5in}}}
\caption{\textbf{Noise-suppression in a short time cross-correlation frame.} (a), (b) and (c) show the correlation pattern, eigenvalue spectra and MDS map, respectively, for a correlation matrix of  short time-epoch of $M=20$ days and $N=194$ stocks of USA, ending on 30/11/2001. The power mapping method~\cite{Guhr_2003} is used to: (i) reduce the noise of the singular correlation matrix ($M<N$) formed by the short time series, or (ii) break the degeneracy of the zero eigenvalues. Two different noise-suppression parameter values, $\epsilon=0.01$ and $\epsilon=0.6$, are used for this purpose. A small value of  $\epsilon=0.01$ is used for (d), (e) and (f). The change in $\lambda_{max}$ as well as the eigenvalue spectrum is clearly visible (the height and spread of the ``emerging spectrum" shown in the inset); the clustering does not change much at this small value. In (g), (h) and (i), when a higher distortion of $\epsilon=0.6$ is given to the correlation frame, the shape of  emerging spectrum as well as the MDS map change drastically. The emerging spectrum for $\epsilon=0.6$ is broader compared to $\epsilon=0.01$. In the MDS plot, the stocks with high correlations come nearer to each other and form more compact and distinct clusters, as compared to $\epsilon=0$ and $\epsilon=0.01$.}
\label{fig:corr_noise}
\end{figure}

Figure~\ref{fig:corr_noise} shows the results of the noise-suppression on the short time cross-correlation matrix using power mapping method \cite{Guhr_2003,Schafer_2010,Schafer_2013,Chakraborti_2018}.   Figure~\ref{fig:corr_noise}(a) shows a correlation frame computed for the short time-epoch $M=20$ days for USA with $N=194$ stocks of S\&P 500 ending on 30/11/2001 (arbitrarily chosen date). The eigenvalue spectrum and MDS map of the correlation frame is shown in figures~\ref{fig:corr_noise}(b) and (c), respectively. As mentioned earlier, for any short time series $M < N$, the highly singular correlation matrices will have $N -M +1$ degenerate eigenvalues at zero. Hence, in our case the eigenvalue spectrum consists of $175$ eigenvalues at zero, followed by 19 distinct positive eigenvalue.   
The non-linear power mapping method removes the degeneracy of eigenvalues at zero, leading to an emerging spectrum \cite{Chakraborti_2018, Pharasi_2018}. Figure~\ref{fig:corr_noise}(d) shows the correlation pattern for $\epsilon=0.01$.  The effect of the small distortion on the corresponding eigenvalue spectrum and MDS map is shown in figures~\ref{fig:corr_noise}(e) and (f), respectively. The effect is less visible on MDS map; $\lambda_{max}$ reduces its value by a small amount from $44.05$ to $43.67$. Next, we use a high value of noise-suppression parameter $\epsilon=0.6$ to reduce considerably the noise of the correlation frame (shown in figure~\ref{fig:corr_noise}(g)). The effect of $\epsilon=0.6$ is clearly visible on the corresponding eigenvalue spectrum and MDS map, as shown in figures~\ref{fig:corr_noise}(h) and (i), respectively. The shape of the eigenvalue spectrum changes completely. The emerging spectrum from $175$ eigenvalues at zero is now non-degenerate in nature, and shows a spread around zero with some negative eigenvalues. Inset of the figures~\ref{fig:corr_noise}(e) and (h) show the emerging spectra in greater details, while for the inset of figure~\ref{fig:corr_noise}(b) the emerging spectrum is absent. Note that, for $\epsilon=0.6$, the value of highest eigenvalue $\lambda_{max}$ decreases by a large amount to $27.27$; the clusters of stocks in the MDS maps are distinct and denser as compare to low noise-suppression ($\epsilon=0.01$) or without noise-suppression ($\epsilon=0$). 

\subsection{Noise-suppression in a similarity matrix among correlation frames over different time-epochs}
\begin{figure}[t!]
\centering
\includegraphics[width=0.235\linewidth]{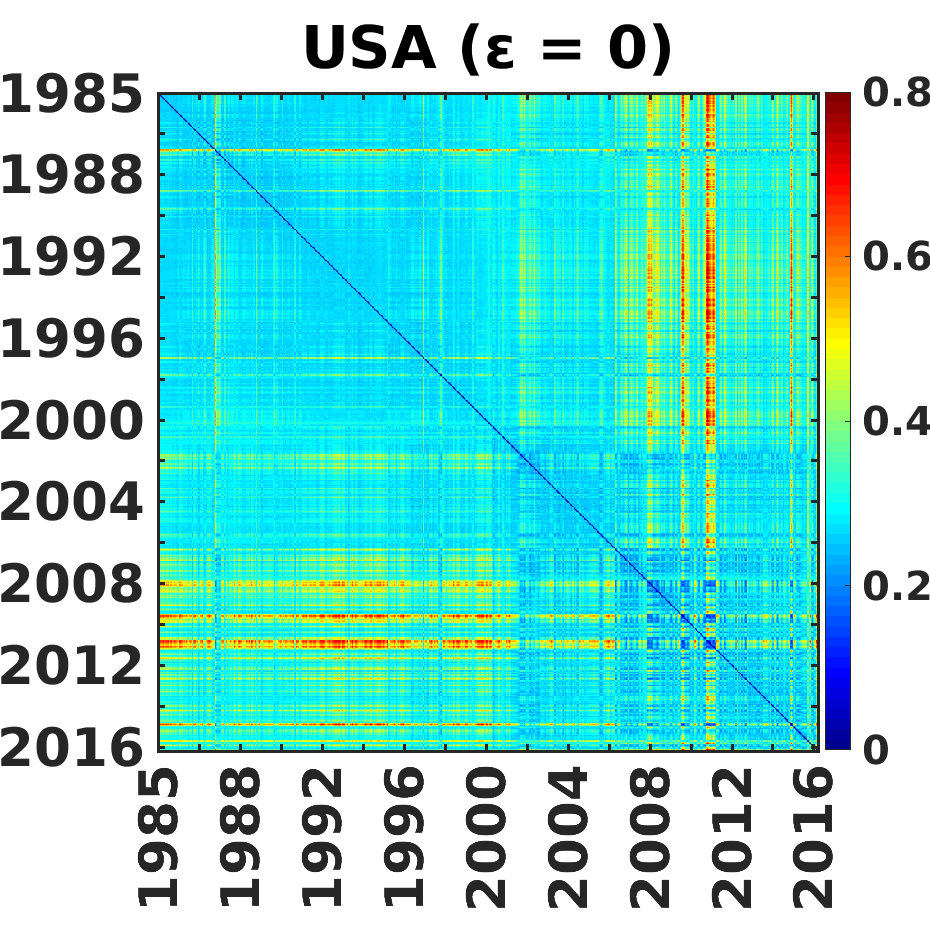}\llap{\parbox[b]{1.6in}{(\textbf{a})\\\rule{0ex}{1.35in}}}
\includegraphics[width=0.252\linewidth]{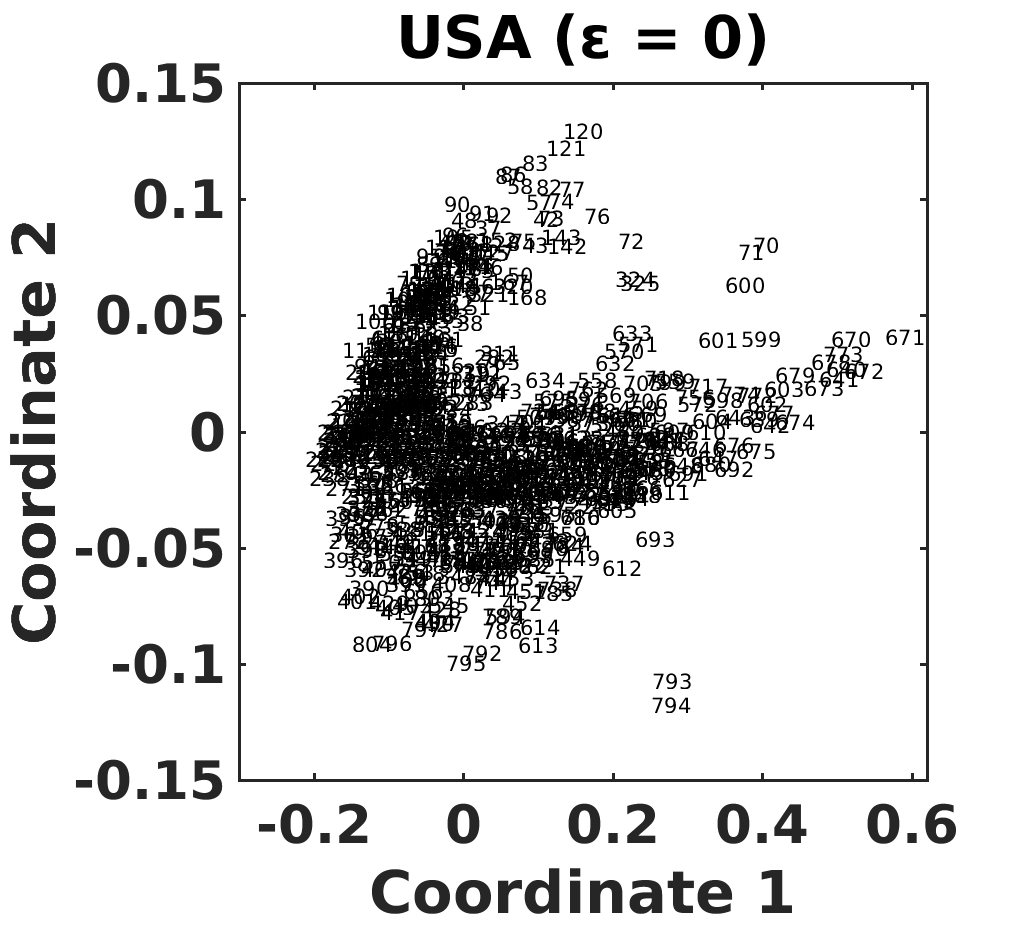}\llap{\parbox[b]{1.5in}{(\textbf{b})\\\rule{0ex}{1.35in}}}
\includegraphics[width=0.235\linewidth]{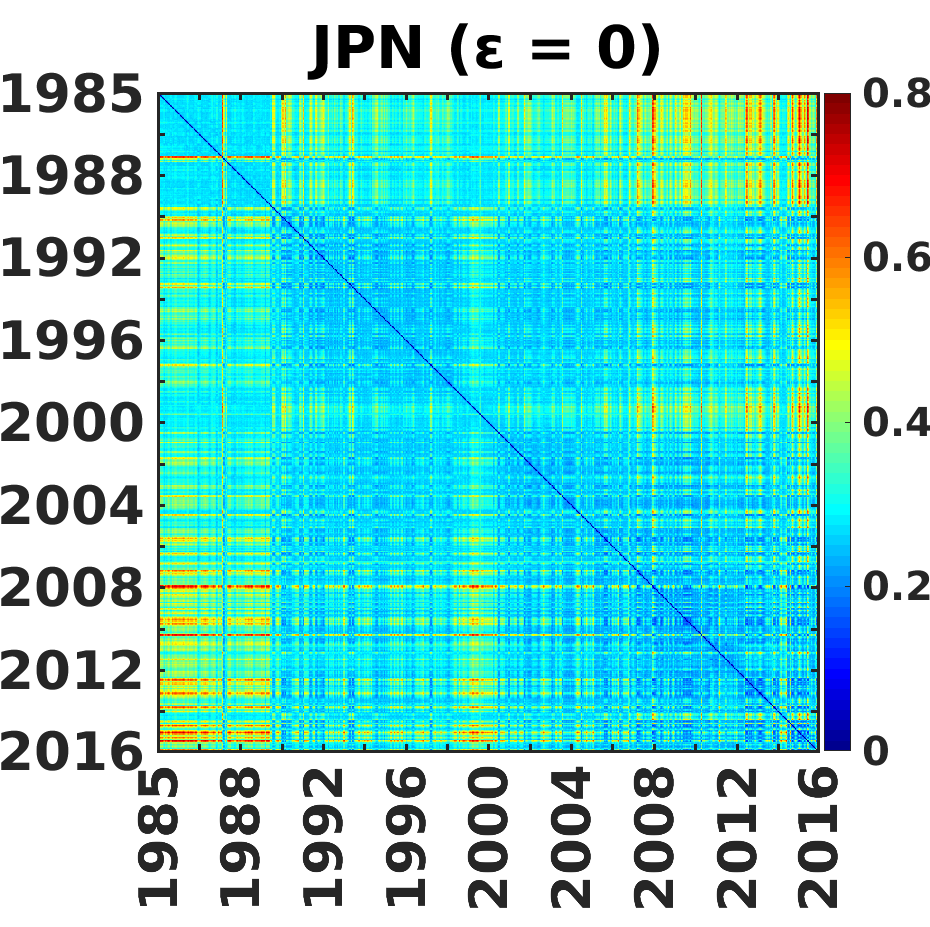}\llap{\parbox[b]{1.6in}{(\textbf{c})\\\rule{0ex}{1.35in}}}
\includegraphics[width=0.252\linewidth]{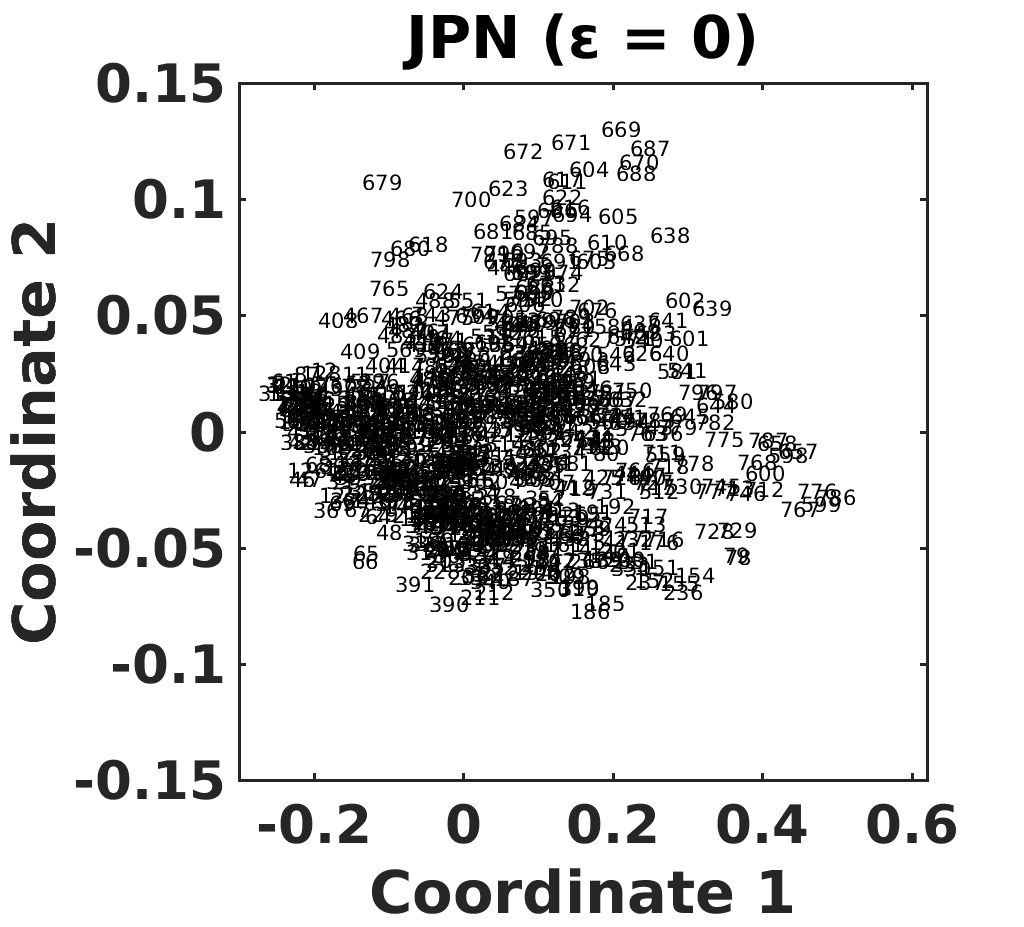}\llap{\parbox[b]{1.5in}{(\textbf{d})\\\rule{0ex}{1.35in}}}
\includegraphics[width=0.235\linewidth]{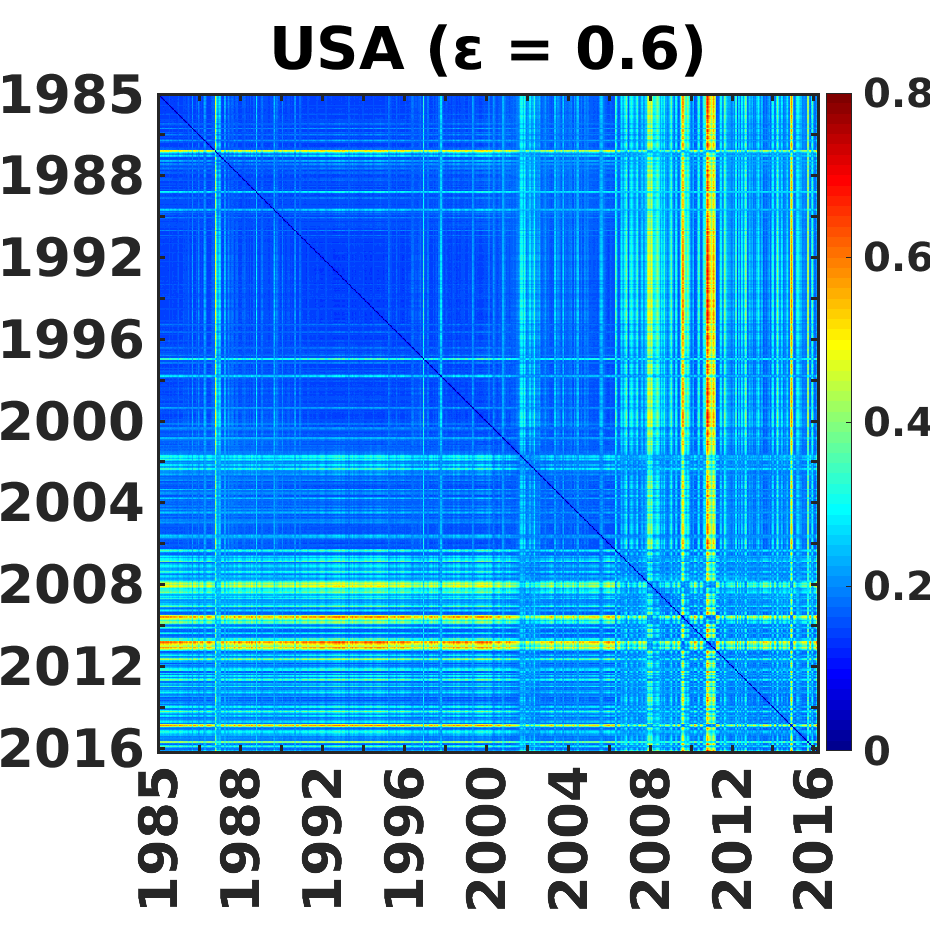}\llap{\parbox[b]{1.6in}{(\textbf{e})\\\rule{0ex}{1.35in}}}
\includegraphics[width=0.252\linewidth]{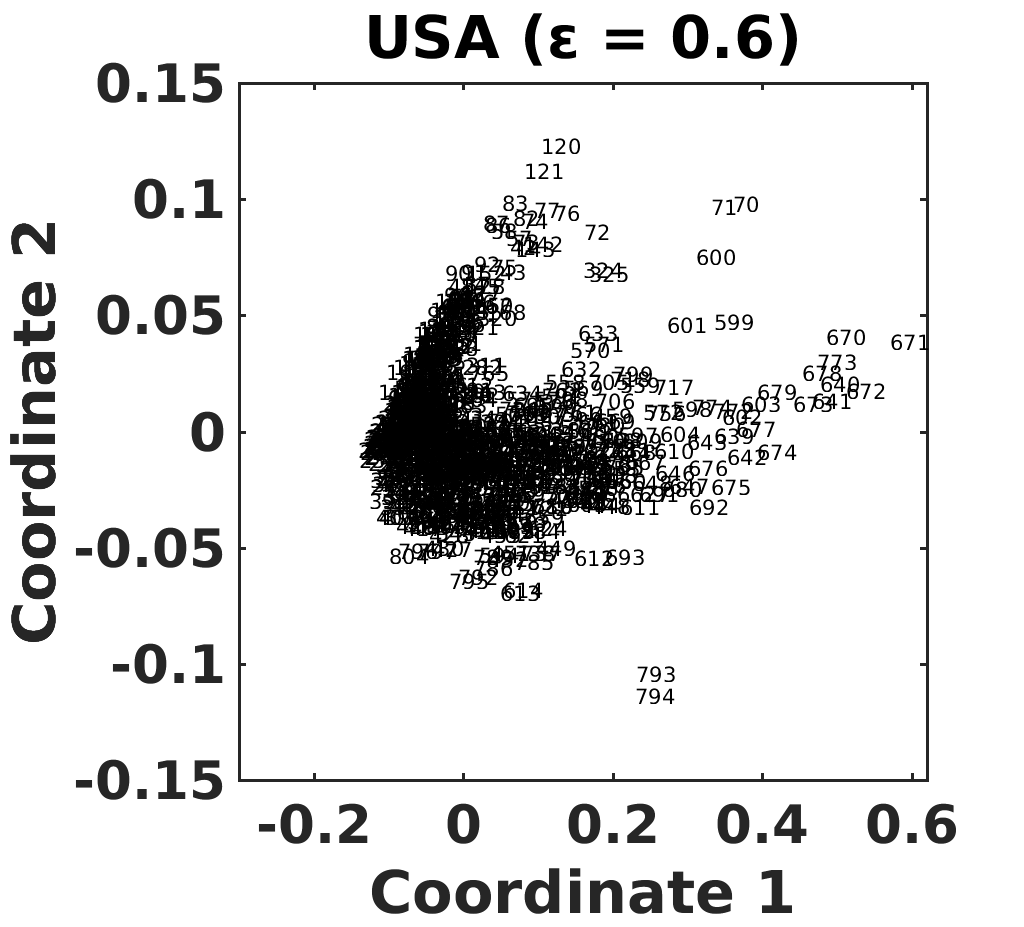}\llap{\parbox[b]{1.5in}{(\textbf{f})\\\rule{0ex}{1.35in}}}
\includegraphics[width=0.235\linewidth]{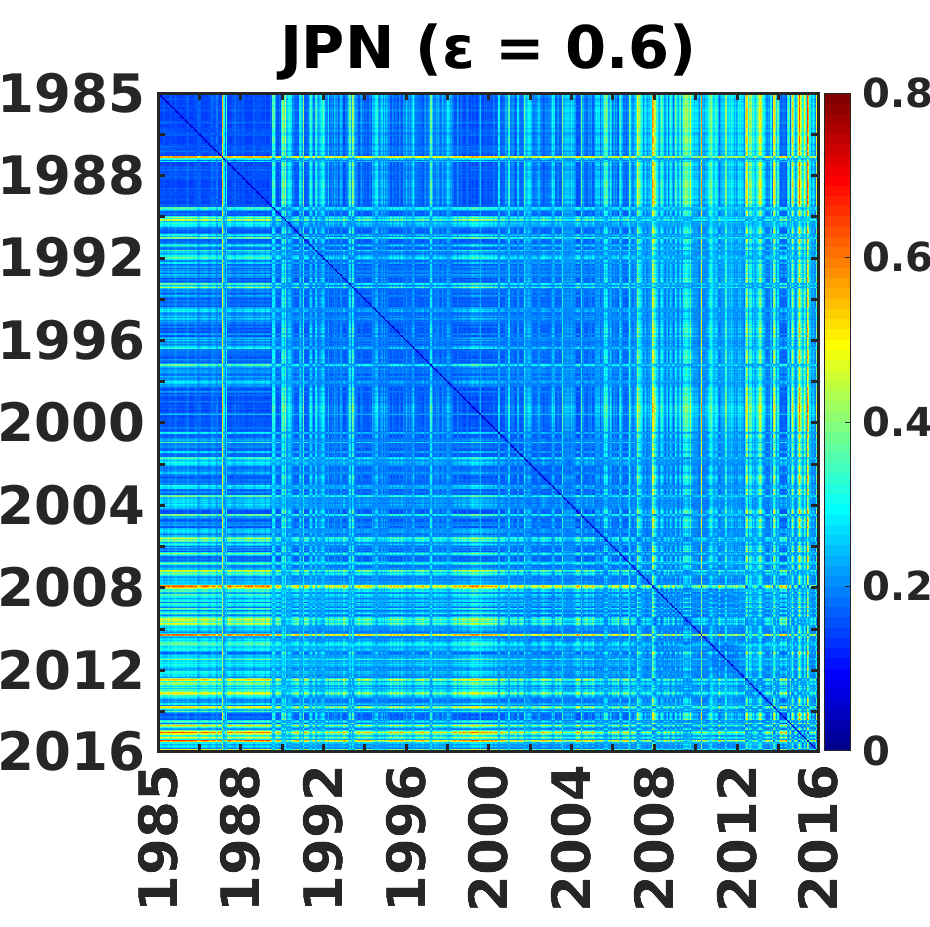}\llap{\parbox[b]{1.6in}{(\textbf{g})\\\rule{0ex}{1.35in}}}
\includegraphics[width=0.252\linewidth]{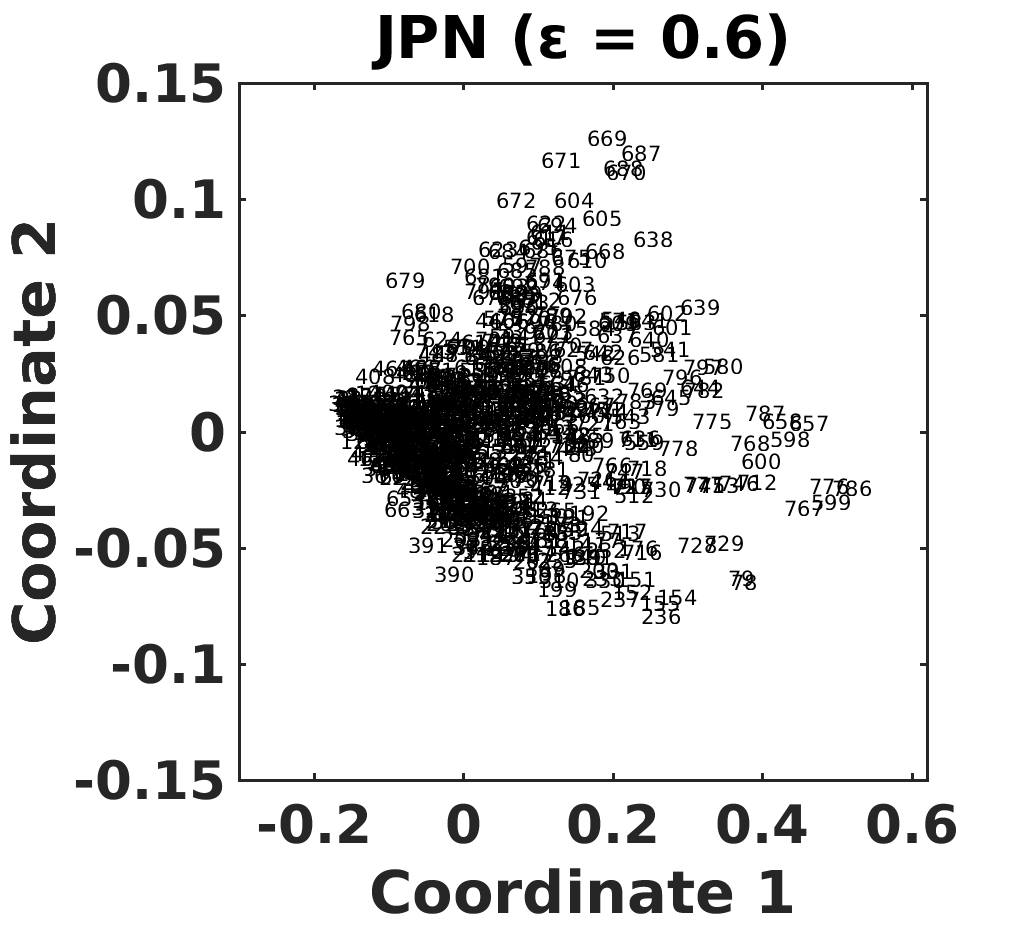}\llap{\parbox[b]{1.5in}{(\textbf{h})\\\rule{0ex}{1.35in}}}
\caption{\textbf{Noise-suppression in a similarity matrix among correlation frames over different time-epochs.} (a) and (e) show the similarity matrices (without noise-suppression $\epsilon=0$ and with noise-suppression $\epsilon=0.6 $) among $805$ correlation frames; (b) and (f) show the corresponding MDS maps for USA. (c) and (g) show the similarity matrices (without noise-suppression $\epsilon=0$ and with noise-suppression $\epsilon=0.6 $) among $798$ correlation frames; (b) and (f) show the corresponding MDS maps for JPN. The similarity matrices give insight of the stock market evolution over $32$ years (1985-2016). Red-yellow strips in the similarity matrices exhibit the crashes of the respective markets. The effect of noise-suppression is clearly visible on both similarity matrices as well as MDS maps.}
\label{fig:SM}
\end{figure}

The noise-suppressed cross-correlation structures of return matrices $\boldsymbol{C}(\tau)$ across different times $\tau=1,...,n$, can be compared based on their similarities.  If there are two correlation matrices $C(\tau_1)$ and $C(\tau_2)$ at different time-epochs $\tau_1$ and $\tau_2$, each computed over a short time-epoch of $M$ days, then to quantify the similarity between the correlation structures, the similarity measure is computed as: $\zeta(\tau_1,\tau_2) \equiv \langle \mid C_{ij} (\tau_1)- C_{ij} (\tau_2) \mid \rangle$, where $\mid ... \mid$ denotes the absolute value and $\langle ... \rangle$ denotes the average over all matrix elements \{$ij$\}  \cite{Munnix_2012}. We then use the MDS map to visualize the information contained in $n \times n$ similarity matrix, where each element is $\zeta(\tau_p,\tau_q)$, where $p,q=1,...n$. 

Interestingly, the noise-suppression applied to individual correlation frames in short time-epochs, has a dramatic effect in the similarity matrix too.
Figure~\ref{fig:SM} shows the effect of noise-suppression on the similarity matrix \cite{Munnix_2012} and the corresponding MDS map. Each correlation frame is computed with $N=194$ stocks of USA; hence, for the time series of length $T=8060$ days during the period  1985-2016, there are $n=805$ correlation frames constructed from short time-epochs of $M=20$ days and shifts of $\Delta \tau= 10$ days ($50\%$ overlapping time-epochs). Similarly, we have $N=165$ stocks of JPN; the time series of length $T=7990$ days in the same period yield $n=798$ correlation frames. The sharp changes in the structural patterns of the similarity matrices become evident at higher $\epsilon=0.6$. It is noteworthy that figure~\ref{fig:SM}(e) shows the block structure for the USA market and reveals the fact that behavior of USA market was relatively calmer till 2002 and it became more volatile afterwards; the red-yellow stripes highlighting the crash periods. Similarly, figure~\ref{fig:SM}(g) shows that the JPN market became more volatile from 1990 onward; also, it went through more critical periods as compared to USA market. Importantly, the MDS maps with the noise-suppression parameter $\epsilon=0.6$ are more compact and denser, which lead to better clustering and determination of optimal number of markets states (see also supplementary figures S2 and S3). \\

\subsection{Determining optimal number of market states}

To determine the number of market states, we find the number of clusters that can group together the noise-suppressed cross-correlation return matrices $\boldsymbol{C}(\tau)$ across different time-epochs $\tau=1,...,n$, based on their similarities\cite{Munnix_2012}. We use the MDS map to visualize the information contained in $n \times n$ similarity matrix, and then use this MDS map with $n$ objects for $k$-means clustering.  The $k$-means clustering, which is a heuristic algorithm, aims to partition $n$ numbers of correlation frames into $k$ clusters or groups in which each object/frame belongs to the cluster with the centroid (nearest mean correlation), serving as a prototype of the cluster. In $k$-means clustering, the value of $k$ can be optimized by different techniques~\cite{Teofilo_1985,Bholowalia_2014}. Here, we propose a new approach for optimizing $k$. We measure the mean and the standard deviation of the intra-cluster distances using an ensemble of fairly large number (say 500) of different initial conditions (choices of random coordinates for the 
$k$-centroids or equivalently random initial clustering of $n$ objects); each set of initial conditions may result in slightly different clustering of the $n$ different correlation frames. If the clusters are distinct (or far apart in coordinate space) then even for different initial conditions, the $k$-means clustering yield same results, yielding a small variance of the intra-cluster distance. The problem of allocating the frames into the different clusters becomes acute when the clusters are very close or overlapping, as the initial conditions can influence the final clustering. So there is a larger variance of the intra-cluster distance. Therefore, the minimum variance or standard deviation for a particular number of clusters displays the robustness of the clustering.
For optimizing the number of clusters, we propose that one should look for \textit{maximum} $k$, which has the \textit{minimum variance} or standard deviation in the intra-cluster distances with different initial conditions. We suppose this is easier than determining the ``elbow point" from the intra-cluster distance versus number of clusters curve \cite{Bholowalia_2014}.

\begin{figure}[]
\centering
\includegraphics[width=0.48\linewidth]{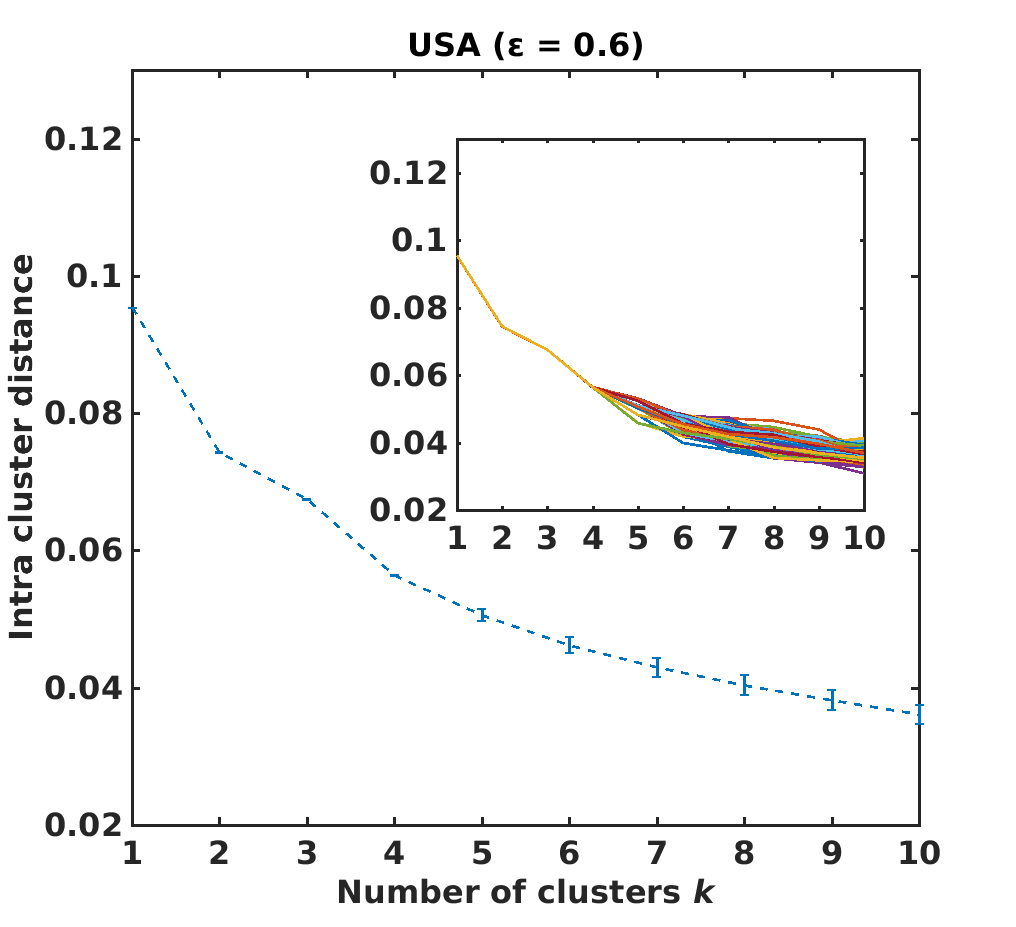}\llap{\parbox[b]{3in}{(\textbf{a})\\\rule{0ex}{2.4in}}}
\includegraphics[width=0.48\linewidth]{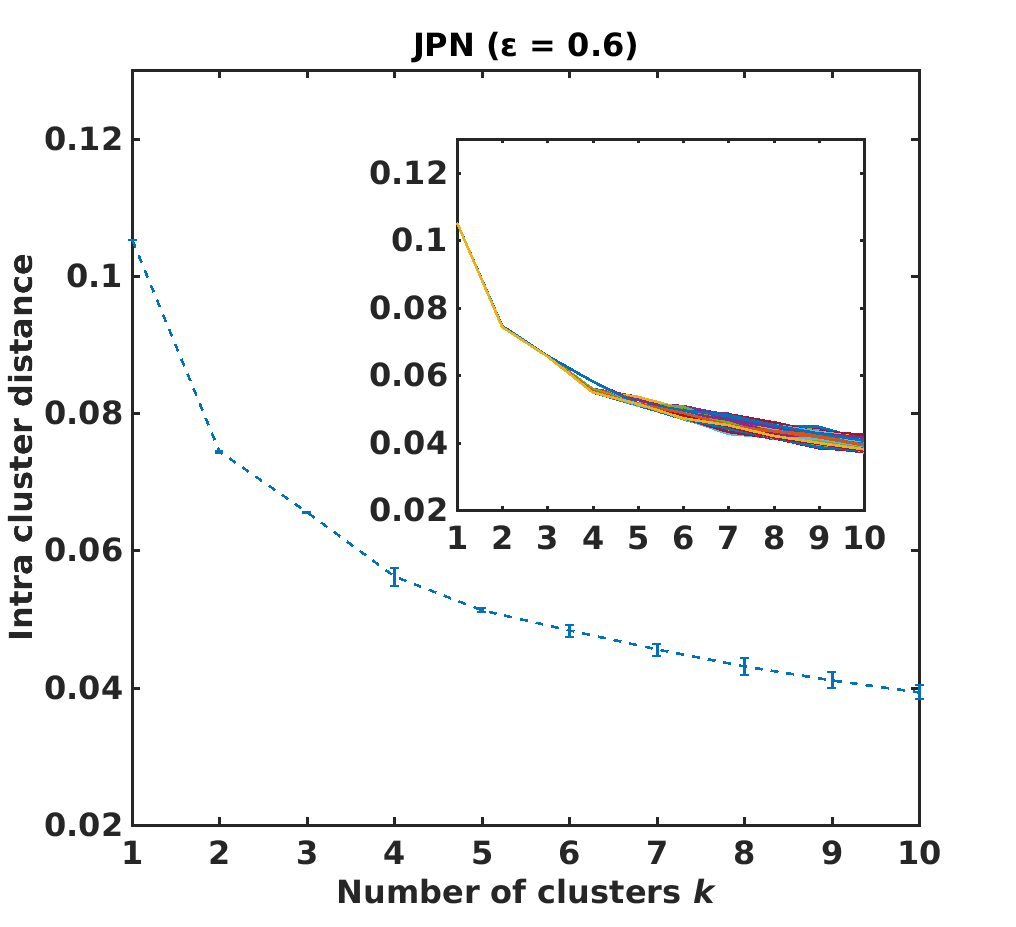}\llap{\parbox[b]{3in}{(\textbf{b})\\\rule{0ex}{2.4in}}} 
\caption{\textbf{Plots of intra-cluster distance as a function of number of clusters}. Results shown for (a) USA and (b) JPN, for the noise-suppression parameter $\epsilon=0.6$. The $k$-means clustering is performed on the MDS map generated from  $805$ noise-suppressed correlation frames of USA and $798$ noise-suppressed correlation frames of JPN, with 500 initial conditions in $k$-means clustering. The errorbars are the standard deviations of the intra-cluster distances arising from the ensemble of 500 random initial conditions to centroids for the initial clustering. The plots show the minima of standard deviations at $k=4$ for USA and $k=5$ for JPN, which indicate the ``optimal'' number of clusters. Inset: Plot of intra-cluster distance vs $k$ for all 500 random initial conditions. Each colored line corresponds to one such initial condition.}
\label{fig:intra-cluster}
\end{figure}

For each cluster, one computes the average/variance of the point-to-centroid distances for all the points belonging to the cluster; the mean/variance of the intra-cluster distances is the mean/variance of the $k$ values obtained from each of the $k$ clusters. Next, we use 500 different initial conditions for the $k$-means clustering, each yielding a slightly different clustering result. One then computes the average as well as the variance (or standard deviation) of the mean intra-cluster distances among the ensemble of 500 runs. Then, the plots of average intra-cluster distance as functions of the number of clusters $k$ for USA and JPN are shown in figures~\ref{fig:intra-cluster}(a) and (b), respectively. The standard deviations of the intra-cluster distances measured for $500$ initial conditions are shown as the error bars. The insets of figures~\ref{fig:intra-cluster}(a) and (b), show the plots for 500 initial conditions.  As mentioned earlier, the value of $k$ is optimized by keeping the standard deviation lowest and the number of clusters highest; note that for $k=1$, the standard deviations are always trivially zero. We find that for USA, the standard deviations are low till $k=4$ and then grow for higher number of clusters; thus, $k=4$ is the optimal number of clusters. For JPN, which is more complex than USA, the standard deviation is low for $k=1,2,3$, increases for $k=4$ and then decreases drastically for $k=5$; beyond that again the standard deviation is higher. Thus, $k=5$ is the optimum number of clusters for JPN. 

\begin{figure}[]
\centering
\includegraphics[width=0.48\linewidth]{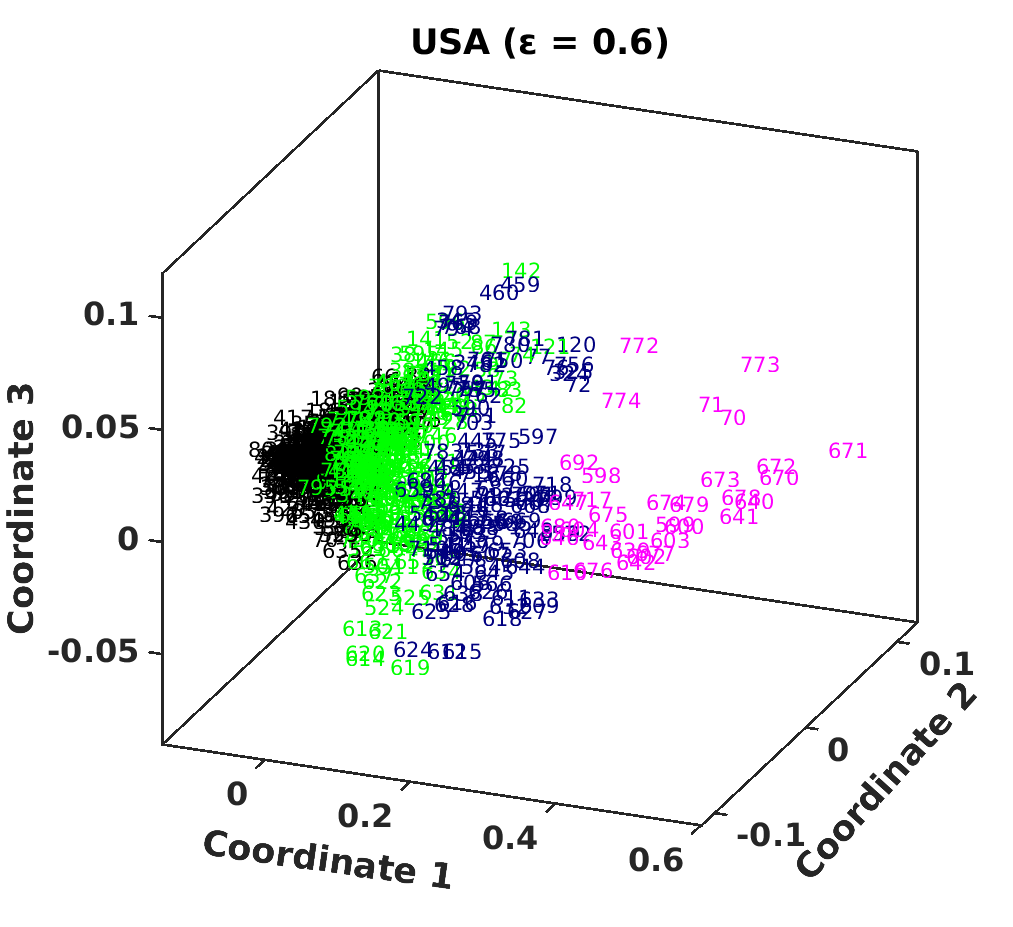}\llap{\parbox[b]{3.2in}{(\textbf{a})\\\rule{0ex}{2.2in}}}
\includegraphics[width=0.48\linewidth]{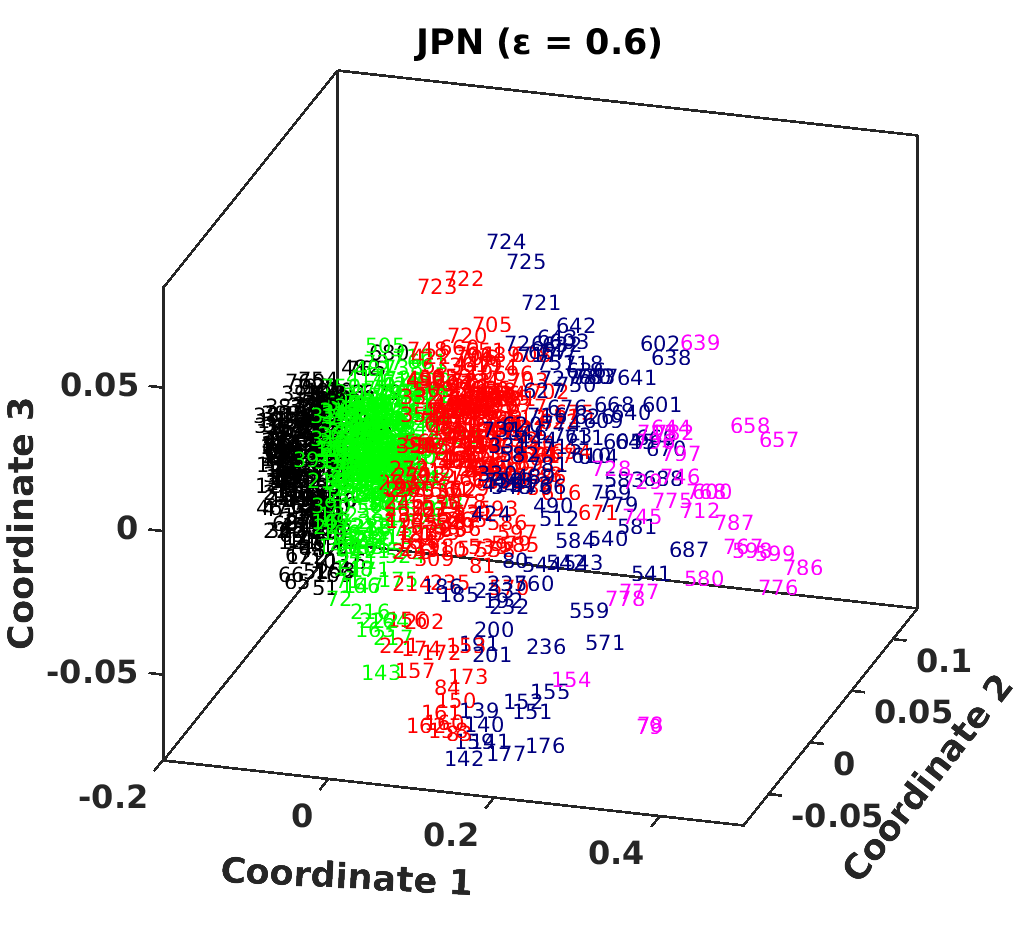}\llap{\parbox[b]{3.2in}{(\textbf{b})\\\rule{0ex}{2.2in}}}
\includegraphics[width=0.24\linewidth]{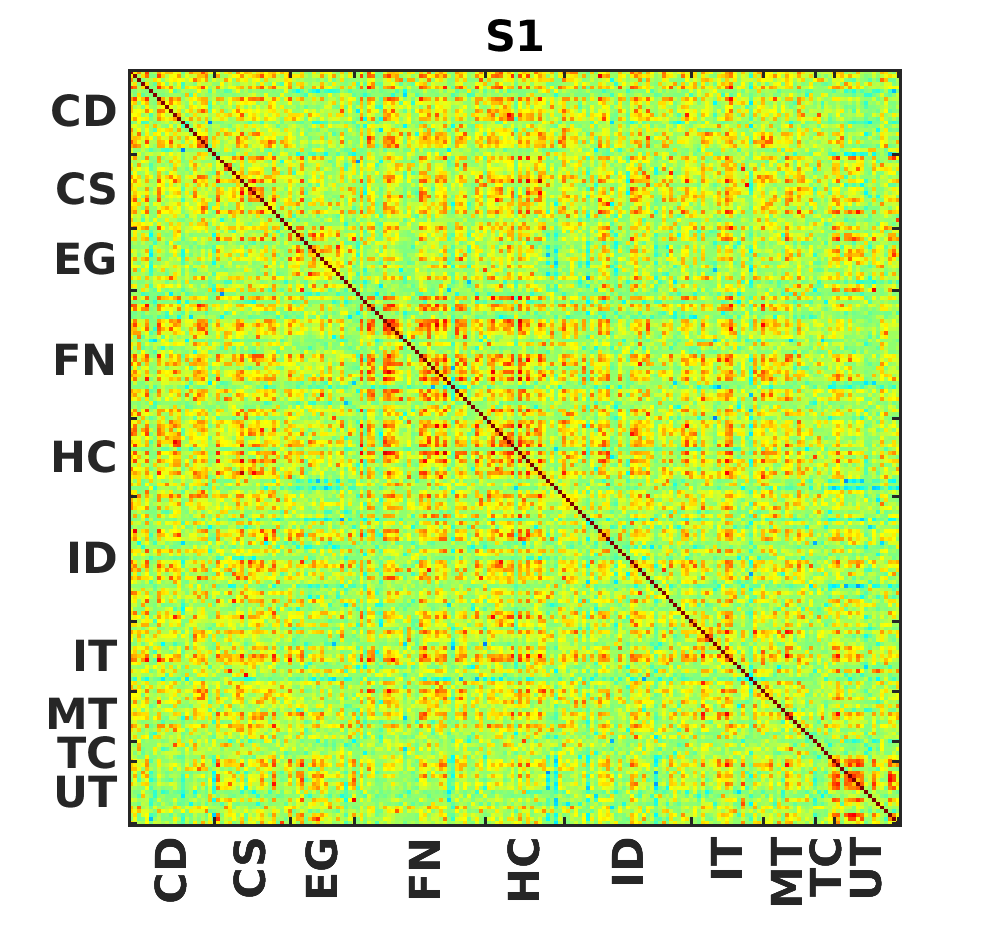}\llap{\parbox[b]{1.6in}{(\textbf{c})\\\rule{0ex}{1.3in}}}
\includegraphics[width=0.24\linewidth]{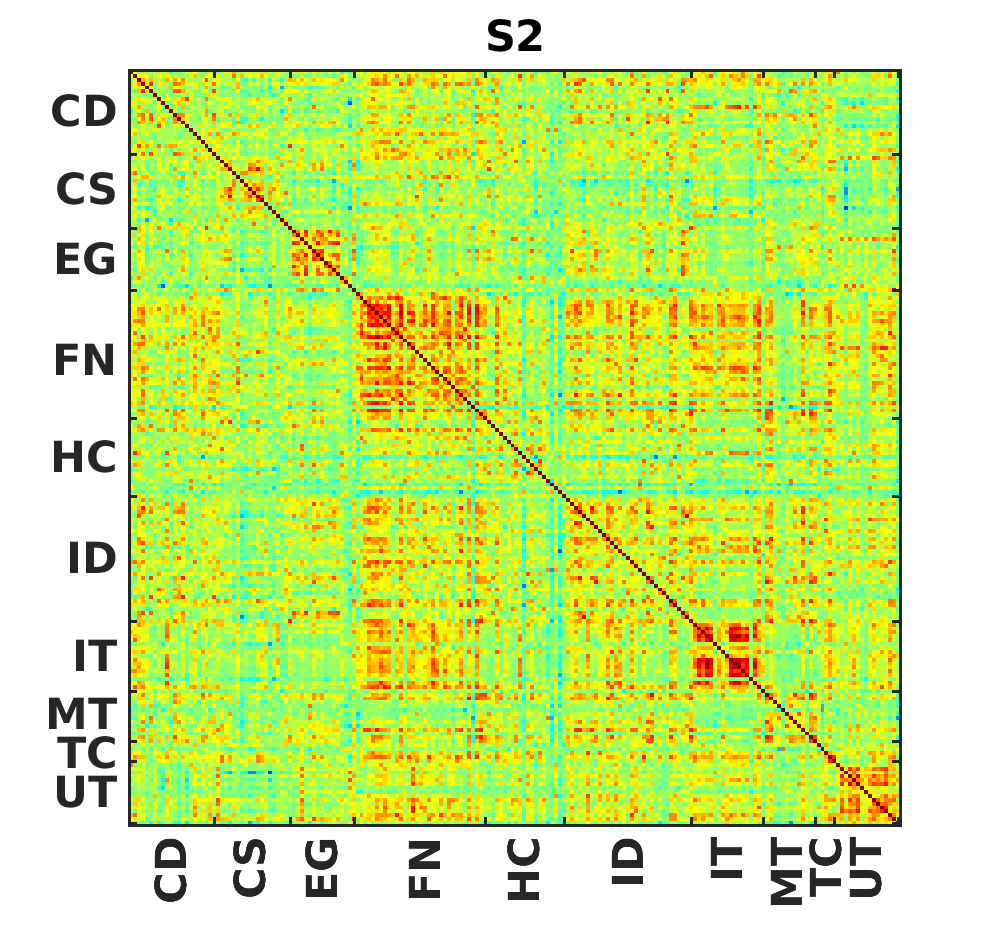}
\includegraphics[width=0.24\linewidth]{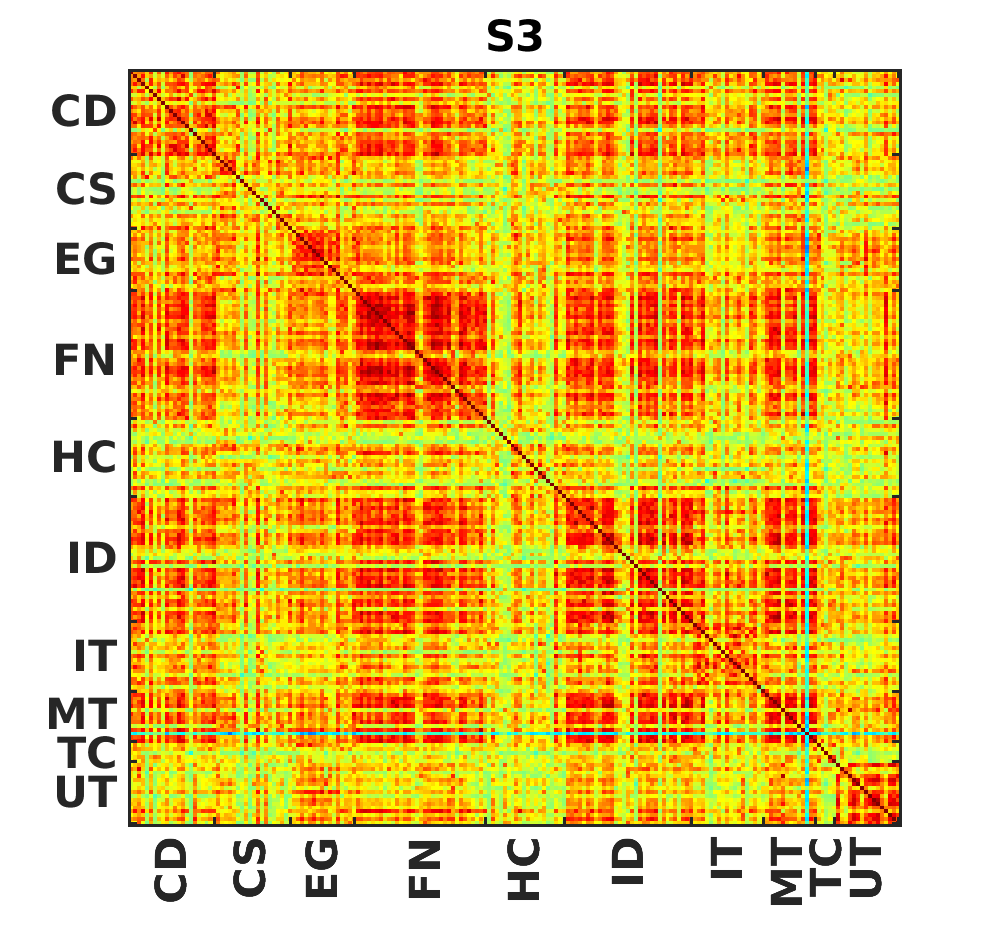}
\includegraphics[width=0.24\linewidth]{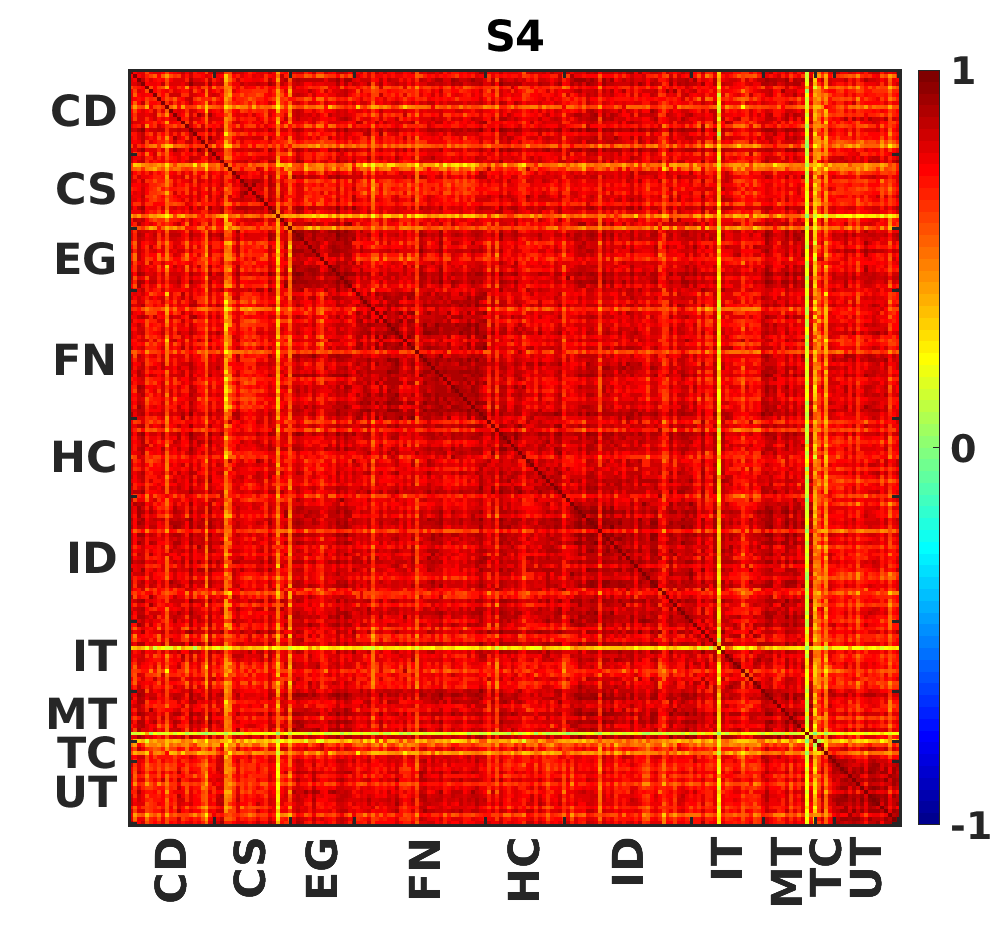}
\includegraphics[width=0.19\linewidth]{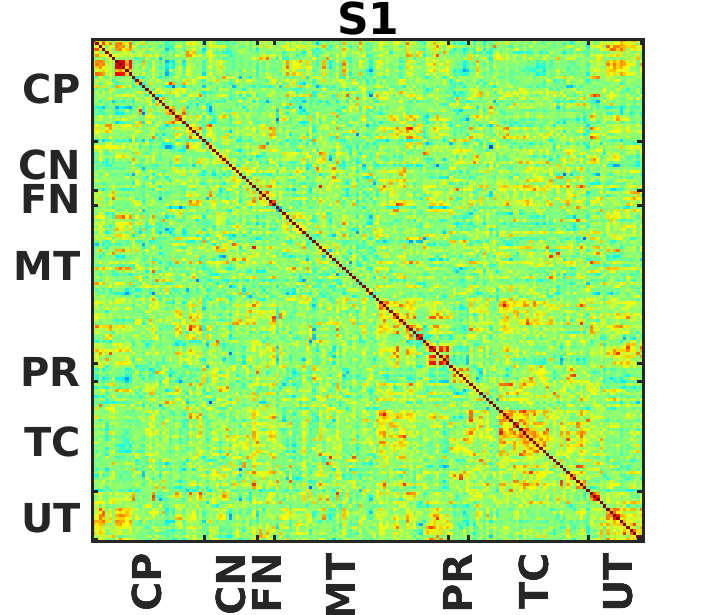}\llap{\parbox[b]{1.2in}{(\textbf{d})\\\rule{0ex}{1.in}}}
\includegraphics[width=0.19\linewidth]{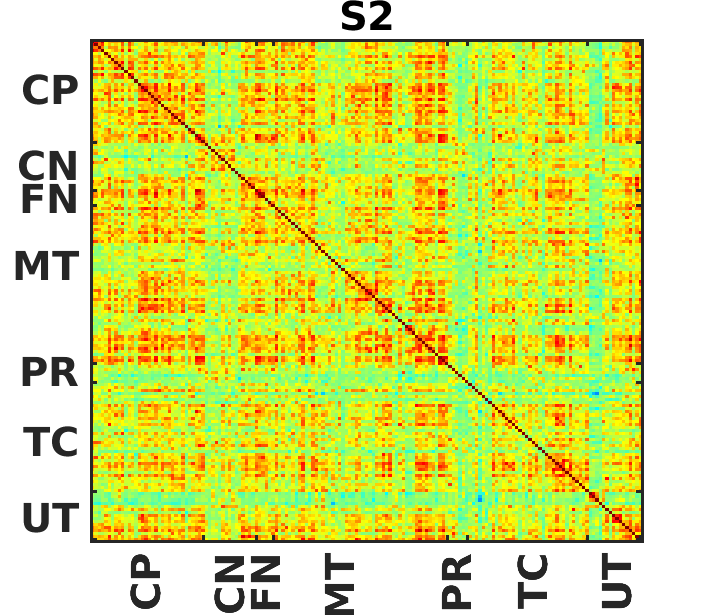}
\includegraphics[width=0.19\linewidth]{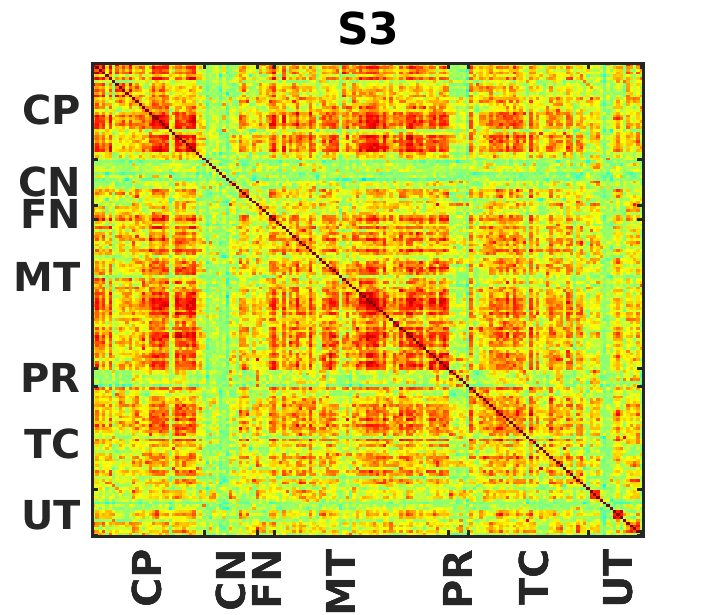}
\includegraphics[width=0.19\linewidth]{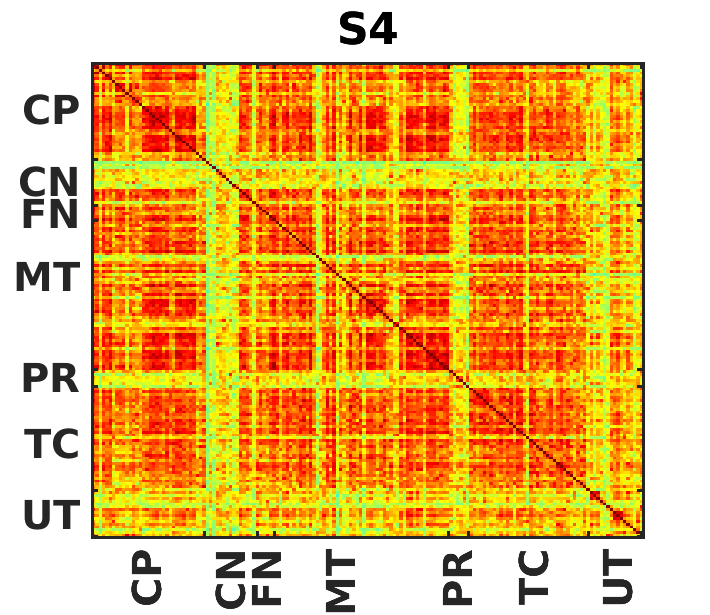}
\includegraphics[width=0.19\linewidth]{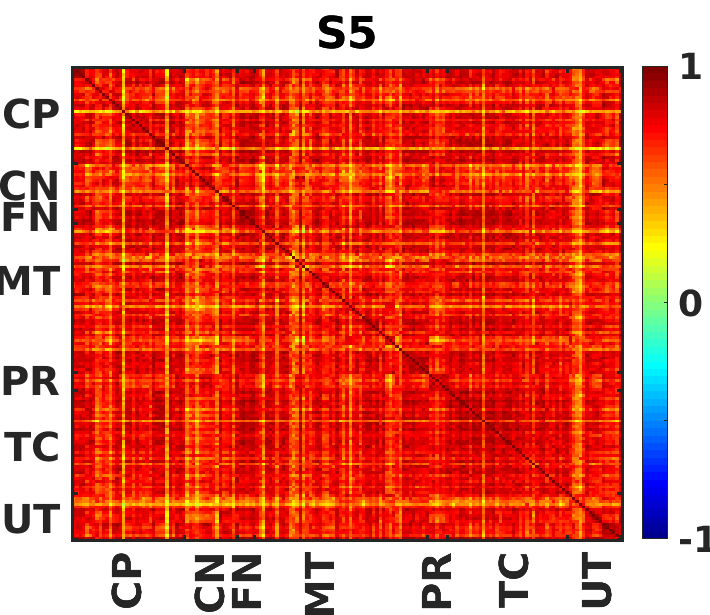}
\caption{\textbf{Market states}. (a) Classification of the USA market into four market states. (b) Classification of the JPN market into five market states.  $k$-means clustering is performed on MDS map constructed from noise suppressed ($\epsilon=0.6$) similarity matrix. The coordinates assigned in the MDS map are the corresponding correlation frames. For USA, we have $805$ correlation frames of time-epoch $M=20$ days with a shift of $\Delta t=10$ days; for JPN, we have $798$ correlation frames for the same. (c) shows the four different states of USA market S1, S2, S3 and S4, where S1 corresponds to a calm state (with low mean correlation) and S4 corresponds to the crash or critical state (with high mean correlation). (d) shows the fives different states of JPN market S1, S2, S3, S4 and S5, where S1 corresponds to the calm state and S5 corresponds to the critical state.}
\label{fig:Kmean_classification}
\end{figure}

The final $k$-means clustering of the correlation frames in the similarity matrix is therefore performed for $k=4$ clusters (USA) and $k=5$ clusters (JPN), as shown in figures~\ref{fig:Kmean_classification}(a) and (b), respectively. We identify the points in each cluster (different colors represent different clusters) with similar correlation patterns and nearby mean correlation as one market state. Based on $k$-means clustering, figure~\ref{fig:Kmean_classification}(c) shows four different market states $S1, S2, S3$ and $S4$ of USA, where S1 corresponds to the calm state (with low mean correlation) and S4 corresponds to the crash or critical state (with high mean correlation); figure~\ref{fig:Kmean_classification}(d) shows five market states  $S1, S2, S3, S4$ and $S5$ of JPN, where S1 corresponds to the calm state and S5 corresponds to the critical state, respectively. The states are arranged in the increasing order of mean correlation. Here, we can also see clear differences structure-wise among the correlation matrices, e.g., there are strong intra-sectoral correlations within the energy, finance and utility sectors, in each of the market states of USA. 

It may also be mentioned that the selection of  noise-suppression parameter $\epsilon=0.6$ is not totally arbitrary. We compared the plots of the average intra-cluster distance as function of the number of clusters for both USA and JPN, using $\epsilon$ ranging from $0.1$ to $0.7$ (shown in supplementary figures S2 and S3). The outcome of the comparison is that $\epsilon=0.6$ yields the best results. \\

\subsection{Co-occurrence probabilities and dynamical transitions of market states} 

\begin{figure}[]
\centering
\includegraphics[width=0.88\linewidth]{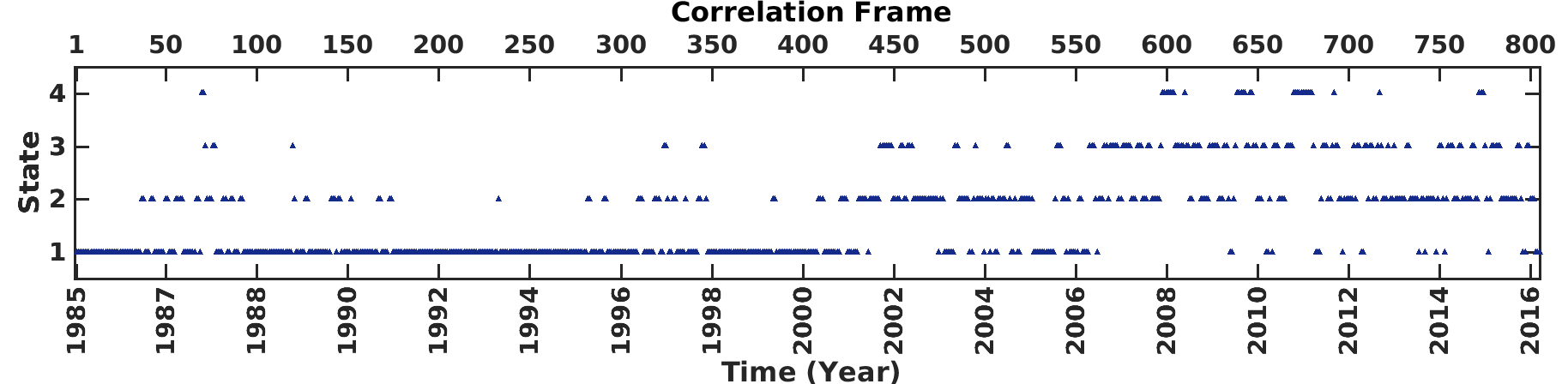}\llap{\parbox[b]{5.5in}{(\textbf{a})\\\rule{0ex}{1.2in}}}
\includegraphics[width=0.88\linewidth]{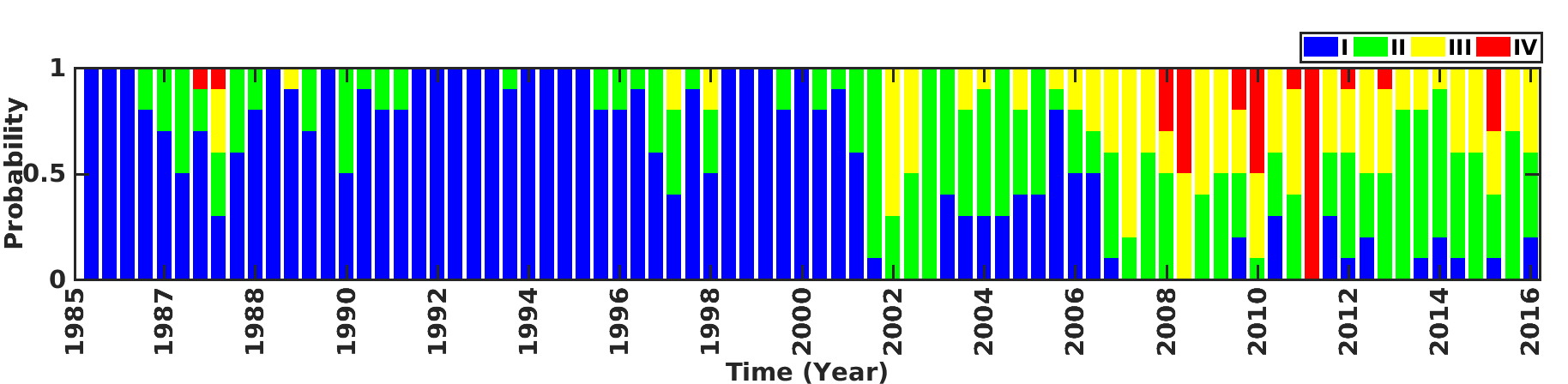}\llap{\parbox[b]{5.5in}{(\textbf{b})\\\rule{0ex}{1.2in}}}\\
\includegraphics[width=0.88\linewidth]{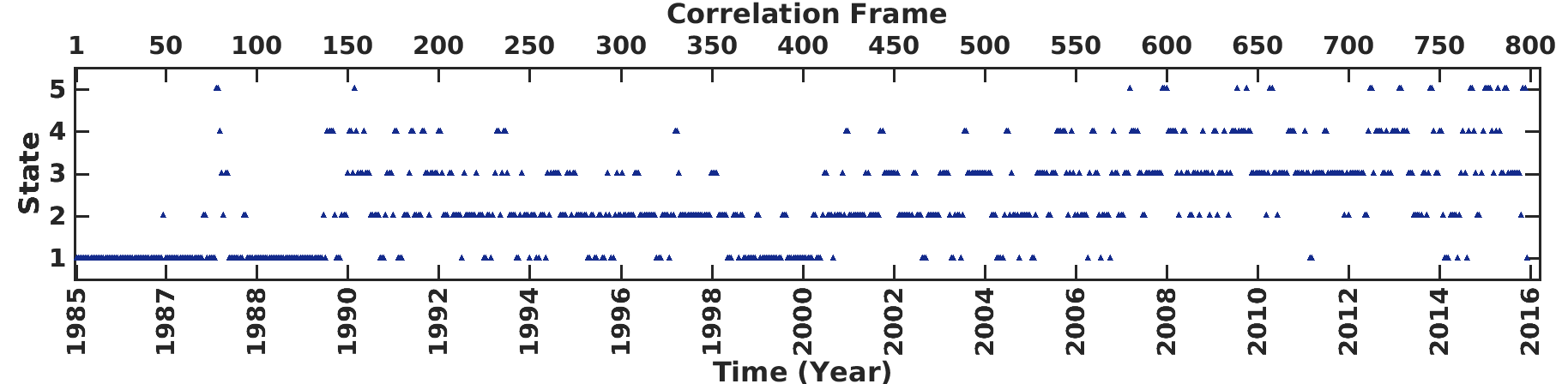}\llap{\parbox[b]{5.5in}{(\textbf{c})\\\rule{0ex}{1.2in}}}
\includegraphics[width=0.88\linewidth]{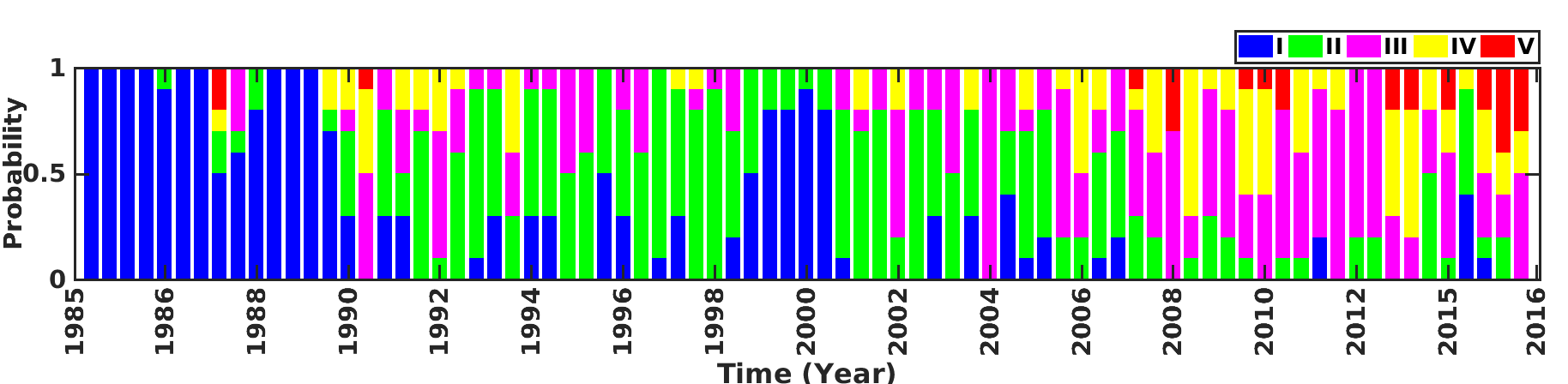}\llap{\parbox[b]{5.5in}{(\textbf{d})\\\rule{0ex}{1.2in}}}
\caption{\textbf{\textbf{Dynamical evolution of market states for USA and JPN}.}
(a) Temporal dynamics of the USA in four different states ($S1, S2, S3$ and $S4$) for the period of  1985-2016. (b) Probability plot of the four market states with each  color length corresponds to the evolution probability of these four states during $110$ days (10 overlapping epochs). (c) and (d) show similar results for JPN with five market states ($S1, S2, S3$, $S4$ and $S5$).}
\label{fig:dynamics1}
\end{figure}

Once the classification of the short-time cross-correlation frames into different market states are complete, one can follow the evolution of the market as dynamical transitions of the different markets states.
Figures~\ref{fig:dynamics1}(a) and (c) show the evolution dynamics of market states of USA and JPN, during  1985-2016. In USA, the market oscillates among the four states $S1, S2, S3$ and $S4$. Often  $S1$ or $S2$ states (with relatively low mean correlations) tend to remain in the same state for a long time; at other times, the market jumps to a higher mean correlation state $ S3 $ or $S4$. Similarly, for JPN the dynamical transitions  among the five market states $S1, S2, S3$, $S4$ and $S5$. The probabilistic plots of the market states dynamics are shown in figures~\ref{fig:dynamics1}(b) and (d), for USA and JPN, respectively. The color length of any market state is the probability of that state computed during $110$ days ($10$ overlapping epochs). Evident from the probability plots: (a) In USA, before $2002$ the market was mostly in state $S1$; the market became more volatile, with more frequent transitions to other states, $2002$ onward, and (b) in JPN, market became more volatile from 1990 onward. The same kind of behavior is also observed from the temporal evolution of the mean correlation (see supplementary figure S1).

Figure~\ref{fig:dynamics2}(a) and (b) show the bar plots of the co-occurrences of the market states for USA and JPN, respectively; the networks representing the transition probabilities (co-occurrences of paired market states) for USA and JPN are respectively shown in figures~\ref{fig:dynamics2}(c) and (d), with corresponding values given in Tables~\ref{table:USA_freq} and ~\ref{table:JPN_freq}. 
The probability of the co-occurrence of paired market states ($S3,S4$) of USA is about $6\%$. If we neglect the diagonal entries of the bar plot, which shows the high probabilities of staying in the same states, then we can safely infer that with the significant transition probability, the state $S3$ of USA acts like a ``precursor'' to the state $S4$ (market crash); similarly, for JPN the state $S4$ acts like a ``precursor'' to the critical state $S5$, with significant transition probability of about $8\%$. Entries just above and below the diagonals of the $3D$ bar plots are also quite high, which show that the transitions primarily happen between immediately adjacent states, and only exceptions of remote transitions being in the cases like the Black Monday crash of 1987, etc.

\begin{figure}[]
\centering
\includegraphics[width=0.4\linewidth]{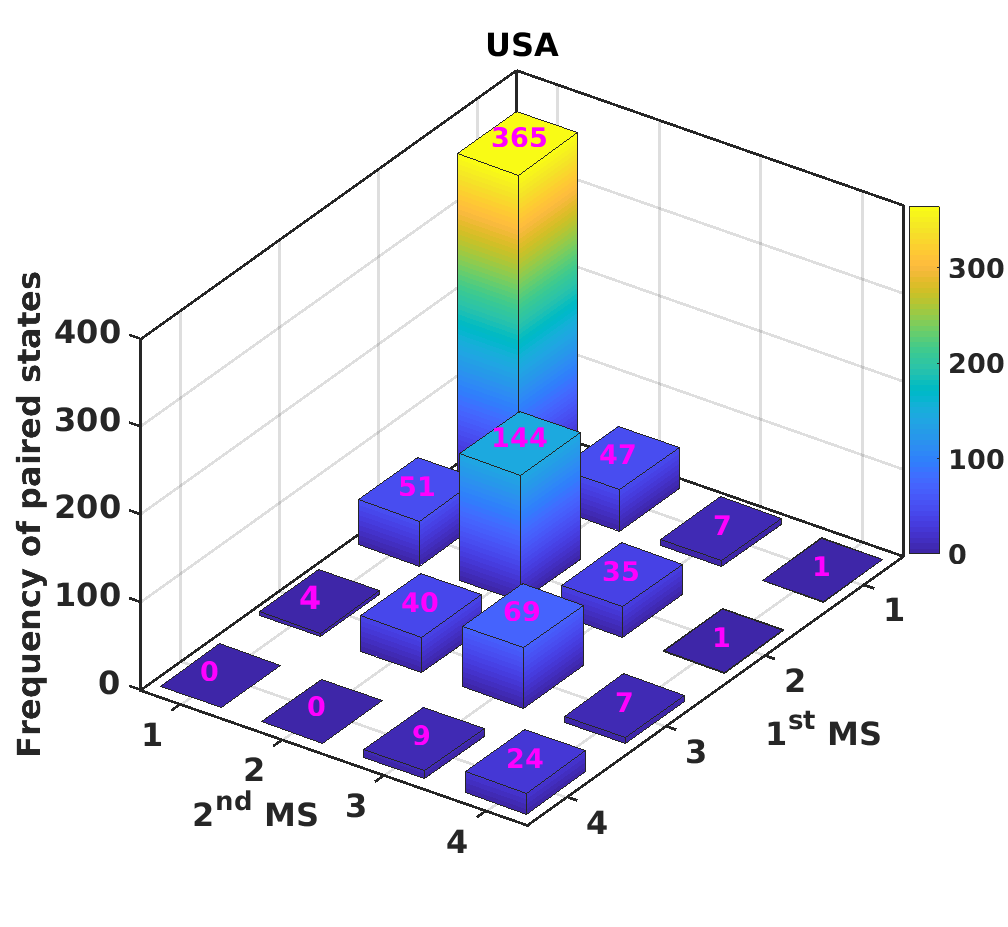}\llap{\parbox[b]{2.4in}{(\textbf{a})\\\rule{0ex}{1.6in}}}\includegraphics[width=0.4\linewidth]{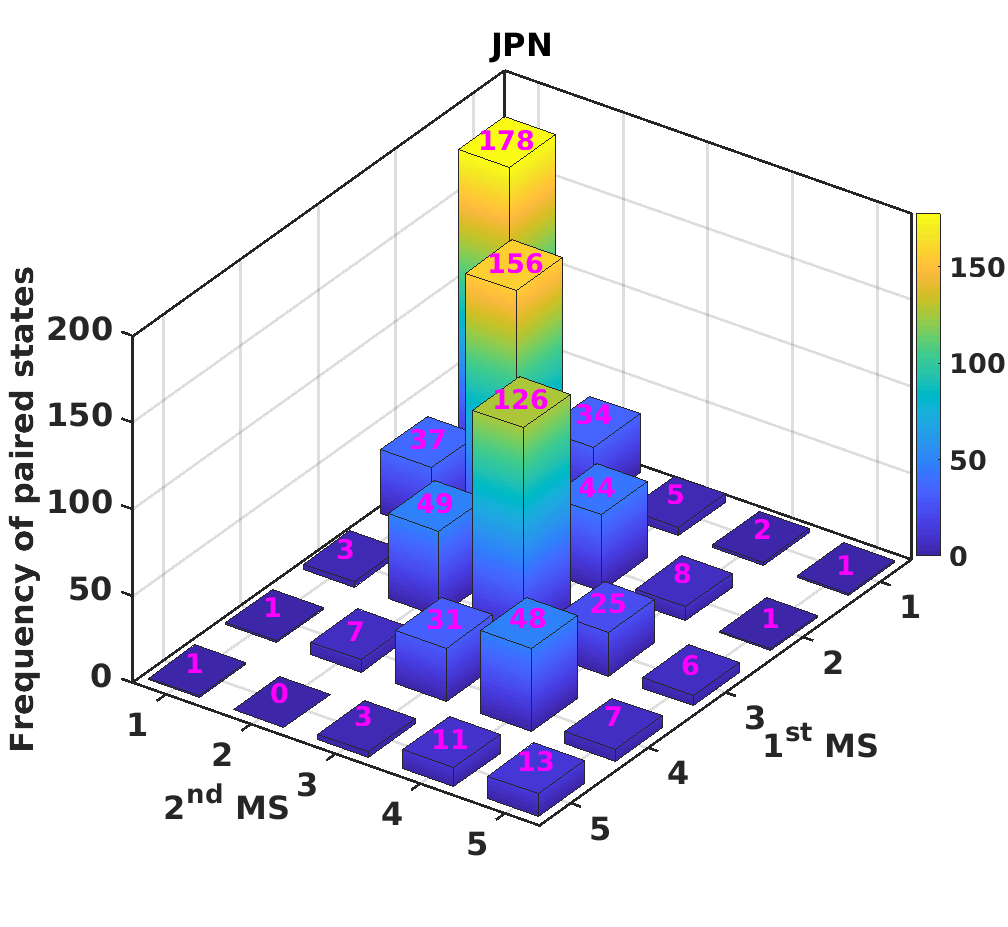}\llap{\parbox[b]{2.4in}{(\textbf{b})\\\rule{0ex}{1.6in}}}
\includegraphics[width=0.4\linewidth]{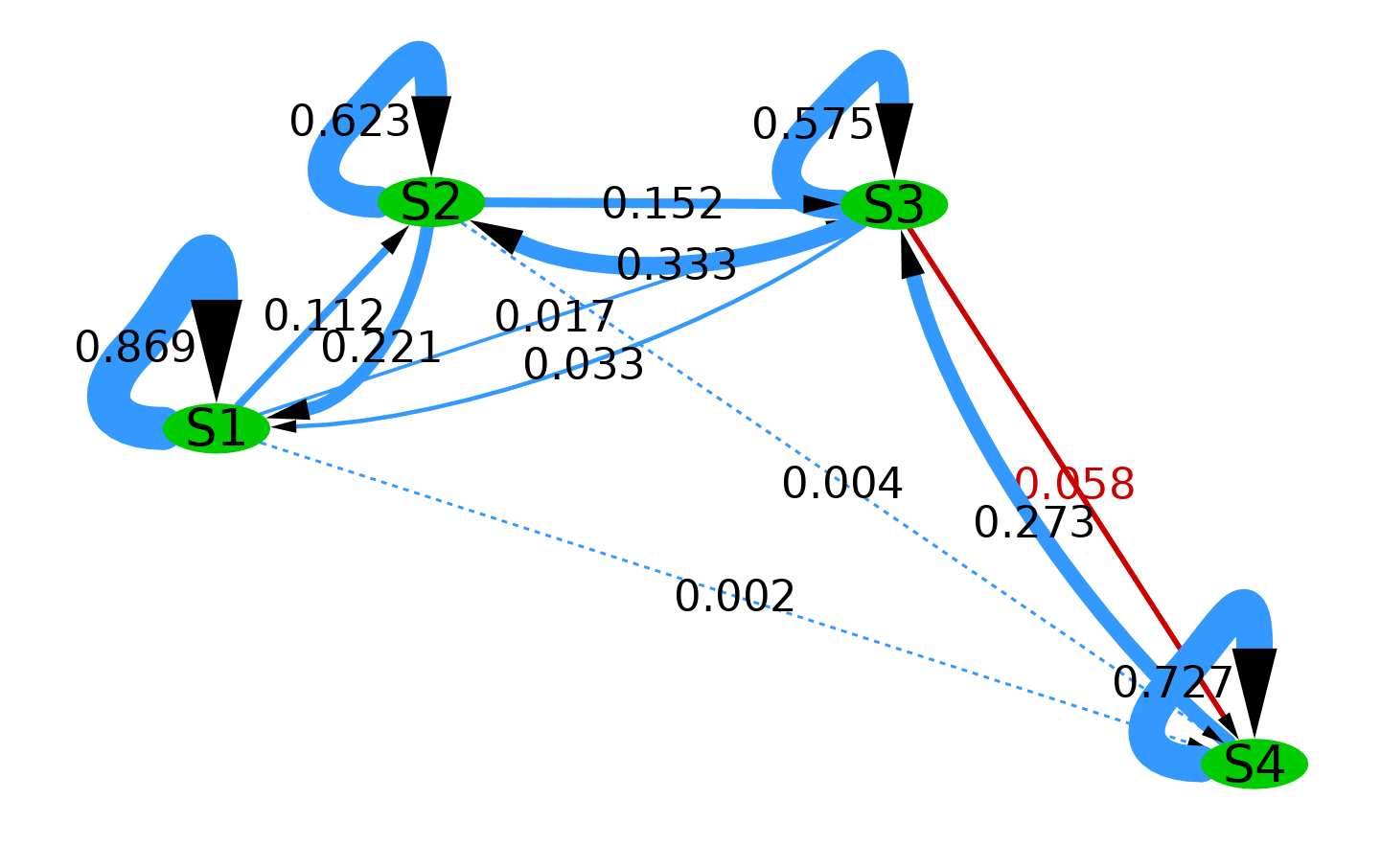}\llap{\parbox[b]{2.4in}{(\textbf{c})\\\rule{0ex}{1.6in}}}\includegraphics[width=0.4\linewidth]{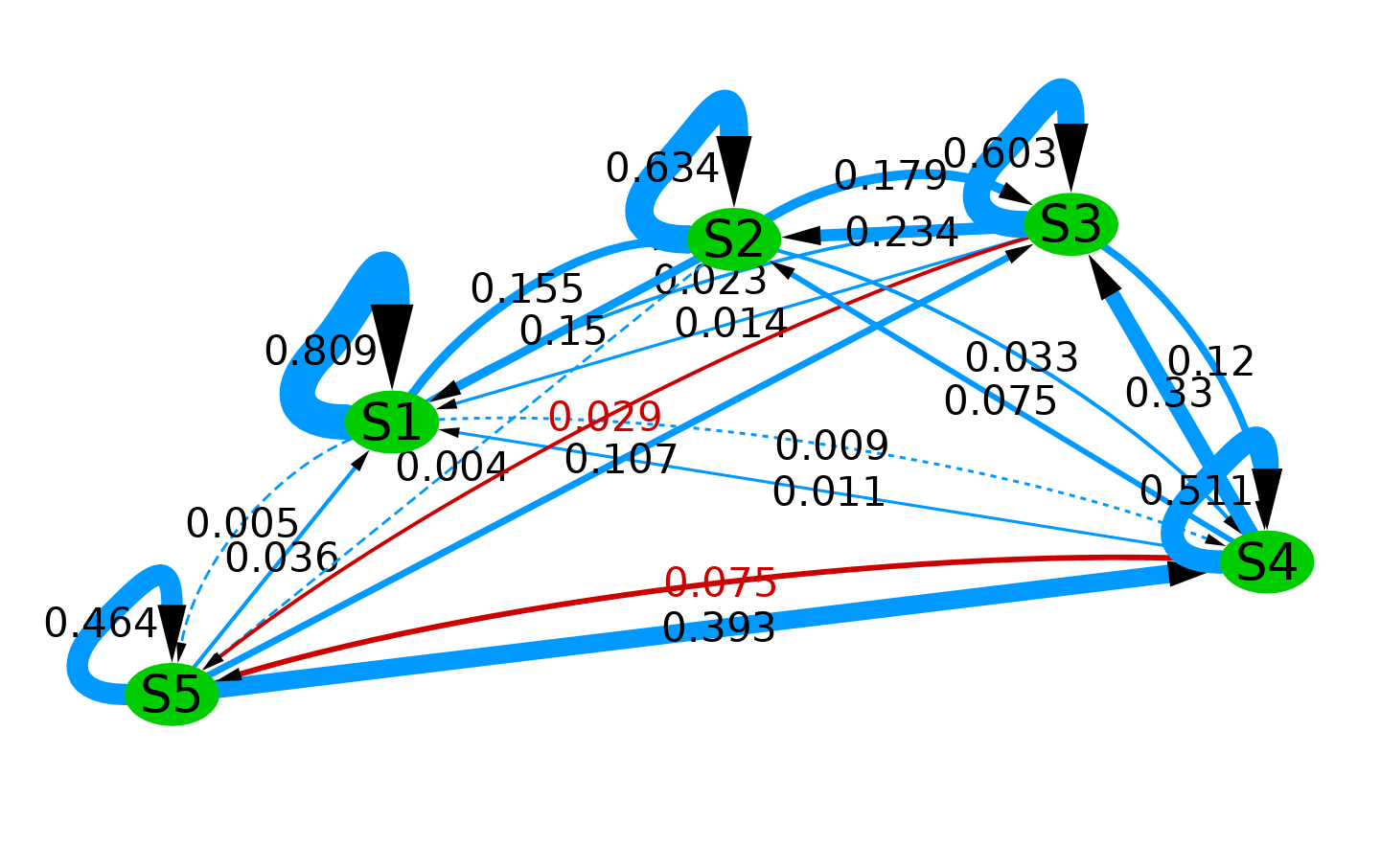}\llap{\parbox[b]{2.4in}{(\textbf{d})\\\rule{0ex}{1.6in}}}
\caption{\textbf{Transition probabilities of market states and determination of long-term precursors of critical states.} 
(a) and (b) $3D$ bar plots of co-occurrence frequencies of paired market states (MS) for USA and JPN, respectively. (c) and (d) represent the networks of transition probabilities between different states for USA and JPN, respectively. See Tables~\ref{table:USA_freq}) and \ref{table:JPN_freq} for the transition probabilities. The probability of the co-occurrence of paired market states ($S3,S4$) of USA is $6\%$, which indicates that state $S3$ of USA acts like a precursor to the market state $S4$ (crashes); similarly, for JPN, the probability of co-occurrence of paired critical market states ($S4,S5$) of JPN is about $8\%$, which indicates that the $S4$ state of JPN acts like a precursor to the critical state $S5$ (crashes).}
\label{fig:dynamics2}
\end{figure}

\begin{table}[]
\centering
\begin{tabular}{|l|l|l|l|l|}
\hline
\begin{tabular}[c]{@{}l@{}}$2^{nd}$ MS $\rightarrow$ \\ $1^{st}$ MS         \\ $\downarrow$\end{tabular} & S1                                                     & S2                                                    & S3                                                    & S4                                                   \\ \hline
S1                                                                     & \begin{tabular}[c]{@{}l@{}}0.869\end{tabular} & \begin{tabular}[c]{@{}l@{}}0.112\end{tabular} & \begin{tabular}[c]{@{}l@{}}0.017\end{tabular}  & \begin{tabular}[c]{@{}l@{}}0.002\end{tabular}  \\ \hline
S2                                                                     & \begin{tabular}[c]{@{}l@{}}0.221\end{tabular}  & \begin{tabular}[c]{@{}l@{}}0.623\end{tabular} & \begin{tabular}[c]{@{}l@{}}0.152\end{tabular} & \begin{tabular}[c]{@{}l@{}}0.004\end{tabular} \\ \hline
S3                                                                     & \begin{tabular}[c]{@{}l@{}}0.033\end{tabular}   & \begin{tabular}[c]{@{}l@{}}0.333\end{tabular} & \begin{tabular}[c]{@{}l@{}}0.575\end{tabular}  & \begin{tabular}[c]{@{}l@{}}0.058\end{tabular} \\ \hline
S4                                                                     &  \begin{tabular}[c]{@{}l@{}}0\end{tabular}  & \begin{tabular}[c]{@{}l@{}}0\end{tabular}  & \begin{tabular}[c]{@{}l@{}}0.273 	\end{tabular} & \begin{tabular}[c]{@{}l@{}}0.727\end{tabular}\\ \hline
\end{tabular}
\caption{ USA:  Co-occurrence probability of four market states (MS) (first is followed by second).}
\label{table:USA_freq}
  \centering
\begin{tabular}{|l|l|l|l|l|l|}
\hline
\begin{tabular}[c]{@{}l@{}}$2^{nd}$ MS  $\rightarrow$  \\ $1^{st}$ MS \\ $\downarrow$\end{tabular} & S1                                                    & S2                                                    & S3                                                    & S4                                                    & S5                                                   \\ \hline
S1                                                                       & \begin{tabular}[c]{@{}l@{}}0.809\end{tabular} & \begin{tabular}[c]{@{}l@{}}0.155\end{tabular} & \begin{tabular}[c]{@{}l@{}}0.023\end{tabular}  & \begin{tabular}[c]{@{}l@{}}0.009\end{tabular}  & \begin{tabular}[c]{@{}l@{}}0.005\end{tabular} \\ \hline
S2                                                                       & \begin{tabular}[c]{@{}l@{}}0.150\end{tabular} & \begin{tabular}[c]{@{}l@{}}0.634\end{tabular} & \begin{tabular}[c]{@{}l@{}}0.179\end{tabular} & \begin{tabular}[c]{@{}l@{}}0.033\end{tabular} & \begin{tabular}[c]{@{}l@{}}0.004\end{tabular} \\ \hline
S3                                                                       & \begin{tabular}[c]{@{}l@{}}0.014\end{tabular}  & \begin{tabular}[c]{@{}l@{}}0.234\end{tabular} & \begin{tabular}[c]{@{}l@{}}0.603\end{tabular} & \begin{tabular}[c]{@{}l@{}}0.120\end{tabular} & \begin{tabular}[c]{@{}l@{}}0.029\end{tabular} \\ \hline
S4                                                                       & \begin{tabular}[c]{@{}l@{}}0.011\end{tabular}  & \begin{tabular}[c]{@{}l@{}}0.075\end{tabular} & \begin{tabular}[c]{@{}l@{}}0.330\end{tabular} & \begin{tabular}[c]{@{}l@{}}0.511\end{tabular} & \begin{tabular}[c]{@{}l@{}}0.075\end{tabular} \\ \hline
S5                                                                       &  \begin{tabular}[c]{@{}l@{}}0.036\end{tabular} &  \begin{tabular}[c]{@{}l@{}}0\end{tabular} & \begin{tabular}[c]{@{}l@{}}0.107\end{tabular}  & \begin{tabular}[c]{@{}l@{}}0.393\end{tabular}  & \begin{tabular}[c]{@{}l@{}}0.464\end{tabular} \\ \hline
\end{tabular}
\caption{JPN: Co-occurrence probability of five market states (MS) (first is followed by second).}
\label{table:JPN_freq}

\label{table:cooccurrence}
\end{table}

Finally, let us test the simple hypothesis whether the system jumps \textit{randomly} from state $S_i$ to $S_j$ with probabilities
$W_{ij}$ or not. Note that, if we simply look at the curves in figures \ref{fig:dynamics2} (c) and (d), it is not obvious that this is indeed the case. However, if we make this hypothesis, we can obtain expressions for the probability that the system should be in one state over long times. This follows from the general theory of Markov chains \cite{ross1996stochastic}, but for the sake of keeping the paper self-contained, 
we briefly explain the details below. 

Let $P_i(n)$ be the probability that the system be in state $i$ after $n$ steps (time-epochs). Using the definition of $W_{ij}$, as well as the assumption that the transition to $j$ depends only on the previous state via $W_{ij}$, and in no way on the previous history, we obtain
\begin{equation}
P_i(n+1)=\sum_j W_{ji}P_j(n),
\label{eq:master}
\end{equation}
where the sum is over all possible states $j$. After long times, it is plausible, and can in fact be proved rigorously, that
the probability distribution becomes independent of $n$; in other words, the distribution reaches an equilibrium 
state $P_i^{(0)}$. The latter then satisfies the equations
\begin{equation}
P_i^{(0)}=\sum_j W_{ji}P_j^{(0)}.
\label{eq:mastereq}
\end{equation}
This can be solved explicitly, if $W_{ij}$ is known. The solution can be proved to be always positive, and can always be normalized such that
\begin{equation}
\sum_iP_i^{(0)}=1,
\end{equation}
so that the numbers $P_i^{(0)}$ can indeed be interpreted as a set of probabilities. 

In the cases where the $W_{ij}$'s are given by Table~\ref{table:USA_freq} (for the USA) or Table~\ref{table:JPN_freq} (for JPN),
it is straightforward to compute the equilibrium distributions: for the USA, one finds:
\begin{equation}
P_1^{(0)}=0.523\qquad P_2^{(0)}=0.288\qquad P_3^{(0)}=0.149\qquad P_4^{(0)}=0.040.
\label{usa_freq}
\end{equation}
For JPN, on the other hand:
\begin{eqnarray}
P_1^{(0)}&=&0.274\qquad P_2^{(0)}=0.308\qquad P_3^{(0)}=0.263\nonumber\\
P_4^{(0)}&=&0.119\qquad P_5^{(0)}=0.036 .
\label{jpn_freq}
\end{eqnarray}
The actual frequencies for the four characteristic market states $S1$, $S2$, $S3$, and $S4$ of USA, obtained from figure~\ref{fig:dynamics1}(a), enable us to compute the probabilities: $0.523, 0.287, 0.149$, and $0.041$, respectively. Similarly, actual frequencies for the five characteristic market states $S1$, $S2$, $S3$, $S4$ and $S5$ of JPN, obtained from figure~\ref{fig:dynamics1}(c), enable us to compute the probabilities: $0.277, 0.308, 0.262, 0.118$ and $0.035$, respectively. These probabilities are indeed very close to  those in Eqs.~\ref{usa_freq} and \ref{jpn_freq}, and therefore our hypothesis is correct.\\

\section{Summary and concluding remarks}

In summary, we have studied the identification of market states and long-term precursors to critical states (crashes) in financial markets, based on the probabilistic occurrences of correlation patterns, determined using noise-suppressed short-time correlation matrices. We analyzed and compared the data of the S\&P 500 (USA) and Nikkei 225 (JPN) stock markets over a 32-year period. We used the power mapping method to reduce the noise of the singular correlation matrices and obtained distinct and denser clusters in the two/three dimensional MDS maps. The effects are prominent also on the similarity matrices and the corresponding MDS maps. The evolution of the market can be followed by the dynamics transitions between the market states.  Using multidimensional scaling maps, we  applied $k$-means clustering to divide the clusters of similar correlation patterns of different time-epochs into $k$ groups or market states. We showed that based on the cluster radii we could have a fairly robust determination of the optimal number of clusters. In each market, the value of optimal number of clusters  was chosen by keeping the standard deviation of the intra-cluster distance `minimum' and  number of clusters `highest'. Thus, based on the modified prescription of finding similar clusters of correlation patterns, we characterized USA by four market states and JPN by five.  One must mention that this method yields the correlation frames that correspond to the critical states (or crashes). We have verified that these indeed correspond to the well-known financial market crashes; also,  specifically studied  the properties of the emerging spectrum and characterization of the critical states (catastrophic instabilities) in Refs.\cite{Chakraborti_2018,Pharasi_2018}.
We also analyzed the co-occurrence probabilities of the paired  market states. We observed that the probability of remaining in the same state is much higher than the transition to a different state. It implies that market states also feel an ``inertia'' -- stay in the same states for a long time. Also, probable transitions are the nearest neighbor transitions and from the co-occurrence table we showed that the probability reduces very fast if one moved away from the diagonal. Hence, the transitions to other states mainly occurred in immediately adjacent states with a few rare intermittent transitions to the remote states. The state adjacent to the critical state (crash) behaved like a long-term precursor for the critical state, and this prescription could be helpful in constructing an early warning system for financial market crashes.

\section*{Acknowledgments}

A.C. and K.S. acknowledge the support by grant number BT/BI/03/004/2003(C) of Govt. of India, Ministry of Science and Technology, Department of Biotechnology, Bioinformatics division, University of Potential Excellence-II grant (Project ID-47) of JNU, New Delhi, and the DST-PURSE grant given to JNU by the Department of Science and Technology, Government of India.
K.S. acknowledges the University Grants Commission (Ministry of Human Research Development, Govt. of India) for her senior research fellowship. H.K.P. and R.C. are grateful for postdoctoral fellowships provided by UNAM-DGAPA. F.L. acknowledges support from the project UNAM-DGAPA-PAPIIT IN103017 and CONACyT CB-254515. A.C., K.S. and T.H.S. acknowledge the support grant by CONACyT through Project FRONTERAS 201, and also support from the project UNAM-DGAPA-PAPIIT IG 100616. 
\section*{References}
\bibliographystyle{iopart-num}
\bibliography{Market_States_NJP}

\providecommand{\newblock}{}
\begin{thebibliography}{10}
\expandafter\ifx\csname url\endcsname\relax
  \def\url#1{{\tt #1}}\fi
\expandafter\ifx\csname urlprefix\endcsname\relax\def\urlprefix{URL }\fi
\providecommand{\eprint}[2][]{\url{#2}}

\bibitem{Vemuri_1978}
Vemuri V 1978 {\em Modeling of Complex Systems: An Introduction\/} (Academic
  Press, New York)

\bibitem{Gellmann_1995}
Gell-Mann M 1995 {\em Complexity\/} {\bf 1} 16--19

\bibitem{Yaneer_2002}
Bar-Yam Y 2002 {\em Encyclopedia of Life Support Systems (EOLSS), UNESCO, EOLSS
  Publishers, Oxford, UK\/}

\bibitem{Mantegna_2007}
Mantegna R~N and Stanley H~E 2007 {\em An introduction to econophysics:
  correlations and complexity in finance\/} (Cambridge University Press,
  Cambridge)

\bibitem{Bouchaud_2003}
Bouchaud J~P and Potters M 2003 {\em {Theory of Financial Risk and Derivative
  Pricing: from Statistical Physics to Risk Management}\/} (Cambridge
  University Press)

\bibitem{Sinha_2010}
Sinha S, Chatterjee A, Chakraborti A and Chakrabarti B~K 2010 {\em
  Econophysics: an introduction\/} (John Wiley \& Sons)

\bibitem{Chakraborti_2011a}
Chakraborti A, Muni~Toke I, Patriarca M and Abergel F 2011 {\em Quantitative
  Finance\/} {\bf 11} 991--1012

\bibitem{Chakraborti_2011b}
Chakraborti A, Muni~Toke I, Patriarca M and Abergel F 2011 {\em Quantitative
  Finance\/} {\bf 11} 1013--1041

\bibitem{Chakraborti_2015}
Chakraborti A, Challet D, Chatterjee A, Marsili M, Zhang Y~C and Chakrabarti
  B~K 2015 {\em Physics Reports\/} {\bf 552} 1--25

\bibitem{Chakraborti_2018}
Chakraborti A, Sharma K, Pharasi H~K, Das S, Chatterjee R and Seligman T~H 2018
  {\em arXiv preprint arXiv:1801.07213\/}

\bibitem{Sornette_2004}
Sornette D 2004 {\em Why Stock Markets Crash: Critical Events in Complex
  Financial Systems\/} (Princeton University Press)

\bibitem{Buchanan_2000}
Buchanan M 2000 {\em Ubiquity: Why Catastrophes Happen\/} (Three Rivers Press,
  New York)

\bibitem{Munnix_2012}
M{\"u}nnix M~C, Shimada T, Sch{\"a}fer R, Leyvraz F, Seligman T~H, Guhr T and
  Stanley H~E 2012 {\em Scientific reports\/} {\bf 2} 644

\bibitem{Desislava_2014}
Chetalova D, Schäfer R and Guhr T 2015 {\em Journal of Statistical Mechanics:
  Theory and Experiment\/} {\bf 2015} P01029

\bibitem{Pharasi_2018}
Pharasi H~K, Sharma K, Chakraborti A and Seligman T~H 2018 Complex market
  dynamics in the light of random matrix theory {\em New Perspectives and
  Challenges in Econophysics and Sociophysics\/} ed Abergel F, Chakrabarti B,
  Chakraborti A, Deo N and Sharma K (Springer New Economic Windows)

\bibitem{Schafer_2013}
Sch{\"a}fer R, Seligman T~H {\em et~al.\/} 2013 {\em Physical Review E\/} {\bf
  88} 032115

\bibitem{Laloux_1999}
Laloux L, Cizeau P, Bouchaud J~P and Potters M 1999 {\em Physical review
  letters\/} {\bf 83} 1467

\bibitem{Plerou_1999}
Plerou V, Gopikrishnan P, Rosenow B, Amaral L~A~N and Stanley H~E 1999 {\em
  Physical review letters\/} {\bf 83} 1471

\bibitem{Guhr_2003}
Guhr T and K{\"a}lber B 2003 {\em Journal of Physics A: Mathematical and
  General\/} {\bf 36} 3009

\bibitem{Bouchaud_2000}
Bouchaud J~P and Potters M 2000 {\em Theory of Financial Risks\/} (Cambridge
  University Press, Cambridge)

\bibitem{Schmitt_2016}
Schmitt T~A, Sch{\"a}fer R, Wied D and Guhr T 2016 {\em Empirical Economics\/}
  {\bf 50} 1091--1109

\bibitem{vinayak_2014}
Vinayak and Seligman T~H 2014 {\em AIP Conference Proceedings\/} {\bf 1575} 196

\bibitem{borg_1997}
Borg I and Groenen P 1997 {\em Modern Multidimensional Scaling: Theory and
  Applications\/} Springer series in statistics (Springer)

\bibitem{Yahoo_finance}
 2017 Yahoo finance database.accessed on 7th july, 2017, using the r open
  source programming language and software environment for statistical
  computing and graphics \urlprefix\url{https://finance.yahoo.co.jp/}

\bibitem{Mantegna_1999}
Mantegna R~N 1999 {\em The European Physical Journal B - Condensed Matter and
  Complex Systems\/} {\bf 11} 193--197

\bibitem{Marcenko_1967}
Mar{\v{c}}enko V~A and Pastur L~A 1967 {\em Mathematics of the USSR-Sbornik\/}
  {\bf 1} 457

\bibitem{Schafer_2010}
Sch{\"a}fer R, Nilsson N~F and Guhr T 2010 {\em Quantitative Finance\/} {\bf
  10} 107--119

\bibitem{Teofilo_1985}
Gonzalez T~F 1985 {\em Theoretical Computer Science\/} {\bf 38} 293 -- 306

\bibitem{Bholowalia_2014}
Bholowalia P and Kumar A 2014 {\em International Journal of Computer
  Applications\/} {\bf 105} 17--24

\bibitem{ross1996stochastic}
Ross S~M 1996 {\em Stochastic processes\/} (Wiley, New York)

\end{thebibliography}

\setcounter{table}{0}
\setcounter{figure}{0}
\renewcommand{\thetable}{S\arabic{table}}
\renewcommand{\thefigure}{S\arabic{figure}}

\section*{Supplementary information}

\begin{figure}[h!]
\centering
\includegraphics[width=0.99\linewidth]{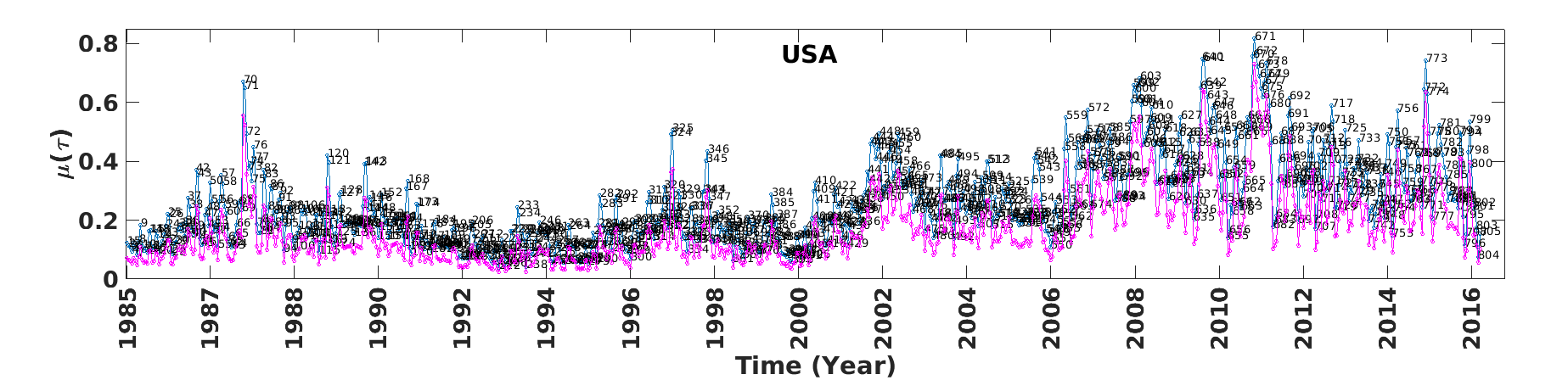}\llap{\parbox[b]{6in}{(\textbf{a})\\\rule{0ex}{1.38in}}}
\includegraphics[width=0.99\linewidth]{FigS1a}\llap{\parbox[b]{6in}{(\textbf{b})\\\rule{0ex}{1.38in}}} 
\caption{Plots of the mean correlation without noise-suppression (blue) and  with high noise-suppression of $\epsilon=0.6$ (magenta). For \textbf{(a)} USA, and \textbf{(b)} JPN. USA market was relatively calm upto $2002$ and became turbulent with high mean correlation from $2002$ onward; JPN market became turbulent $1990$ onward.}\label{fig:Mean_eps}
\end{figure}
\begin{figure}[ht]
\centering
\includegraphics[width=0.4\linewidth]{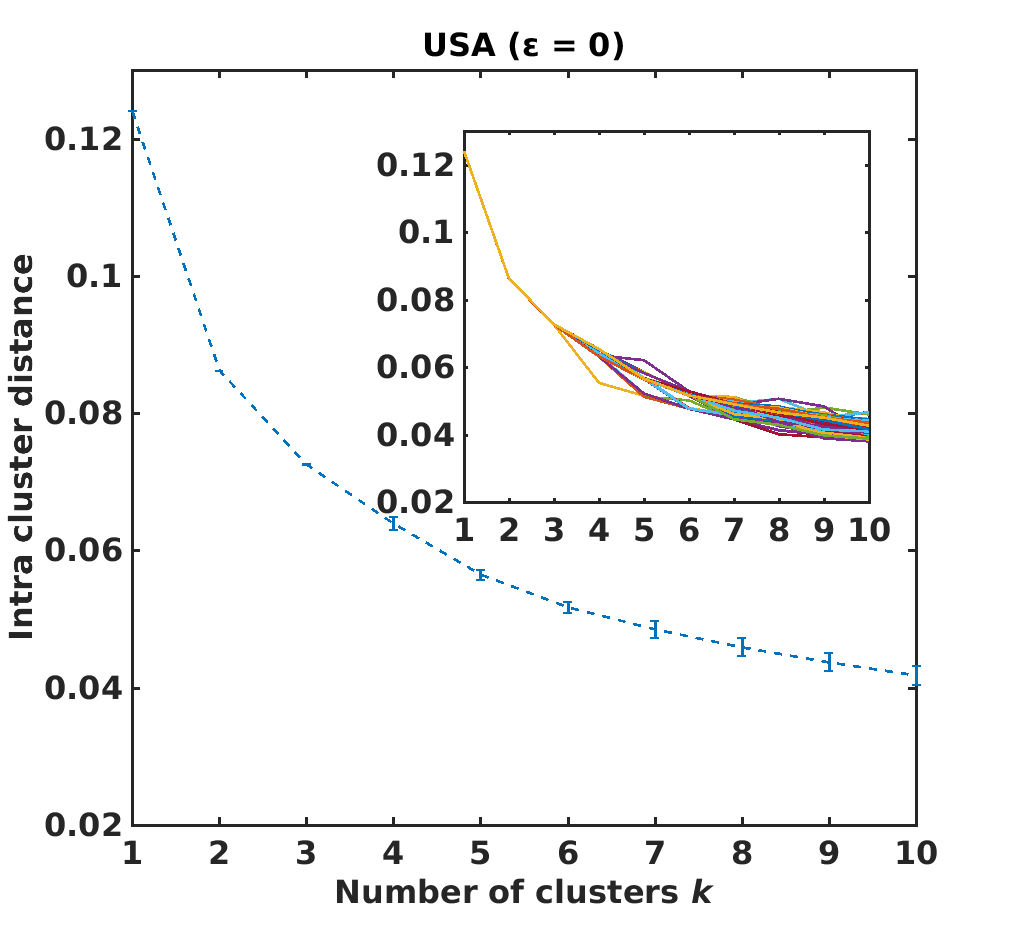} 
\includegraphics[width=0.4\linewidth]{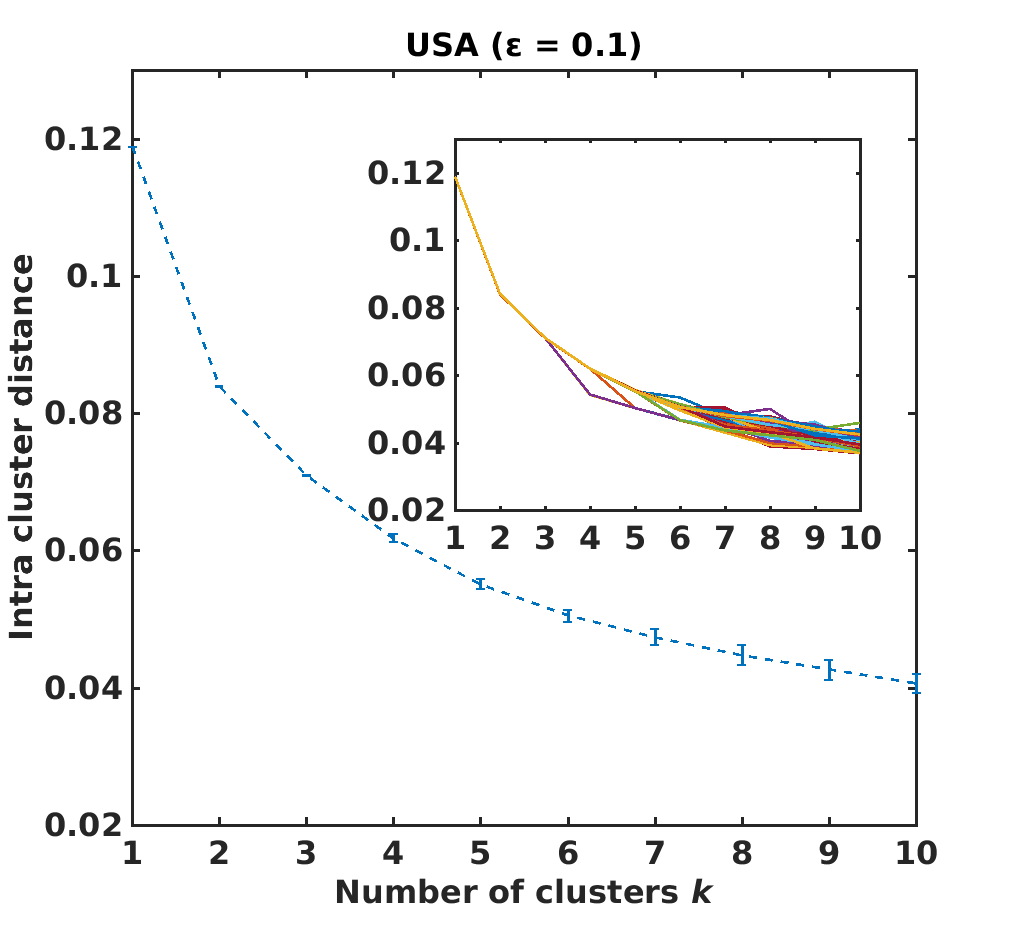} \\
\includegraphics[width=0.4\linewidth]{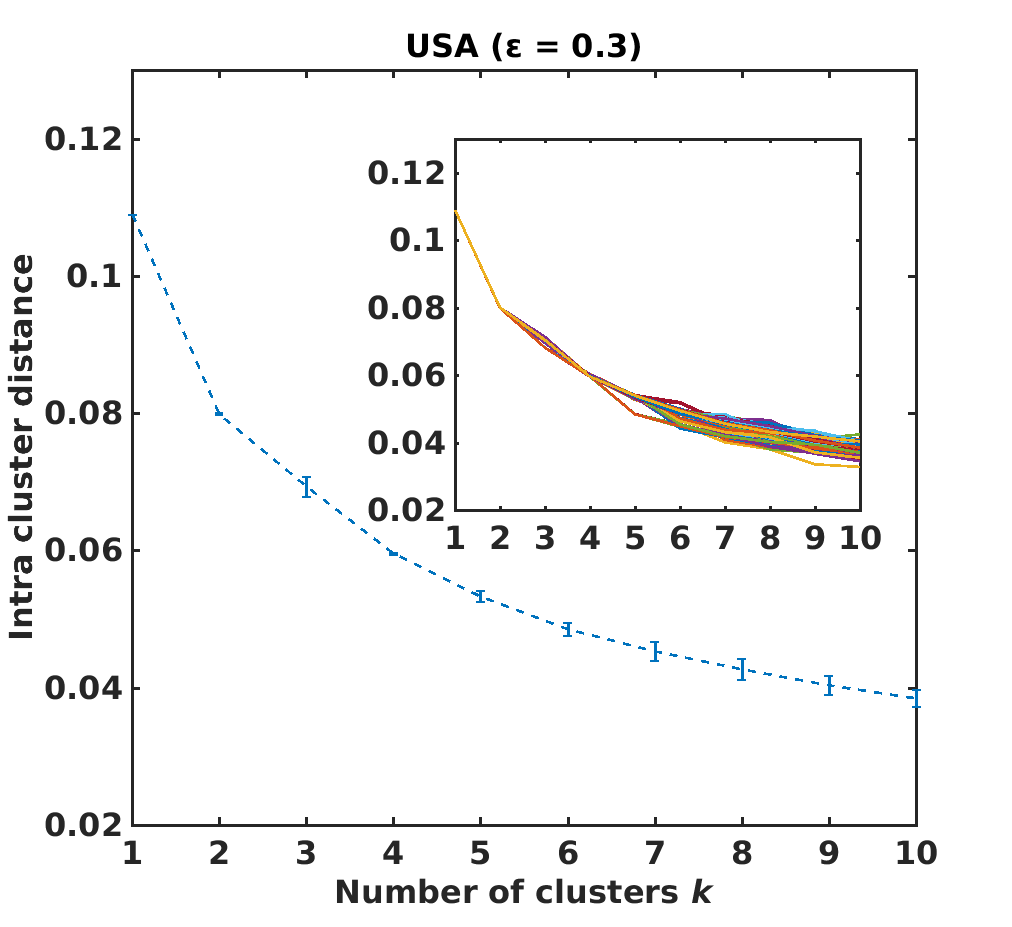} 
\includegraphics[width=0.4\linewidth]{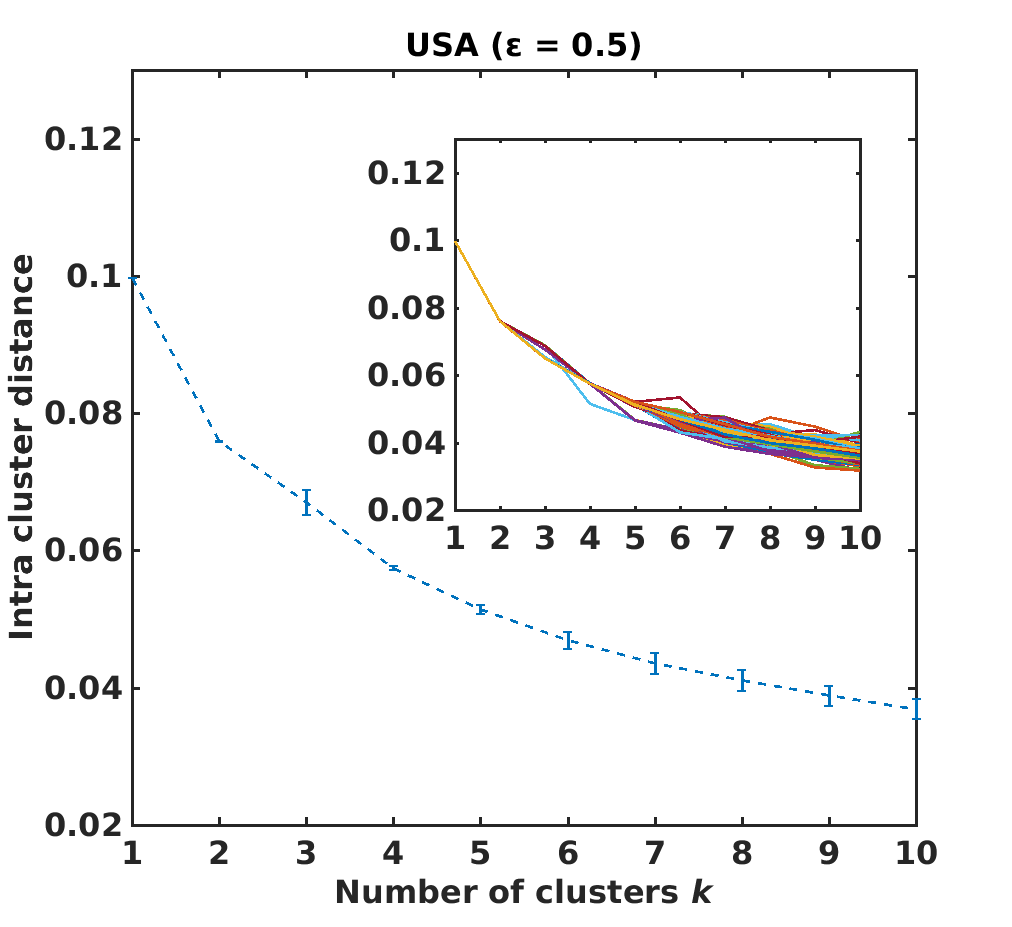} \\
\includegraphics[width=0.4\linewidth]{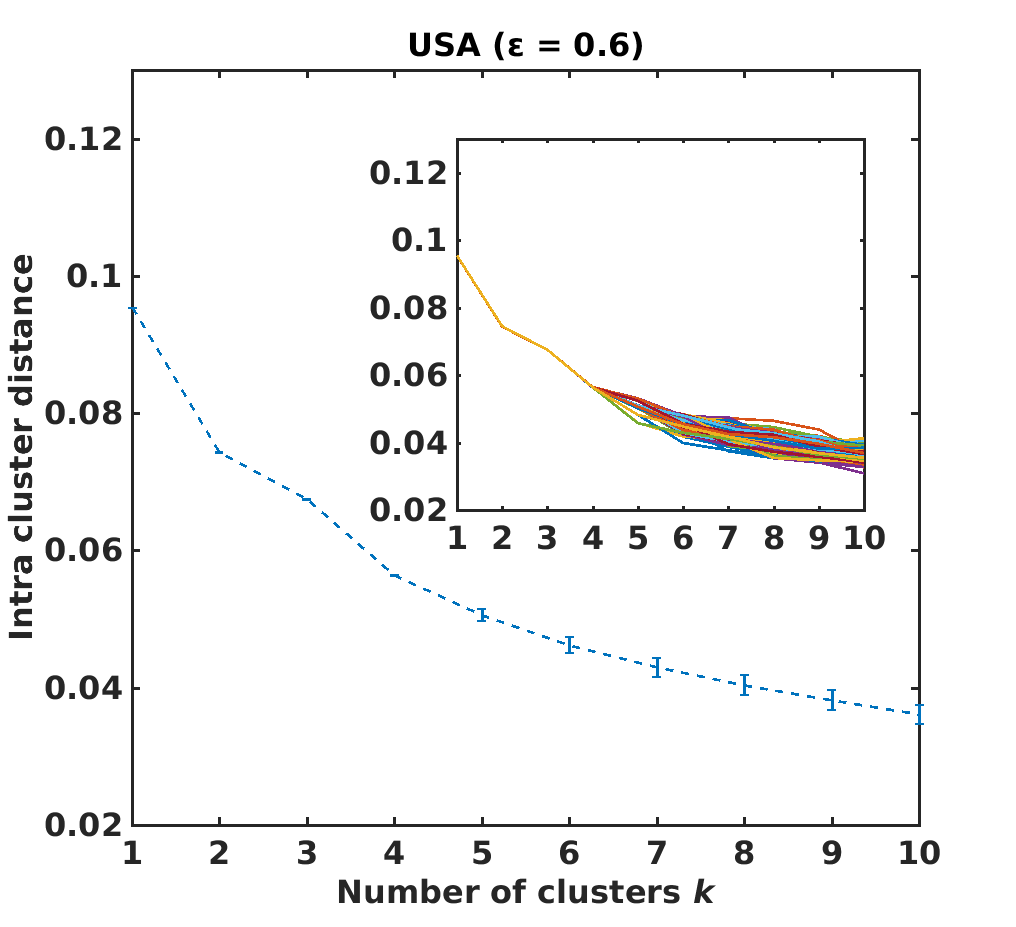} 
\includegraphics[width=0.4\linewidth]{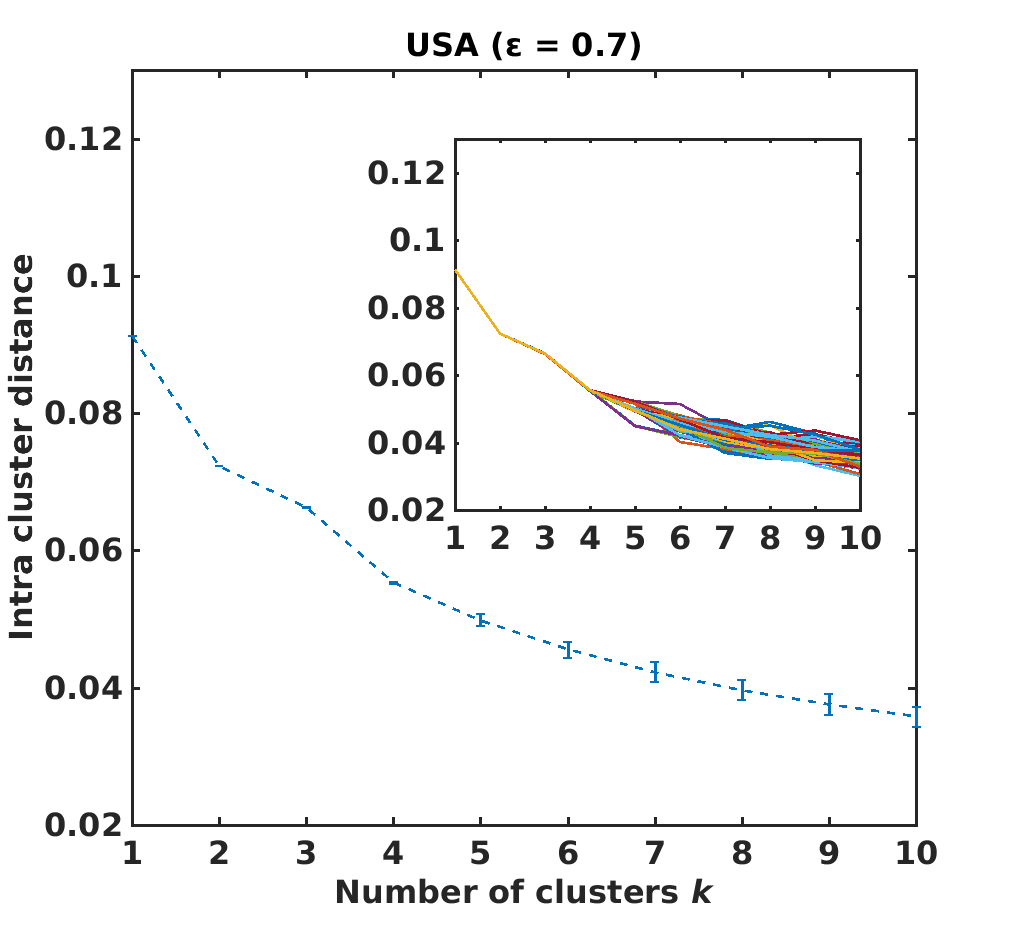} 
\caption{ Plots of the intracluster distance as a function of number of clusters $k$ of USA, for different value of noise-suppression parameter $0.1 \leqslant \epsilon \leqslant 0.7$ using $k$-means clustering. We used an ensemble of  $500$ random generated seeds for analyzing  the robustness of different clusters in the $k$-means clustering. The errorbars are the deviation of the measure of intra-cluster distances arise due to different random seeds. The points lie on the boundary of different clusters are subjected to change the association with the cluster for different initial condition to the centroids in the $k$-mean clustering. It changes the measure of intra-cluster distance among clusters. Inset shows different color lines corresponds to different seed. The value is optimized by keeping the standard deviation `lowest' and number of cluster `highest', simultaneously, for the intra-cluster distance. The results are best for $\epsilon=0.6$ and show minimum deviation for $k=4$ (max) and it grows for $k>4$. }
\label{fig:corr_noise_supp1}
\end{figure}
\begin{figure}[ht]
\centering
\includegraphics[width=0.4\linewidth]{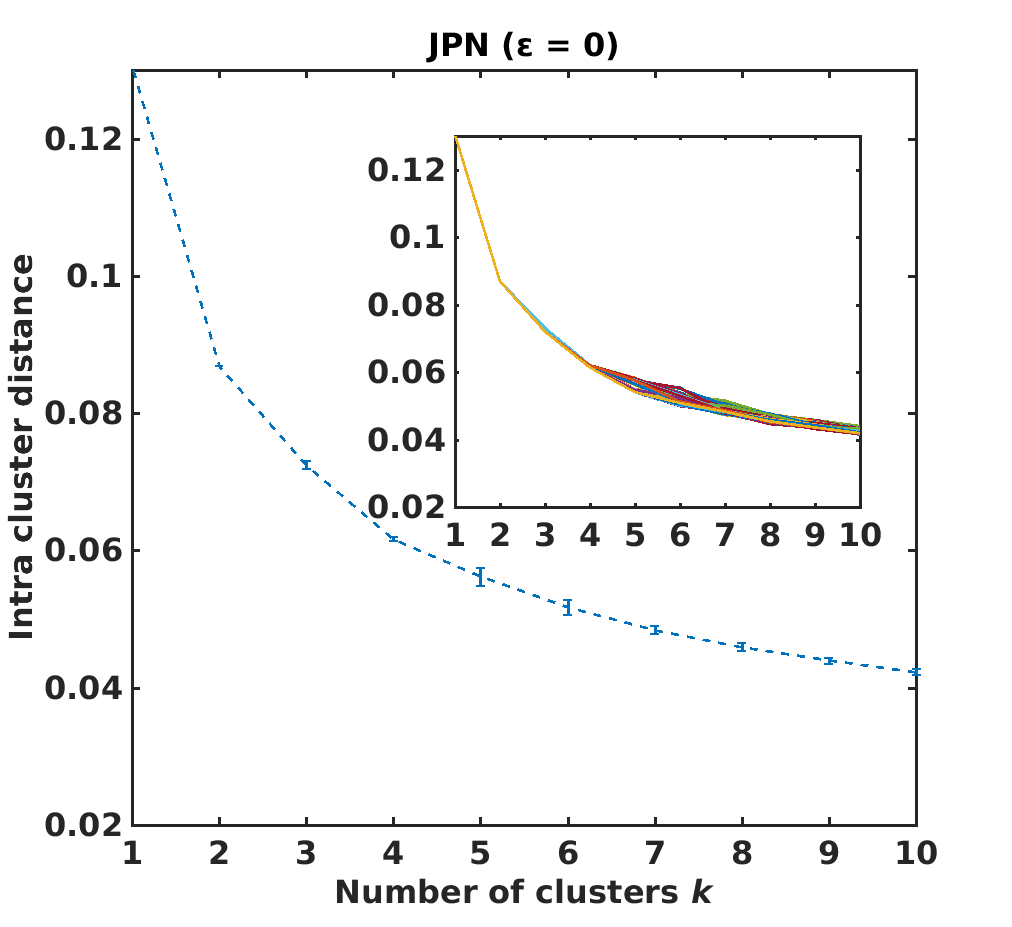} 
\includegraphics[width=0.4\linewidth]{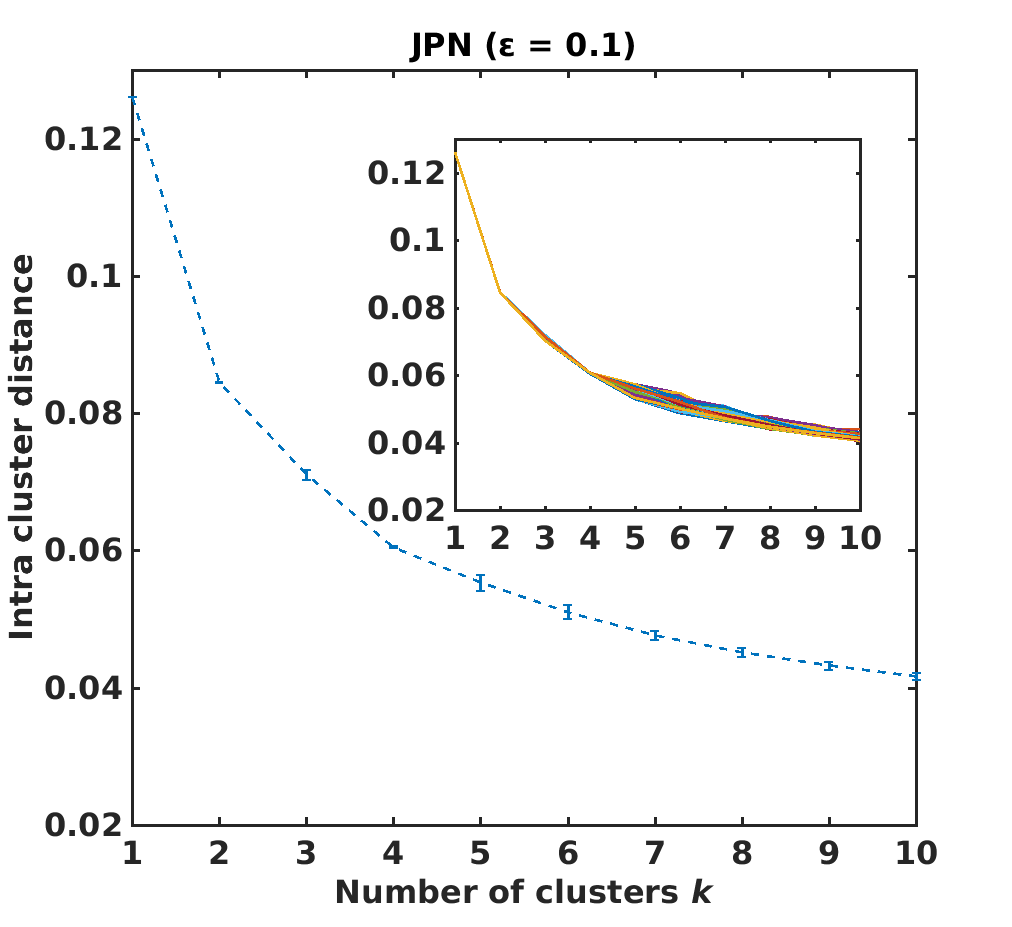} \\
\includegraphics[width=0.4\linewidth]{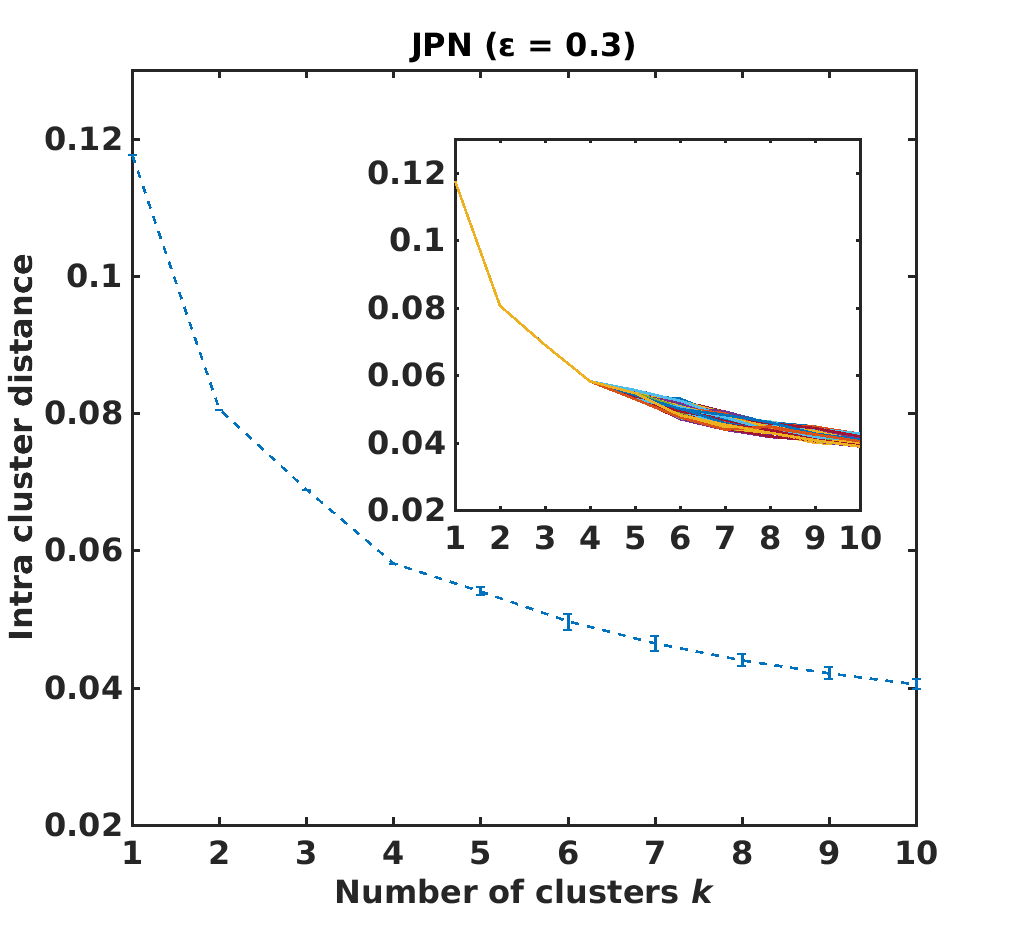} 
\includegraphics[width=0.4\linewidth]{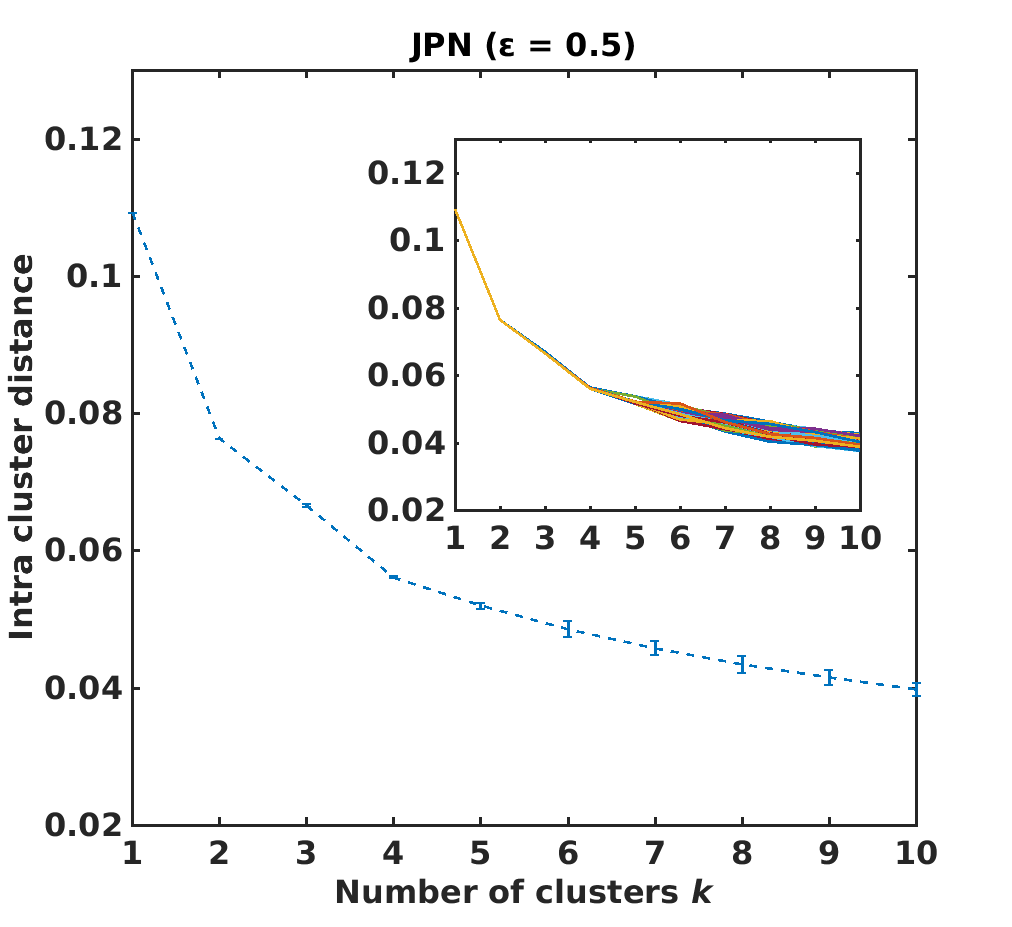} \\
\includegraphics[width=0.4\linewidth]{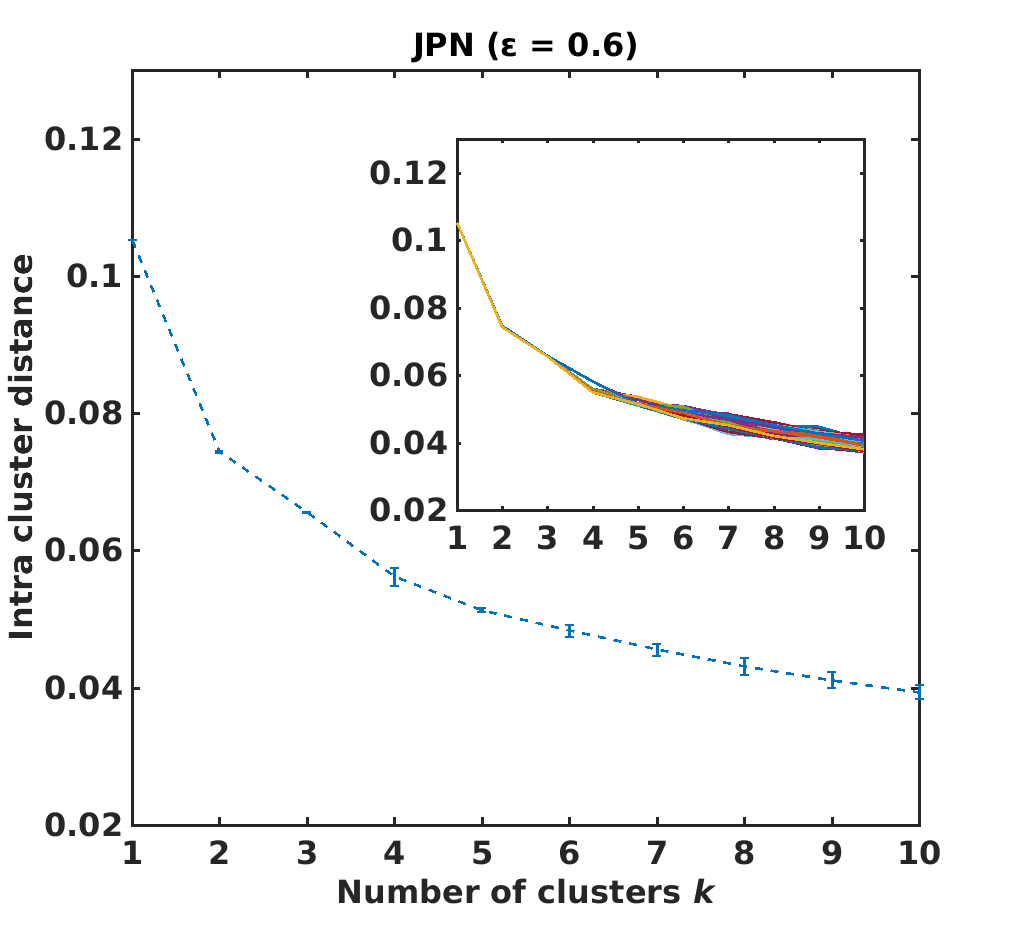} 
\includegraphics[width=0.4\linewidth]{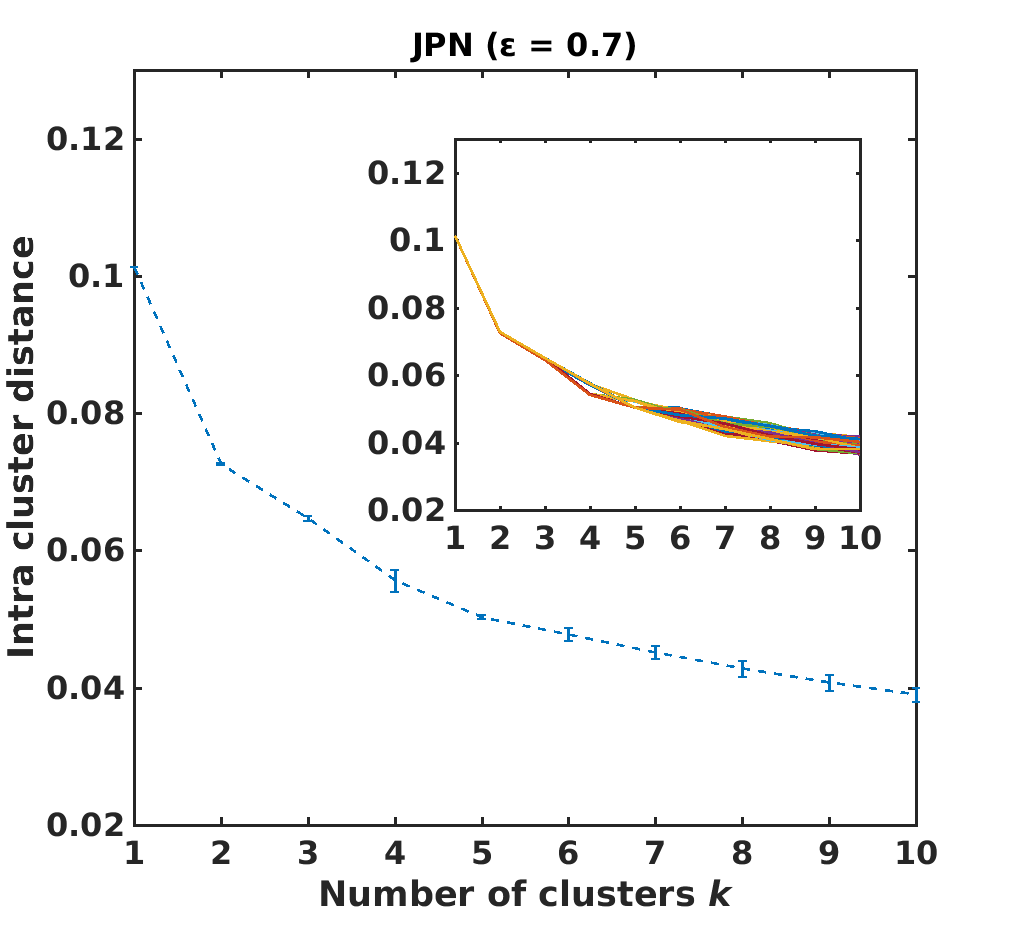} 
\caption{Plots of the intra-cluster distance as a function of number of clusters $k$ of JPN, for different value of nonlinear suppression parameter $0.1 \leqslant \epsilon \leqslant 0.7$ using $k$-means clustering. We used an ensemble of $500$ random generated seeds for analyzing  the robustness of different clusters in the $k$-means clustering. The errorbars are the deviation of the measure of intra-cluster distances arise due to different random seeds. The points lie on the boundary of different clusters are subjected to change the association with the cluster for different initial condition to the centroids in the $k$-mean clustering. It changes the measure of intra-cluster distance among clusters. Inset shows different color lines corresponds to different seed. The value is optimized by keeping the standard deviation `lowest' and number of cluster `highest', simultaneously, for the intra-cluster distance. The results are best for $\epsilon=0.6$ and show minimum deviation for $k=5$ (max) and it grows for $k>5$.}
\label{fig:corr_noise_supp2}
\end{figure}

\begin{table}[h]
\centering
\caption{List of all stocks of USA market (S\&P 500) considered for the analysis. The first column has the serial number,
the second column has the abbreviation, the third column has the full name of the stock, and the
fourth column specifies the sector as given in the S\&P 500.}
\label{USA_Table_stocks}
\begin{tabular}{|l|l|l|l|}
\hline
\textbf{S.No.} & \textbf{Code} & \textbf{Company Name}     & \textbf{Sector}        \\ \hline
1              & CMCSA         & Comcast Corp.             & Consumer Discretionary \\ \hline
2              & DIS           & Walt Disney Co.           & Consumer Discretionary \\ \hline
3              & F             & Ford Motor                & Consumer Discretionary \\ \hline
4              & GPC           & Genuine Parts             & Consumer Discretionary \\ \hline
5              & GPS           & Gap (The)                 & Consumer Discretionary \\ \hline
6              & GT            & Goodyear Tire \& Rubber   & Consumer Discretionary \\ \hline
7              & HAS           & Hasbro Inc.               & Consumer Discretionary \\ \hline
8              & HD            & Home Depot                & Consumer Discretionary \\ \hline
9              & HRB           & Block H\&R                & Consumer Discretionary \\ \hline
10             & IPG           & Interpublic Group         & Consumer Discretionary \\ \hline
11             & JCP           & Penney (J.C.)             & Consumer Discretionary \\ \hline
12             & JWN           & Nordstrom                 & Consumer Discretionary \\ \hline
13             & LEG           & Leggett \& Platt          & Consumer Discretionary \\ \hline
14             & LEN           & Lennar Corp.              & Consumer Discretionary \\ \hline
15             & LOW           & Lowe's Cos.               & Consumer Discretionary \\ \hline
16             & MAT           & Mattel Inc.               & Consumer Discretionary \\ \hline
17             & MCD           & McDonald's Corp.          & Consumer Discretionary \\ \hline
18             & NKE           & NIKE Inc.                 & Consumer Discretionary \\ \hline
19             & SHW           & Sherwin-Williams          & Consumer Discretionary \\ \hline
20             & TGT           & Target Corp.              & Consumer Discretionary \\ \hline
21             & VFC           & V.F. Corp.                & Consumer Discretionary \\ \hline
22             & WHR           & Whirlpool Corp.           & Consumer Discretionary \\ \hline
23             & ADM           & Archer-Daniels-Midland Co & Consumer Staples       \\ \hline
24             & AVP           & Avon Products             & Consumer Staples       \\ \hline
25             & CAG           & ConAgra Foods Inc.        & Consumer Staples       \\ \hline
26             & CL            & Colgate-Palmolive         & Consumer Staples       \\ \hline
27             & CPB           & Campbell Soup             & Consumer Staples       \\ \hline
28             & CVS           & CVS Caremark Corp.        & Consumer Staples       \\ \hline
29             & GIS           & General Mills             & Consumer Staples       \\ \hline
30             & HRL           & Hormel Foods Corp.        & Consumer Staples       \\ \hline
31             & HSY           & The Hershey Company       & Consumer Staples       \\ \hline
32             & K             & Kellogg Co.               & Consumer Staples       \\ \hline
33             & KMB           & Kimberly-Clark            & Consumer Staples       \\ \hline
34             & KO            & Coca Cola Co.             & Consumer Staples       \\ \hline
35             & KR            & Kroger Co.                & Consumer Staples       \\ \hline
36             & MKC           & McCormick \& Co.          & Consumer Staples       \\ \hline
\end{tabular}
\end{table}
\begin{table}[]
\centering
\begin{tabular}{|l|l|l|l|}
\hline
37 & MO   & Altria Group Inc                  & Consumer Staples \\ \hline
38 & SYY  & Sysco Corp.                       & Consumer Staples \\ \hline
39 & TAP  & Molson Coors Brewing Company      & Consumer Staples \\ \hline
40 & TSN  & Tyson Foods                       & Consumer Staples \\ \hline
41 & WMT  & Wal-Mart Stores                   & Consumer Staples \\ \hline
42 & APA  & Apache Corporation                & Energy           \\ \hline
43 & COP  & ConocoPhillips                    & Energy           \\ \hline
44 & CVX  & Chevron Corp.                     & Energy           \\ \hline
45 & ESV  & Ensco plc                         & Energy           \\ \hline
46 & HAL  & Halliburton Co.                   & Energy           \\ \hline
47 & HES  & Hess Corporation                  & Energy           \\ \hline
48 & HP   & Helmerich \& Payne                & Energy           \\ \hline
49 & MRO  & Marathon Oil Corp.                & Energy           \\ \hline
50 & MUR  & Murphy Oil                        & Energy           \\ \hline
51 & NBL  & Noble Energy Inc                  & Energy           \\ \hline
52 & NBR  & Nabors Industries Ltd.            & Energy           \\ \hline
53 & SLB  & Schlumberger Ltd.                 & Energy           \\ \hline
54 & TSO  & Tesoro Petroleum Co.              & Energy           \\ \hline
55 & VLO  & Valero Energy                     & Energy           \\ \hline
56 & WMB  & Williams Cos.                     & Energy           \\ \hline
57 & XOM  & Exxon Mobil Corp.                 & Energy           \\ \hline
58 & AFL  & AFLAC Inc                         & Financials       \\ \hline
59 & AIG  & American Intl Group Inc           & Financials       \\ \hline
60 & AON  & Aon plc                           & Financials       \\ \hline
61 & AXP  & American Express Co               & Financials       \\ \hline
62 & BAC  & Bank of America Corp              & Financials       \\ \hline
63 & BBT  & BB\&T Corporation                 & Financials       \\ \hline
64 & BEN  & Franklin Resources                & Financials       \\ \hline
65 & BK   & The Bank of New York Mellon Corp. & Financials       \\ \hline
66 & C    & Citigroup Inc.                    & Financials       \\ \hline
67 & CB   & Chubb Corp.                       & Financials       \\ \hline
68 & CINF & Cincinnati Financial              & Financials       \\ \hline
69 & CMA  & Comerica Inc.                     & Financials       \\ \hline
70 & EFX  & Equifax Inc.                      & Financials       \\ \hline
71 & FHN  & First Horizon National            & Financials       \\ \hline
72 & HBAN & Huntington Bancshares             & Financials       \\ \hline
73 & HCN  & Health Care REIT                  & Financials       \\ \hline
74 & HST  & Host Hotels \& Resorts            & Financials       \\ \hline
75 & JPM  & JPMorgan Chase \& Co.             & Financials       \\ \hline
76 & L    & Loews Corp.                       & Financials       \\ \hline
\end{tabular}
\end{table}

\begin{table}[]
\centering
\begin{tabular}{|l|l|l|l|}
\hline
77  & LM   & Legg Mason                   & Financials  \\ \hline
78  & LNC  & Lincoln National             & Financials  \\ \hline
79  & LUK  & Leucadia National Corp.      & Financials  \\ \hline
80  & MMC  & Marsh \& McLennan            & Financials  \\ \hline
81  & MTB  & M\&T Bank Corp.              & Financials  \\ \hline
82  & PSA  & Public Storage               & Financials  \\ \hline
83  & SLM  & SLM Corporation              & Financials  \\ \hline
84  & TMK  & Torchmark Corp.              & Financials  \\ \hline
85  & TRV  & The Travelers Companies Inc. & Financials  \\ \hline
86  & USB  & U.S. Bancorp                 & Financials  \\ \hline
87  & VNO  & Vornado Realty Trust         & Financials  \\ \hline
88  & WFC  & Wells Fargo                  & Financials  \\ \hline
89  & WY   & Weyerhaeuser Corp.           & Financials  \\ \hline
90  & ZION & Zions Bancorp                & Financials  \\ \hline
91  & ABT  & Abbott Laboratories          & Health Care \\ \hline
92  & AET  & Aetna Inc                    & Health Care \\ \hline
93  & AMGN & Amgen Inc                    & Health Care \\ \hline
94  & BAX  & Baxter International Inc.    & Health Care \\ \hline
95  & BCR  & Bard (C.R.) Inc.             & Health Care \\ \hline
96  & BDX  & Becton Dickinson             & Health Care \\ \hline
97  & BMY  & Bristol-Myers Squibb         & Health Care \\ \hline
98  & CAH  & Cardinal Health Inc.         & Health Care \\ \hline
99  & CI   & CIGNA Corp.                  & Health Care \\ \hline
100 & HUM  & Humana Inc.                  & Health Care \\ \hline
101 & JNJ  & Johnson \& Johnson           & Health Care \\ \hline
102 & LLY  & Lilly (Eli) \& Co.           & Health Care \\ \hline
103 & MDT  & Medtronic Inc.               & Health Care \\ \hline
104 & MRK  & Merck \& Co.                 & Health Care \\ \hline
105 & MYL  & Mylan Inc.                   & Health Care \\ \hline
106 & SYK  & Stryker Corp.                & Health Care \\ \hline
107 & THC  & Tenet Healthcare Corp.       & Health Care \\ \hline
108 & TMO  & Thermo Fisher Scientific     & Health Care \\ \hline
109 & UNH  & United Health Group Inc.     & Health Care \\ \hline
110 & VAR  & Varian Medical Systems       & Health Care \\ \hline
111 & AVY  & Avery Dennison Corp          & Industrials \\ \hline
112 & BA   & Boeing Company               & Industrials \\ \hline
113 & CAT  & Caterpillar Inc.             & Industrials \\ \hline
114 & CMI  & Cummins Inc.                 & Industrials \\ \hline
115 & CSX  & CSX Corp.                    & Industrials \\ \hline
116 & CTAS & Cintas Corporation           & Industrials \\ \hline
117 & DE   & Deere \& Co.                 & Industrials \\ \hline
\end{tabular}
\end{table}
\begin{table}[]
\centering
\begin{tabular}{|l|l|l|l|}
\hline
118 & DHR  & Danaher Corp.               & Industrials            \\ \hline
119 & DNB  & Dun \& Bradstreet           & Industrials            \\ \hline
120 & DOV  & Dover Corp.                 & Industrials            \\ \hline
121 & EMR  & Emerson Electric            & Industrials            \\ \hline
122 & ETN  & Eaton Corp.                 & Industrials            \\ \hline
123 & EXPD & Expeditors Int'l            & Industrials            \\ \hline
124 & FDX  & FedEx Corporation           & Industrials            \\ \hline
125 & FLS  & Flowserve Corporation       & Industrials            \\ \hline
126 & GD   & General Dynamics            & Industrials            \\ \hline
127 & GE   & General Electric            & Industrials            \\ \hline
128 & GLW  & Corning Inc.                & Industrials            \\ \hline
129 & GWW  & Grainger (W.W.) Inc.        & Industrials            \\ \hline
130 & HON  & Honeywell Int'l Inc.        & Industrials            \\ \hline
131 & IR   & Ingersoll-Rand PLC          & Industrials            \\ \hline
132 & ITW  & Illinois Tool Works         & Industrials            \\ \hline
133 & JEC  & Jacobs Engineering Group    & Industrials            \\ \hline
134 & LMT  & Lockheed Martin Corp.       & Industrials            \\ \hline
135 & LUV  & Southwest Airlines          & Industrials            \\ \hline
136 & MAS  & Masco Corp.                 & Industrials            \\ \hline
137 & MMM  & 3M Co.                      & Industrials            \\ \hline
138 & ROK  & Rockwell Automation Inc.    & Industrials            \\ \hline
139 & RTN  & Raytheon Co.                & Industrials            \\ \hline
140 & TXT  & Textron Inc.                & Industrials            \\ \hline
141 & UNP  & Union Pacific               & Industrials            \\ \hline
142 & UTX  & United Technologies         & Industrials            \\ \hline
143 & AAPL & Apple Inc.                  & Information Technology \\ \hline
144 & ADI  & Analog Devices Inc          & Information Technology \\ \hline
145 & ADP  & Automatic Data Processing   & Information Technology \\ \hline
146 & AMAT & Applied Materials Inc       & Information Technology \\ \hline
147 & AMD  & Advanced Micro Devices      & Information Technology \\ \hline
148 & CA   & CA, Inc.                    & Information Technology \\ \hline
149 & HPQ  & Hewlett-Packard             & Information Technology \\ \hline
150 & HRS  & Harris Corporation          & Information Technology \\ \hline
151 & IBM  & International Bus. Machines & Information Technology \\ \hline
152 & INTC & Intel Corp.                 & Information Technology \\ \hline
153 & KLAC & KLA-Tencor Corp.            & Information Technology \\ \hline
154 & LRCX & Lam Research                & Information Technology \\ \hline
155 & MSI  & Motorola Solutions Inc.     & Information Technology \\ \hline
156 & MU   & Micron Technology           & Information Technology \\ \hline
157 & TSS  & Total System Services       & Information Technology \\ \hline
158 & TXN  & Texas Instruments           & Information Technology \\ \hline
\end{tabular}
\end{table}
\begin{table}[]
\centering
\begin{tabular}{|l|l|l|l|}
\hline
159 & WDC & Western Digital                  & Information Technology      \\ \hline
160 & XRX & Xerox Corp.                      & Information Technology      \\ \hline
161 & AA  & Alcoa Inc                        & Materials                   \\ \hline
162 & APD & Air Products \& Chemicals Inc    & Materials                   \\ \hline
163 & BLL & Ball Corp                        & Materials                   \\ \hline
164 & BMS & Bemis Company                    & Materials                   \\ \hline
165 & CLF & Cliffs Natural Resources         & Materials                   \\ \hline
166 & DD  & Du Pont (E.I.)                   & Materials                   \\ \hline
167 & ECL & Ecolab Inc.                      & Materials                   \\ \hline
168 & FMC & FMC Corporation                  & Materials                   \\ \hline
169 & IFF & International Flav/Frag          & Materials                   \\ \hline
170 & IP  & International Paper              & Materials                   \\ \hline
171 & NEM & Newmont Mining Corp. (Hldg. Co.) & Materials                   \\ \hline
172 & PPG & PPG Industries                   & Materials                   \\ \hline
173 & VMC & Vulcan Materials                 & Materials                   \\ \hline
174 & CTL & CenturyLink Inc                  & Telecommunications Services \\ \hline
175 & FTR & Frontier Communications          & Telecommunications Services \\ \hline
176 & S   & Sprint Nextel Corp.              & Telecommunications Services \\ \hline
177 & T   & AT\&T Inc                        & Telecommunications Services \\ \hline
178 & VZ  & Verizon Communications           & Telecommunications Services \\ \hline
179 & AEP & American Electric Power          & Utilities                   \\ \hline
180 & CMS & CMS Energy                       & Utilities                   \\ \hline
181 & CNP & CenterPoint Energy               & Utilities                   \\ \hline
182 & D   & Dominion Resources               & Utilities                   \\ \hline
183 & DTE & DTE Energy Co.                   & Utilities                   \\ \hline
184 & ED  & Consolidated Edison              & Utilities                   \\ \hline
185 & EIX & Edison Int'l                     & Utilities                   \\ \hline
186 & EQT & EQT Corporation                  & Utilities                   \\ \hline
187 & ETR & Entergy Corp.                    & Utilities                   \\ \hline
188 & EXC & Exelon Corp.                     & Utilities                   \\ \hline
189 & NEE & NextEra Energy Resources         & Utilities                   \\ \hline
190 & NI  & NiSource Inc.                    & Utilities                   \\ \hline
191 & PNW & Pinnacle West Capital            & Utilities                   \\ \hline
192 & SO  & Southern Co.                     & Utilities                   \\ \hline
193 & WEC & Wisconsin Energy Corporation     & Utilities                   \\ \hline
194 & XEL & Xcel Energy Inc                  & Utilities                   \\ \hline
\end{tabular}
\end{table}
\begin{table}[]
\centering
\caption{List of all stocks of Japan market (Nikkei 225) considered for the analysis. The first column has the serial number,
the second column has the abbreviation, the third column has the full name of the stock, and the
fourth column specifies the sector as given in the Nikkei 225.}
\label{JPN_Table}
\begin{tabular}{|l|l|l|l|}
\hline
\textbf{S. No.} & \textbf{Code} & \textbf{Company Name}                    & \textbf{Sector} \\ \hline
1               & S-8801        & MITSUI FUDOSAN CO., LTD.                 & Capital Goods   \\ \hline
2               & S-8802        & MITSUBISHI ESTATE CO., LTD.              & Capital Goods   \\ \hline
3               & S-8804        & TOKYO TATEMONO CO., LTD.                 & Capital Goods   \\ \hline
4               & S-8830        & SUMITOMO REALTY \& DEVELOPMENT CO., LTD. & Capital Goods   \\ \hline
5               & S-7003        & MITSUI ENG. \& SHIPBUILD. CO., LTD.      & Capital Goods   \\ \hline
6               & S-7012        & KAWASAKI HEAVY IND., LTD.                & Capital Goods   \\ \hline
7               & S-9202        & ANA HOLDINGS INC.                        & Capital Goods   \\ \hline
8               & S-1801        & TAISEI CORP.                             & Capital Goods   \\ \hline
9               & S-1802        & OBAYASHI CORP.                           & Capital Goods   \\ \hline
10              & S-1803        & SHIMIZU CORP.                            & Capital Goods   \\ \hline
11              & S-1808        & HASEKO CORP.                             & Capital Goods   \\ \hline
12              & S-1812        & KAJIMA CORP.                             & Capital Goods   \\ \hline
13              & S-1925        & DAIWA HOUSE IND. CO., LTD.               & Capital Goods   \\ \hline
14              & S-1928        & SEKISUI HOUSE, LTD.                      & Capital Goods   \\ \hline
15              & S-1963        & JGC CORP.                                & Capital Goods   \\ \hline
16              & S-5631        & THE JAPAN STEEL WORKS, LTD.              & Capital Goods   \\ \hline
17              & S-6103        & OKUMA CORP.                              & Capital Goods   \\ \hline
18              & S-6113        & AMADA HOLDINGS CO., LTD.                 & Capital Goods   \\ \hline
19              & S-6301        & KOMATSU LTD.                             & Capital Goods   \\ \hline
20              & S-6302        & SUMITOMO HEAVY IND., LTD.                & Capital Goods   \\ \hline
21              & S-6305        & HITACHI CONST. MACH. CO., LTD.           & Capital Goods   \\ \hline
22              & S-6326        & KUBOTA CORP.                             & Capital Goods   \\ \hline
23              & S-6361        & EBARA CORP.                              & Capital Goods   \\ \hline
24              & S-6366        & CHIYODA CORP.                            & Capital Goods   \\ \hline
25              & S-6367        & DAIKIN INDUSTRIES, LTD.                  & Capital Goods   \\ \hline
26              & S-6471        & NSK LTD.                                 & Capital Goods   \\ \hline
27              & S-6472        & NTN CORP.                                & Capital Goods   \\ \hline
28              & S-6473        & JTEKT CORP.                              & Capital Goods   \\ \hline
29              & S-7004        & HITACHI ZOSEN CORP.                      & Capital Goods   \\ \hline
30              & S-7011        & MITSUBISHI HEAVY IND., LTD.              & Capital Goods   \\ \hline
31              & S-7013        & IHI CORP.                                & Capital Goods   \\ \hline
32              & S-7911        & TOPPAN PRINTING CO., LTD.                & Capital Goods   \\ \hline
33              & S-7912        & DAI NIPPON PRINTING CO., LTD.            & Capital Goods   \\ \hline
34              & S-7951        & YAMAHA CORP.                             & Capital Goods   \\ \hline
35              & S-1332        & NIPPON SUISAN KAISHA, LTD.               & Consumer Goods  \\ \hline
36              & S-2002        & NISSHIN SEIFUN GROUP INC.                & Consumer Goods  \\ \hline
\end{tabular}
\end{table}

\begin{table}[]
\centering
\begin{tabular}{|l|l|l|l|}
\hline
37 & S-2282 & NH FOODS LTD.                   & Consumer Goods \\ \hline
38 & S-2501 & SAPPORO HOLDINGS LTD.           & Consumer Goods \\ \hline
39 & S-2502 & ASAHI GROUP HOLDINGS, LTD.      & Consumer Goods \\ \hline
40 & S-2503 & KIRIN HOLDINGS CO., LTD.        & Consumer Goods \\ \hline
41 & S-2531 & TAKARA HOLDINGS INC.            & Consumer Goods \\ \hline
42 & S-2801 & KIKKOMAN CORP.                  & Consumer Goods \\ \hline
43 & S-2802 & AJINOMOTO CO., INC.             & Consumer Goods \\ \hline
44 & S-2871 & NICHIREI CORP.                  & Consumer Goods \\ \hline
45 & S-8233 & TAKASHIMAYA CO., LTD.           & Consumer Goods \\ \hline
46 & S-8252 & MARUI GROUP CO., LTD.           & Consumer Goods \\ \hline
47 & S-8267 & AEON CO., LTD.                  & Consumer Goods \\ \hline
48 & S-9602 & TOHO CO., LTD                   & Consumer Goods \\ \hline
49 & S-9681 & TOKYO DOME CORP.                & Consumer Goods \\ \hline
50 & S-9735 & SECOM CO., LTD.                 & Consumer Goods \\ \hline
51 & S-8331 & THE CHIBA BANK, LTD.            & Finance        \\ \hline
52 & S-8355 & THE SHIZUOKA BANK, LTD.         & Finance        \\ \hline
53 & S-8253 & CREDIT SAISON CO., LTD.         & Finance        \\ \hline
54 & S-8601 & DAIWA SECURITIES GROUP INC.     & Finance        \\ \hline
55 & S-8604 & NOMURA HOLDINGS, INC.           & Finance        \\ \hline
56 & S-3405 & KURARAY CO., LTD.               & Materials      \\ \hline
57 & S-3407 & ASAHI KASEI CORP.               & Materials      \\ \hline
58 & S-4004 & SHOWA DENKO K.K.                & Materials      \\ \hline
59 & S-4005 & SUMITOMO CHEMICAL CO., LTD.     & Materials      \\ \hline
60 & S-4021 & NISSAN CHEMICAL IND., LTD.      & Materials      \\ \hline
61 & S-4042 & TOSOH CORP.                     & Materials      \\ \hline
62 & S-4043 & TOKUYAMA CORP.                  & Materials      \\ \hline
63 & S-4061 & DENKA CO., LTD.                 & Materials      \\ \hline
64 & S-4063 & SHIN-ETSU CHEMICAL CO., LTD.    & Materials      \\ \hline
65 & S-4183 & MITSUI CHEMICALS, INC.          & Materials      \\ \hline
66 & S-4208 & UBE INDUSTRIES, LTD.            & Materials      \\ \hline
67 & S-4272 & NIPPON KAYAKU CO., LTD.         & Materials      \\ \hline
68 & S-4452 & KAO CORP.                       & Materials      \\ \hline
69 & S-4901 & FUJIFILM HOLDINGS CORP.         & Materials      \\ \hline
70 & S-4911 & SHISEIDO CO., LTD.              & Materials      \\ \hline
71 & S-6988 & NITTO DENKO CORP.               & Materials      \\ \hline
72 & S-5002 & SHOWA SHELL SEKIYU K.K.         & Materials      \\ \hline
73 & S-5201 & ASAHI GLASS CO., LTD.           & Materials      \\ \hline
74 & S-5202 & NIPPON SHEET GLASS CO., LTD.    & Materials      \\ \hline
75 & S-5214 & NIPPON ELECTRIC GLASS CO., LTD. & Materials      \\ \hline
76 & S-5232 & SUMITOMO OSAKA CEMENT CO., LTD. & Materials      \\ \hline
\end{tabular}
\end{table}

\begin{table}[]
\centering
\begin{tabular}{|l|l|l|l|}
\hline
77  & S-5233 & TAIHEIYO CEMENT CORP.                & Materials       \\ \hline
78  & S-5301 & TOKAI CARBON CO., LTD.               & Materials       \\ \hline
79  & S-5332 & TOTO LTD.                            & Materials       \\ \hline
80  & S-5333 & NGK INSULATORS, LTD.                 & Materials       \\ \hline
81  & S-5706 & MITSUI MINING \& SMELTING CO.        & Materials       \\ \hline
82  & S-5707 & TOHO ZINC CO., LTD.                  & Materials       \\ \hline
83  & S-5711 & MITSUBISHI MATERIALS CORP.           & Materials       \\ \hline
84  & S-5713 & SUMITOMO METAL MINING CO., LTD.      & Materials       \\ \hline
85  & S-5714 & DOWA HOLDINGS CO., LTD.              & Materials       \\ \hline
86  & S-5715 & FURUKAWA CO., LTD.                   & Materials       \\ \hline
87  & S-5801 & FURUKAWA ELECTRIC CO., LTD.          & Materials       \\ \hline
88  & S-5802 & SUMITOMO ELECTRIC IND., LTD.         & Materials       \\ \hline
89  & S-5803 & FUJIKURA LTD.                        & Materials       \\ \hline
90  & S-5901 & TOYO SEIKAN GROUP HOLDINGS, LTD.     & Materials       \\ \hline
91  & S-3865 & HOKUETSU KISHU PAPER CO., LTD.       & Materials       \\ \hline
92  & S-3861 & OJI HOLDINGS CORP.                   & Materials       \\ \hline
93  & S-5101 & THE YOKOHAMA RUBBER CO., LTD.        & Materials       \\ \hline
94  & S-5108 & BRIDGESTONE CORP.                    & Materials       \\ \hline
95  & S-5401 & NIPPON STEEL \& SUMITOMO METAL CORP. & Materials       \\ \hline
96  & S-5406 & KOBE STEEL, LTD.                     & Materials       \\ \hline
97  & S-5541 & PACIFIC METALS CO., LTD.             & Materials       \\ \hline
98  & S-3101 & TOYOBO CO., LTD.                     & Materials       \\ \hline
99  & S-3103 & UNITIKA, LTD.                        & Materials       \\ \hline
100 & S-3401 & TEIJIN LTD.                          & Materials       \\ \hline
101 & S-3402 & TORAY INDUSTRIES, INC.               & Materials       \\ \hline
102 & S-8001 & ITOCHU CORP.                         & Materials       \\ \hline
103 & S-8002 & MARUBENI CORP.                       & Materials       \\ \hline
104 & S-8015 & TOYOTA TSUSHO CORP.                  & Materials       \\ \hline
105 & S-8031 & MITSUI \& CO., LTD.                  & Materials       \\ \hline
106 & S-8053 & SUMITOMO CORP.                       & Materials       \\ \hline
107 & S-8058 & MITSUBISHI CORP.                     & Materials       \\ \hline
108 & S-4151 & KYOWA HAKKO KIRIN CO., LTD.          & Pharmaceuticals \\ \hline
109 & S-4503 & ASTELLAS PHARMA INC.                 & Pharmaceuticals \\ \hline
110 & S-4506 & SUMITOMO DAINIPPON PHARMA CO., LTD.  & Pharmaceuticals \\ \hline
111 & S-4507 & SHIONOGI \& CO., LTD.                & Pharmaceuticals \\ \hline
112 & S-4519 & CHUGAI PHARMACEUTICAL CO., LTD.      & Pharmaceuticals \\ \hline
113 & S-4523 & EISAI CO., LTD.                      & Pharmaceuticals \\ \hline
114 & S-7201 & NISSAN MOTOR CO., LTD.               & Technology      \\ \hline
115 & S-7202 & ISUZU MOTORS LTD.                    & Technology      \\ \hline
116 & S-7205 & HINO MOTORS, LTD.                    & Technology      \\ \hline
\end{tabular}
\end{table}

\begin{table}[]
\centering
\begin{tabular}{|l|l|l|l|}
\hline
117 & S-7261 & MAZDA MOTOR CORP.                        & Technology                  \\ \hline
118 & S-7267 & HONDA MOTOR CO., LTD.                    & Technology                  \\ \hline
119 & S-7270 & SUBARU CORP.                             & Technology                  \\ \hline
120 & S-7272 & YAMAHA MOTOR CO., LTD.                   & Technology                  \\ \hline
121 & S-3105 & NISSHINBO HOLDINGS INC.                  & Technology                  \\ \hline
122 & S-6479 & MINEBEA MITSUMI INC.                     & Technology                  \\ \hline
123 & S-6501 & HITACHI, LTD.                            & Technology                  \\ \hline
124 & S-6502 & TOSHIBA CORP.                            & Technology                  \\ \hline
125 & S-6503 & MITSUBISHI ELECTRIC CORP.                & Technology                  \\ \hline
126 & S-6504 & FUJI ELECTRIC CO., LTD.                  & Technology                  \\ \hline
127 & S-6506 & YASKAWA ELECTRIC CORP.                   & Technology                  \\ \hline
128 & S-6508 & MEIDENSHA CORP.                          & Technology                  \\ \hline
129 & S-6701 & NEC CORP.                                & Technology                  \\ \hline
130 & S-6702 & FUJITSU LTD.                             & Technology                  \\ \hline
131 & S-6703 & OKI ELECTRIC IND. CO., LTD.              & Technology                  \\ \hline
132 & S-6752 & PANASONIC CORP.                          & Technology                  \\ \hline
133 & S-6758 & SONY CORP.                               & Technology                  \\ \hline
134 & S-6762 & TDK CORP.                                & Technology                  \\ \hline
135 & S-6770 & ALPS ELECTRIC CO., LTD.                  & Technology                  \\ \hline
136 & S-6773 & PIONEER CORP.                            & Technology                  \\ \hline
137 & S-6841 & YOKOGAWA ELECTRIC CORP.                  & Technology                  \\ \hline
138 & S-6902 & DENSO CORP.                              & Technology                  \\ \hline
139 & S-6952 & CASIO COMPUTER CO., LTD.                 & Technology                  \\ \hline
140 & S-6954 & FANUC CORP.                              & Technology                  \\ \hline
141 & S-6971 & KYOCERA CORP.                            & Technology                  \\ \hline
142 & S-6976 & TAIYO YUDEN CO., LTD.                    & Technology                  \\ \hline
143 & S-7752 & RICOH CO., LTD.                          & Technology                  \\ \hline
144 & S-8035 & TOKYO ELECTRON LTD.                      & Technology                  \\ \hline
145 & S-4543 & TERUMO CORP.                             & Technology                  \\ \hline
146 & S-4902 & KONICA MINOLTA, INC.                     & Technology                  \\ \hline
147 & S-7731 & NIKON CORP.                              & Technology                  \\ \hline
148 & S-7733 & OLYMPUS CORP.                            & Technology                  \\ \hline
149 & S-7762 & CITIZEN WATCH CO., LTD.                  & Technology                  \\ \hline
150 & S-9501 & TOKYO ELECTRIC POWER COMPANY  & Transportation \& Utilities \\ \hline
151 & S-9502 & CHUBU ELECTRIC POWER CO., INC.           & Transportation \& Utilities \\ \hline
152 & S-9503 & THE KANSAI ELECTRIC POWER CO., INC.      & Transportation \& Utilities \\ \hline
153 & S-9531 & TOKYO GAS CO., LTD.                      & Transportation \& Utilities \\ \hline
154 & S-9532 & OSAKA GAS CO., LTD.                      & Transportation \& Utilities \\ \hline
155 & S-9062 & NIPPON EXPRESS CO., LTD.                 & Transportation \& Utilities \\ \hline
156 & S-9064 & YAMATO HOLDINGS CO., LTD.                & Transportation \& Utilities \\ \hline
\end{tabular}
\end{table}

\begin{table}[]
\centering
\begin{tabular}{|l|l|l|l|}
\hline
157 & S-9101 & NIPPON YUSEN K.K.                 & Transportation \& Utilities \\ \hline
158 & S-9104 & MITSUI O.S.K.LINES, LTD.          & Transportation \& Utilities \\ \hline
159 & S-9107 & KAWASAKI KISEN KAISHA, LTD.       & Transportation \& Utilities \\ \hline
160 & S-9001 & TOBU RAILWAY CO., LTD.            & Transportation \& Utilities \\ \hline
161 & S-9005 & TOKYU CORP.                       & Transportation \& Utilities \\ \hline
162 & S-9007 & ODAKYU ELECTRIC RAILWAY CO., LTD. & Transportation \& Utilities \\ \hline
163 & S-9008 & KEIO CORP.                        & Transportation \& Utilities \\ \hline
164 & S-9009 & KEISEI ELECTRIC RAILWAY CO., LTD. & Transportation \& Utilities \\ \hline
165 & S-9301 & MITSUBISHI LOGISTICS CORP.        & Transportation \& Utilities \\ \hline
\end{tabular}
\end{table}
\end{document}